\documentclass[11pt]{article}

\usepackage[a4paper, margin=27mm]{geometry}
\usepackage{amsmath, amssymb}
\usepackage{graphicx}
\usepackage{booktabs}
\usepackage{multirow}
\usepackage{array}
\usepackage{subcaption}
\usepackage{microtype}
\usepackage[hidelinks]{hyperref}
\usepackage{xcolor}
\usepackage{tikz}

\graphicspath{{figures/}}

\newcommand{\sweeplsd}{SweepLSD}
\newcommand{\degr}{\ensuremath{^\circ}}

\title{\textbf{\sweeplsd{}: A One-Pass, $O(\text{width})$-Memory Line Segment Detector\\ with an Integer-Only Streaming Core and a Real-Time FPGA Realization}}
\author{Yoshiyasu Shimizu\\[2pt]
  \normalsize Independent researcher%
  \\[2pt] \normalsize \texttt{yoshiyasu.shimizu.research@gmail.com}
}
\date{}

\begin{document}
\maketitle

\begin{abstract}
We present \sweeplsd{}, a line segment detector that reads the image exactly
once, top to bottom, and emits each segment within a few rows of its last
pixel passing the scan line. Every stage, including connected-component
labeling and the final line test, processes the image as a row stream:
intermediate memory is $O(\text{width})$ rather than $O(\text{pixels})$, and
the per-pixel core is integer-only. We give the first complete description of
the algorithm, designed in the author's 2014 master's thesis but never
published, together with an open-source C++17 implementation with
individually measured refinements and an FPGA realization --- held bit-exact
against the software in its hardware configuration --- detecting segments in
live 1080p30 video on 2009-era silicon without frame buffer or external
memory. On structure-rich public 4K photographs downscaled to Full-HD, one CPU
thread detects segments in $\sim$11\,ms --- $4.6\times$/$5.2\times$/$25\times$
faster than the original authors' implementations of ELSED, EDLines, and LSD ---
with the tightest frame-time distribution and the best per-segment direction
accuracy of the four detectors and, alone among them, curve rejection by
design, while trailing ELSED in F-score on synthetic ground truth. A
Manhattan-frame vanishing-point study on York Urban and NYU-VP scores every
detector under a selection/evaluation-separated best-estimator-per-detector
protocol, under which \sweeplsd{} leads on NYU-VP by $\approx$0.3\degr{} and
trails by $0.1$\degr{} on York Urban, with the fastest end-to-end pipeline of the four detectors on
both. A single-frame camera-attitude application, evaluated from 540p to 4K on
synthetic scenes with exact ground truth and on EuRoC and TUM-VI, matches the
baselines' accuracy at a fraction of their memory, and drives a 4K horizon
lock to $0.06$\degr{} median attitude error at 32\,ms median per frame. Code,
benchmarks, and evaluation harnesses are MIT-licensed at
\url{https://github.com/yosh-shimizu/sweeplsd}.
\end{abstract}

\smallskip
\noindent\textbf{Keywords}\enspace line segment detection \textperiodcentered\
real-time image processing \textperiodcentered\ single-pass streaming
\textperiodcentered\ FPGA \textperiodcentered\ embedded vision
\textperiodcentered\ low-memory

\section{Introduction}
\label{sec:intro}

Line segments are a compact, geometrically meaningful image description used in
camera calibration~\cite{caprile1990}, vanishing-point
estimation~\cite{denis2008,kluger2020}, wireframe parsing~\cite{huang2018},
SLAM~\cite{gomezojeda2019}, and industrial inspection. The dominant detectors --- LSD~\cite{vongioi2010,vongioi2012},
EDLines~\cite{akinlar2011}, and the faster ELSED~\cite{suarez2022} --- are accurate
and reasonably fast, but all hold the full image and several full-resolution
intermediates in memory (gradient-magnitude, orientation, and edge or
visited-pixel maps; Section~\ref{sec:memlat} measures the resulting
footprints) and revisit pixels many times (region growing in LSD;
edge drawing, linking, and least-squares fitting in EDLines and ELSED). This is a poor match for two settings: (i) hardware
pipelines (FPGA/ASIC), where external-memory bandwidth is the scarce resource and
a raster-streaming datapath with a few line buffers is the natural architecture,
and (ii) throughput- and latency-critical CPU pipelines, where cache locality
dominates.

\sweeplsd{} takes the opposite design point. The image is treated as a
\emph{row stream}: every stage of the pipeline --- Gaussian smoothing, gradient,
edge thinning, endpoint-candidate detection, connected-component labeling, and even the
final ``is this a line?'' test --- is expressed so that it needs only a few rows
of state. The image flows through once, top to bottom, and each segment is
finalized the moment its last pixel passes the scan line (labeling trails the
sweep by 7 rows). Memory for intermediates is $O(\text{width})$ --- every
$O(\text{width})$-memory claim in this paper refers to this detector-owned
intermediate state, exclusive of the input image itself, which a streaming
consumer never stores; the per-pixel
core is integer-only, with floating point confined to a once-per-segment
finalization step. The same property that makes the design hardware-friendly
also makes it fast on CPUs: the working set stays in L1/L2 cache, and the row
kernels are written so that mainstream optimizing compilers (GCC, Clang)
auto-vectorize them --- the implementation contains no SIMD intrinsics.

\begin{figure}[t]
  \centering
  \includegraphics[width=\linewidth]{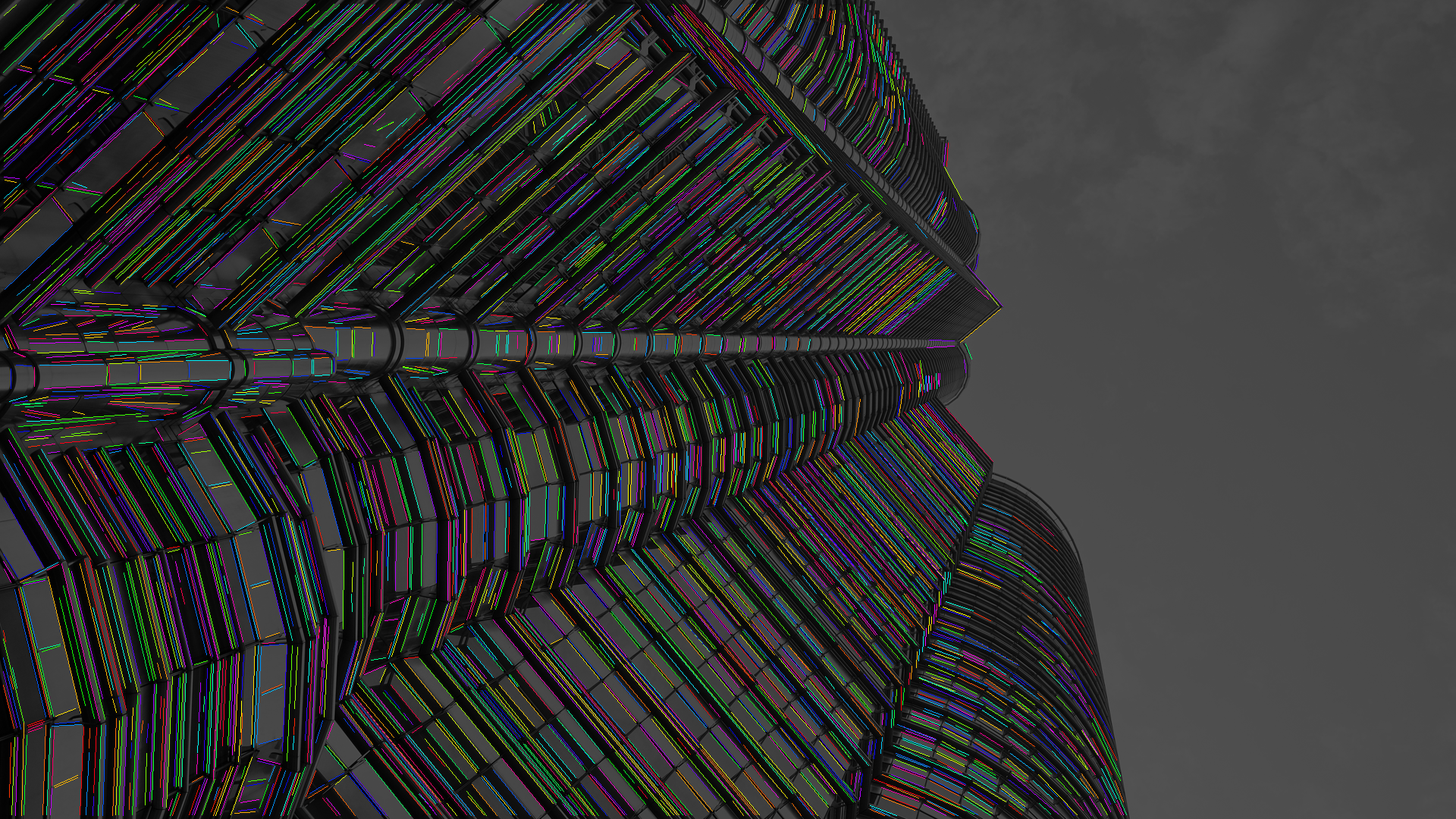}
  \caption{\sweeplsd{} on a Full-HD photograph: 3{,}110 segments in
  $\sim$12\,ms (one detector sweep, single thread, i7-8700K).}
  \label{fig:hero}
\end{figure}

The algorithm was designed in the author's 2014 master's
thesis~\cite{shimizu2014} under the name OPLSD (``one-pass LSD''). The thesis
was never published; the method was presented once, as a poster at a Japanese
workshop with no archival proceedings, and no implementation was
released. As
far as the literature is concerned, the algorithm --- including several
components that are not standard practice, such as its endpoint-candidate stage
and its contact-driven, in-stream-judging labeler --- has therefore never been
described. This paper closes that gap under the new name \sweeplsd{}, after
its sweeping raster scan (Figure~\ref{fig:hero}), and makes four
contributions:

\begin{enumerate}
\item \textbf{The first complete public description of the algorithm}
  (Section~\ref{sec:method}), normative for the accompanying open-source
  implementation (C++17, zero dependencies, MIT license). The library ships two
  drivers over the same per-pixel kernels --- a multi-pass driver that is easy
  to read and a streaming one-pass driver faithful to the hardware design
  intent --- tested to produce identical output.
\item \textbf{Measured refinements} of the 2014 design, described in place in
  Section~\ref{sec:method}: sub-pixel non-maximum
  suppression (NMS), a noise-adaptive streaming hysteresis, curve rejection,
  a strict NMS tie-break, and projection-extreme endpoints. Every refinement preserves the streaming $O(\text{width})$ property
  and was adopted only after measuring its effect --- four for a measured
  gain, while the fifth, projection-extreme endpoints, is retained as a
  $0.2$\,ms robustness guard that a dedicated junction probe shows to be
  behaviorally neutral on straight-bar scenes
  (Appendix~\ref{app:ablation}); each can be disabled in
  the library, and with all of them off the implementation reproduces the
  2014 behavior bit for bit. Appendix~\ref{app:ablation} ablates them
  individually --- cumulatively from the 2014 pipeline and leave-one-out from
  the shipped configuration --- across detection quality, per-segment
  geometry, downstream vanishing-point accuracy, and runtime.
\item \textbf{A comprehensive evaluation} (Sections~\ref{sec:eval}
  and~\ref{sec:apps}) against the original authors' implementations of LSD,
  EDLines, and ELSED: timing and frame-time dispersion, memory and emission
  latency, synthetic-ground-truth accuracy, orientation isotropy, curve
  rejection, endpoint accuracy, repeatability under flips and
  $90$\degr{}-multiple rotations, a single-frame camera-attitude study swept
  from 540p to 4K and generalized to EuRoC and TUM-VI, and a downstream
  Manhattan-frame vanishing-point study on York Urban~\cite{denis2008} and
  NYU-VP~\cite{kluger2020,silberman2012} (``NYU'' for short below). The
  vanishing-point study includes an error-source decomposition and a
  selection/evaluation-separated ``best-estimator-per-detector'' protocol, so
  that no detector is ranked under an estimator configuration chosen to suit
  another. We also state plainly
  where \sweeplsd{} loses: ELSED leads the synthetic-ground-truth F-score at
  every noise level,
  and \sweeplsd{}'s contrast-gated edge model misses soft, low-contrast
  structure that LSD and EDLines recover.
\item \textbf{A hardware realization} (Section~\ref{sec:fpga}): the detector
  as synthesizable high-level-synthesis (HLS) C++ and as hand-written
  portable Verilog, held
  \emph{bit-exact} against the software over a 123-photograph Full-HD corpus
  (in the hardware configuration, which omits the software-only sub-pixel
  NMS), and run live --- HDMI in, detect, overlay,
  HDMI out at 1080p30 --- on a 2009-era Spartan-6 with no frame buffer and no
  external memory, lossless on all but the two densest corpus frames (the
  overload limit is quantified there). The FPGA
  form the thesis aimed at, left as future work in 2014, is thereby
  demonstrated, and the streaming claims of Section~\ref{sec:onepass} are
  validated in silicon.
\end{enumerate}

\section{Related work}
\label{sec:related}

\textbf{Line segment detection.} Burns et al.~\cite{burns1986} introduced
gradient-orientation-based line-support regions. LSD~\cite{vongioi2010,vongioi2012}
grows regions of pixels sharing a gradient orientation and validates each
candidate with an a-contrario NFA (number of false alarms)
test~\cite{desolneux2000}, giving excellent
false-detection control at the cost of many passes over full-image buffers.
EDLines~\cite{akinlar2011} builds on Edge Drawing~\cite{topal2012}
(published in full the following year): it draws
connected edge chains, fits lines by least squares, and applies the same NFA
validation; it is several times faster than LSD. ELSED~\cite{suarez2022} pushed
the speed of the drawing approach further and is among the fastest
published CPU detectors; it runs in every head-to-head comparison in this
paper. The classical family is broader still --- parameter-free Canny-guided
detection (CannyLines~\cite{lu2015}), globally optimal Hough-domain dynamic
programming (MCMLSD~\cite{almazan2017}), probabilistic modeling of
rasterization units (Linelet~\cite{cho2018}), active grouping
(AG3line~\cite{zhang2021}), and multi-scale perceptual grouping
(MPG-LSD~\cite{wang2024}) --- trading speed for accuracy or false-alarm
control in varying proportions. Learned detectors form a second, fast-moving family: wireframe parsers that
regress junctions and lines end-to-end (L-CNN~\cite{zhou2019},
HAWP~\cite{xue2020,xue2023}), transformer set-prediction detectors
(LETR~\cite{xu2021}, DT-LSD~\cite{janampa2024}), M-LSD~\cite{gu2022} for
mobile devices, DeepLSD~\cite{pautrat2023}, which predicts a gradient
field with a network and hands it to a handcrafted detector, joint detection-and-description networks
(SOLD$^2$~\cite{pautrat2021}), ScaleLSD~\cite{ke2025}, trained
self-supervised on 10M unlabeled images, and MiLSD~\cite{panahi2026}, which
fits a learned detector into a sub-megabyte microcontroller budget. They
improve robustness on difficult imagery --- ScaleLSD detects more segments than
LSD itself --- but run network inference over frame-sized feature maps,
typically on a GPU or NPU. None of the methods reviewed above, classical or learned, is a
bounded-row streaming design: each operates on frame-sized images, edge maps,
or feature maps --- $O(\text{pixels})$ state revisited across stages.
\sweeplsd{} differs in kind, not degree: it is a raster-streaming design where
no stage may look more than a few rows behind the scan line --- which is also
why Canny's hysteresis~\cite{canny1986}, which is not one-pass-friendly, is
replaced by a streaming alternative (Section~\ref{sec:edge}).

\textbf{Streaming connected components.} Single-pass connected-components
analysis, in which per-component features are accumulated during one raster
scan and finalized when a component ends, is established in the FPGA
literature~\cite{bailey2007} and its architectures continue to be
refined~\cite{bailey2019}. \sweeplsd{}'s labeling stage
(Section~\ref{sec:labeling}) is of this family but differs in what it
computes: components are edge \emph{runs}, delimited by endpoint candidates,
closure is driven by endpoint contacts rather than component death, and each
closed run is judged in-stream --- its moments reduce to an eigenvalue ratio
that accepts or rejects --- and
emitted as a line segment while the scan continues.

\textbf{Hardware line detection.} Classical FPGA line extraction centers on the
Hough transform, whose voting accumulator maps naturally onto on-chip
RAM~\cite{bailey2011,lu2013} but yields infinite \emph{lines} rather than
segments, and only once the frame's votes have been accumulated. Closer to
\sweeplsd{} are FPGA detectors of \emph{line segments} that
stream-adapt LSD --- Zhou et al.~\cite{zhou2018} and, most directly, Manabe et
al.~\cite{manabe2022}, whose stream-based detector emits segments
frame-buffer-free on a Zynq-7000 at VGA/60\,fps. \sweeplsd{} shares that
streaming, external-memory-free stance but reaches it by a different route ---
edge runs delimited by endpoint candidates, grouped by union--find labeling,
and closed by endpoint contacts, rather than a stream-adapted region grow --- and, from a 2009-era Spartan-6,
sustains $\sim$3$\times$ the pixel rate (1080p30 / 720p60); the wider line
buffers that 1080p needs are simply the $O(\text{width})$ cost.
Segment-detection hardware is active beyond FPGAs, too: a fast segment
detector has been accelerated with approximate computing~\cite{ossimitz2021},
and --- concurrently with this paper --- Jalilvand et
al.~\cite{jalilvand2026} synthesized (45\,nm; no hardware run or software
baseline comparison reported) a fully pipelined line-segment ASIC around a
three-row register line buffer with a deterministic $5W{+}27$-cycle output
latency, corroborating from the ASIC side the design point argued here:
raster streaming, a few rows of state, in-scan emission.
\sweeplsd{} is further held \emph{bit-exact} across software, HLS, and
register-transfer-level (RTL) code
(Section~\ref{sec:fpga}) --- a contract the prior FPGA detectors do not report.
Table~\ref{tab:hw} places the three approaches side by side.

\begin{table}[t]
  \centering\footnotesize
  \caption{\sweeplsd{}'s hardware realization against representative prior FPGA
  line detectors, by design characteristic (resource counts are in
  Table~\ref{tab:fpga}). All three keep their working state on-chip; they differ
  in what they emit, and when; ``---'' = not reported. ``3-way bit-exact'' is
  decided at the emitted integer run records --- the hardware data path's
  output --- with the once-per-segment floating-point finalization shared as
  host code (Section~\ref{sec:fpga}).}
  \label{tab:hw}
  \begin{tabular}{@{}l >{\raggedright\arraybackslash}p{2.5cm} >{\raggedright\arraybackslash}p{2.3cm} >{\raggedright\arraybackslash}p{2.4cm}@{}}
    \toprule
     & \sweeplsd{} (this work) & Manabe 2022 \cite{manabe2022} & Hough/FPGA \cite{lu2013} \\
    \midrule
    Output          & finite segments            & finite segments            & infinite lines $(\rho,\theta)$ \\
    Processing      & row stream                 & row stream                 & frame accumulator \\
    Ext.\ memory    & none                       & none                       & none (on-chip) \\
    Resolution      & 1080p30 / 720p60           & VGA @60\,fps               & 1024$\times$768 \\
    Device (era)    & Spartan-6 (2009)           & Zynq-7000 (2011)           & Cyclone IV (2009) \\
    Mechanism       & endpoint runs, union--find  & stream-adapted region grow & vote accumulation \\
    3-way bit-exact & yes (SW/HLS/RTL)           & ---                        & --- \\
    \bottomrule
  \end{tabular}
\end{table}

\textbf{Evaluation methodology.} Detector papers usually score against
hand-labeled line annotations. These are incomplete and geometrically
coarse~\cite{lin2024}, so, in the spirit of downstream evaluation, we instead measure (i) strict
one-to-one matching against \emph{synthetic} ground truth, and (ii) the accuracy
of a calibrated Manhattan frame estimated from each detector's segments on York
Urban~\cite{denis2008} and NYU~\cite{kluger2020}, against rotation ground truth
(Section~\ref{sec:vp}).

\section{The \sweeplsd{} algorithm}
\label{sec:method}

This section is a complete, self-contained description of the detector. It is
normative for the open-source implementation, which follows it exactly; the
2014 thesis is cited for provenance, not required reading. One design
constraint governs everything: every stage must consume and produce \emph{row
streams} with a small fixed vertical support, and nothing may ever require
revisiting an earlier row. Default parameter values below are for 8-bit
grayscale input (the design was calibrated on $1920\times1080$ photographs);
Appendix~\ref{app:params} collects every default in one place.

The description integrates everything the shipped detector does: \sweeplsd{}
is one detector in one configuration, and every number in this paper is
measured on it. The exceptions --- an optional gap-tolerant segment linker
(off by default) and the hardware configuration of Section~\ref{sec:fpga},
which omits sub-pixel NMS --- are always named where they
occur.\footnote{Provenance, once and for the record: the pipeline is the
algorithm as designed in the 2014 thesis~\cite{shimizu2014}, refined in the
present work by a strict NMS tie-break, sub-pixel NMS, streaming hysteresis,
projection-extreme endpoints, and curve rejection --- each
adopted only after measuring its effect on synthetic-ground-truth F-score,
real-photo inspection, and the downstream vanishing-point error of
Section~\ref{sec:vp}. The library can disable each refinement individually,
and with all of them off it reproduces the 2014 behavior.}

\subsection{Pipeline overview}

\begin{figure}[t]
  \centering
  \begin{subfigure}[b]{0.32\linewidth}
    \includegraphics[width=\linewidth]{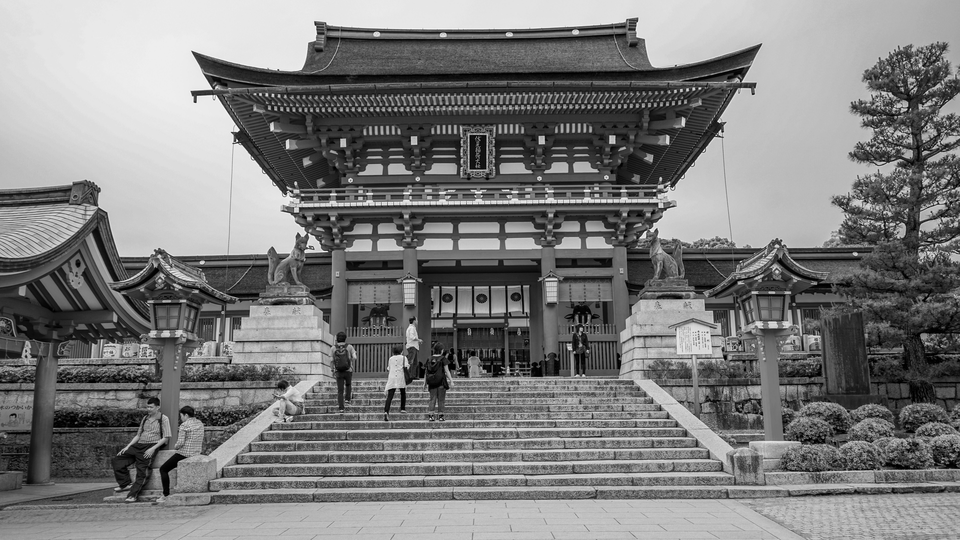}
    \caption{input (8-bit gray)}
  \end{subfigure}\hfill
  \begin{subfigure}[b]{0.32\linewidth}
    \includegraphics[width=\linewidth]{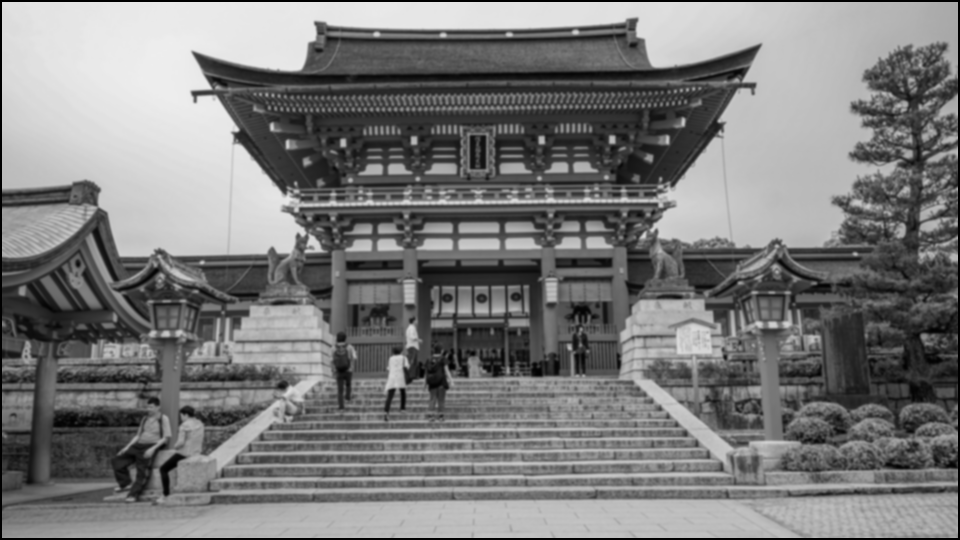}
    \caption{1. Gaussian $5\times5$, integer}
  \end{subfigure}\hfill
  \begin{subfigure}[b]{0.32\linewidth}
    \includegraphics[width=\linewidth]{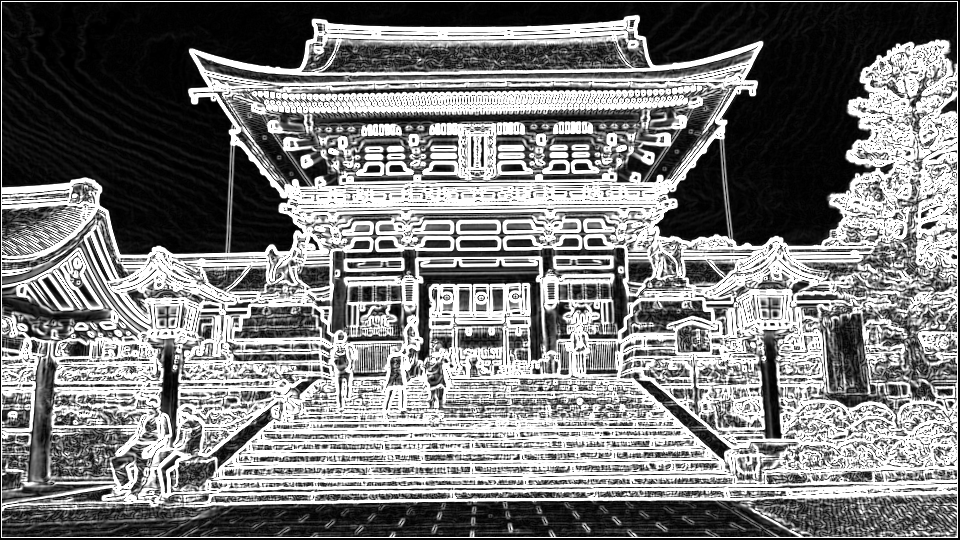}
    \caption{2. gradient power ($2\times2$)}
  \end{subfigure}\\[2pt]
  \begin{subfigure}[b]{0.32\linewidth}
    \includegraphics[width=\linewidth]{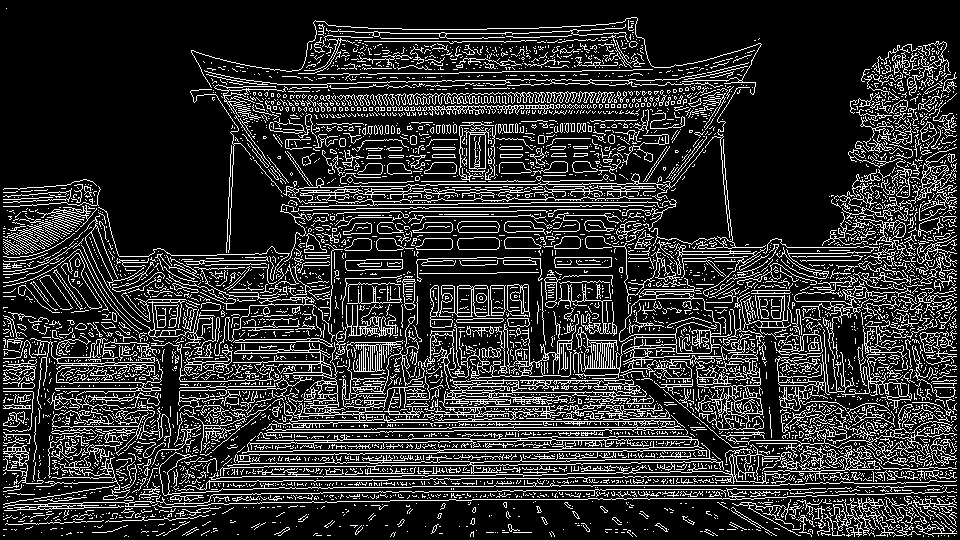}
    \caption{3. threshold + NMS}
  \end{subfigure}\hfill
  \begin{subfigure}[b]{0.32\linewidth}
    \includegraphics[width=\linewidth]{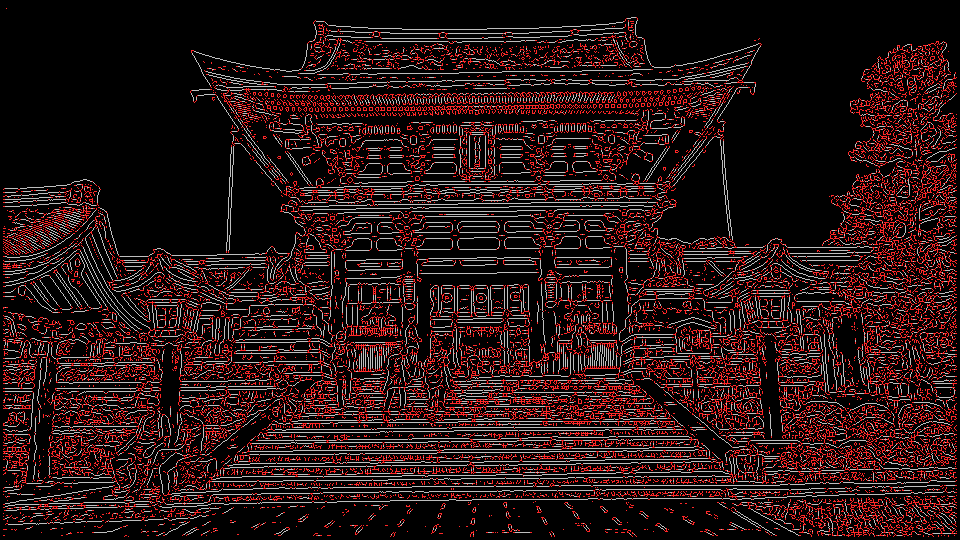}
    \caption{4. endpoint candidates (red)}
  \end{subfigure}\hfill
  \begin{subfigure}[b]{0.32\linewidth}
    \includegraphics[width=\linewidth]{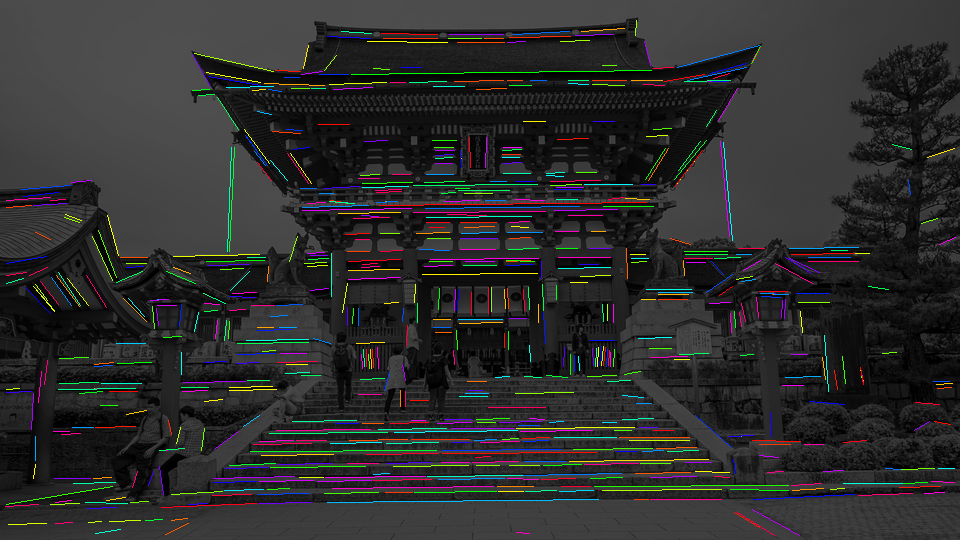}
    \caption{5. labeled + judged}
  \end{subfigure}
  \caption{The five pipeline stages on a $960\times540$ photograph; every image
  is an actual intermediate dumped from the implementation.}
  \label{fig:stages}
\end{figure}

\sweeplsd{} consists of five stages (Figure~\ref{fig:stages}): (1) integer
Gaussian smoothing, (2) a $2\times2$ gradient with two-way direction
quantization, and (3) thresholded direction-aware non-maximum suppression,
which together produce an edge map thinned to 1 pixel along the quantized
gradient axis (Section~\ref{sec:edge}); (4) an \emph{endpoint-candidate} test that marks the
pixels where a segment starts, ends, branches, or turns a sharp corner
(Section~\ref{sec:endpoints}); and (5) a streaming connected-component
labeler that groups the edge pixels between endpoint candidates, carries
scatter moments per label, and judges each closed run in-stream
(Sections~\ref{sec:labeling}--\ref{sec:judge}). Section~\ref{sec:onepass}
assembles the stages into the one-pass whole and derives its memory and
latency properties.

\subsection{Edge extraction}
\label{sec:edge}

\textbf{Gaussian smoothing.} A separable $5\times5$ blur with integer weights
$(16, 64, 96, 64, 16)$, applied vertically then horizontally with a single
final rescale by a 10-bit right shift ($\gg 10$). Since the full 2-D weight sum is $256^2 = 2^{16}$,
the $\gg 10$ leaves the smoothed image at $64\times$ the input scale, which
preserves fractional precision for the gradient stage at zero cost.

\textbf{Gradient.} A $2\times2$ operator on the smoothed image: with the
block $\{p_{00}, p_{10}; p_{01}, p_{11}\}$ ($x$ rightward, $y$ downward),
$d_x = (p_{10} + p_{11}) - (p_{00} + p_{01})$ and
$d_y = (p_{01} + p_{11}) - (p_{00} + p_{10})$. The
gradient \emph{power} is the L1 approximation $g = (|d_x| + |d_y| + 1)/2$ in
integer arithmetic (the $+1$ rounds the halving to nearest) ---
no square root --- and the direction is quantized to just two classes,
\emph{horizontal} if $|d_x| > |d_y|$, else \emph{vertical} (an exact tie
counts as vertical). Two classes are
enough because the only consumer of the direction is the non-maximum
suppression, which needs to know the dominant axis and nothing more; no
arctangent is ever computed. The isotropy cost of this coarse quantization
is measured in Section~\ref{sec:isotropy}. (The $2\times2$ stencil lives on pixel
\emph{corners}, half a pixel off the pixel-center lattice; output coordinates
account for this offset, as does canonical LSD for its own $2\times2$
gradient.)

\textbf{Thresholding and non-maximum suppression.} In the base rule, a pixel
is an edge if and only if
$g \ge g_{\mathrm{th}}$ (default $256$ at the $64\times$ scale, i.e.\ about 4
gray levels of L1 gradient) \emph{and} $g$ is a local maximum along the
quantized gradient axis --- left/right neighbors for the horizontal class,
up/down for the vertical. On exact plateau ties, keeping both competitors yields
2-pixel-thick edges and duplicate labels downstream, so
the maximum test is strict against the left (horizontal class) or upper
(vertical class) neighbor and non-strict against the other ($g > g_{-}$ and
$g \ge g_{+}$), thinning a plateau to its single leftmost or topmost pixel.
Both tests are branch-free; the output is an edge map thinned to 1 pixel
along the quantized gradient axis.

Two refinements extend this stage. First, at each surviving edge pixel a
parabolic interpolation of $g$ along the NMS axis yields a sub-pixel
offset --- the three-point vertex
$\delta = (g_{+} - g_{-}) \,/\, (2\,(2g - g_{-} - g_{+}))$ in $1/16$-px
fixed point, truncated toward zero, clamped to $\pm 1/2$\,px, and zero when
the denominator is not positive --- which the labeler
(Section~\ref{sec:labeling}) adds into its coordinate moments; lateral
endpoint error drops accordingly. Second, the threshold is
hysteretic: Canny's hysteresis is not one-pass-friendly, so instead
extraction runs at a \emph{low} threshold --- adapted to the image noise
floor with $O(1)$ state, a decayed running histogram of gradient
power\footnote{The full rule: a 64-bin power histogram (bin width 64,
saturating), decayed by $\times(1-2^{-8})$ once per row and then fed every
fourth pixel of the row; the low threshold for row $y$ is twice the
midpoint of the histogram's 80th-percentile bin, computed from rows up to
$y-2$ (one row of slack so hardware can scan the percentile), clamped to
$[120, g_{\mathrm{th}}]$; the first two rows use the clamp floor.} ---
while each edge pixel records a ``strong'' flag indicating whether it also
cleared
$g_{\mathrm{th}}$; the line judgment (Section~\ref{sec:judge}) accepts a
segment only if enough member pixels were strong. This raises recall on clean
images without flooding noisy ones. (With hysteresis active, the operative
edge rule is therefore: low threshold \emph{and} local maximum;
$g_{\mathrm{th}}$ moves to the judge as the strong-evidence gate.)

The first and last three rows and columns are cleared from the edge map: the
$2\times2$ stencil straddling the frame boundary
otherwise manufactures a spurious full-frame ring of false edges, and the
3-px band also keeps the $5\times5$ endpoint stencil of every retained pixel
inside the image.

\subsection{Endpoint candidates}
\label{sec:endpoints}

This stage has no counterpart in LSD or the Edge-Drawing family, and it is
what spares the labeler (Section~\ref{sec:labeling}) from ever tracing
chains. Its job is to turn the edge map into \emph{runs}: every pixel where a
segment could start, end, branch, or turn a corner is marked as an
\emph{endpoint candidate}. Candidates act as cut points --- they are never labeled --- so
the edge map falls apart into runs between candidates by construction. (Edge
pixels that are not candidates are called \emph{interior} below.) The
test is a pure $5\times5$ stencil, evaluated independently at every edge
pixel: no state, no walking.

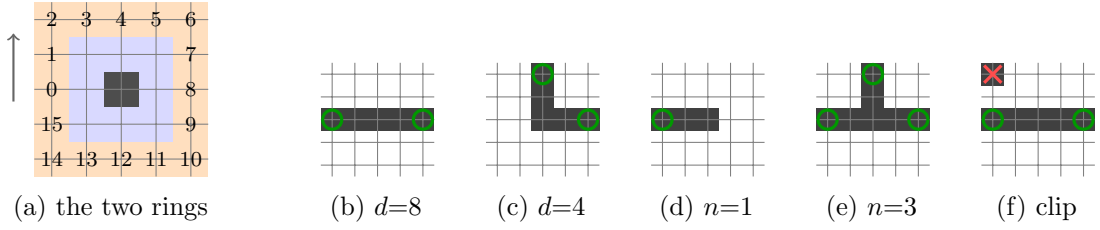
\begin{figure}[t]
  \centering
  \begin{subfigure}[b]{0.30\linewidth}
    \centering
    \begin{tikzpicture}[scale=0.46, font=\scriptsize]
      \foreach \dx in {-2,...,2} \foreach \dy in {-2,...,2} {
        \pgfmathtruncatemacro{\cheb}{max(abs(\dx),abs(\dy))}
        \ifnum\cheb=2 \fill[orange!25] (\dx-0.5,-\dy-0.5) rectangle (\dx+0.5,-\dy+0.5); \fi
        \ifnum\cheb=1 \fill[blue!15] (\dx-0.5,-\dy-0.5) rectangle (\dx+0.5,-\dy+0.5); \fi
      }
      \fill[black!70] (-0.5,-0.5) rectangle (0.5,0.5);
      \draw[black!50] (-2.5,-2.5) grid (2.5,2.5);
      \node at (-2, 0) {0};  \node at (-2, 1) {1};  \node at (-2, 2) {2};
      \node at (-1, 2) {3};  \node at ( 0, 2) {4};  \node at ( 1, 2) {5};
      \node at ( 2, 2) {6};  \node at ( 2, 1) {7};  \node at ( 2, 0) {8};
      \node at ( 2,-1) {9};  \node at ( 2,-2) {10}; \node at ( 1,-2) {11};
      \node at ( 0,-2) {12}; \node at (-1,-2) {13}; \node at (-2,-2) {14};
      \node at (-2,-1) {15};
      \draw[->, thick, black!60] (-3.1,-0.35) -- (-3.1,1.6);
    \end{tikzpicture}
    \caption{the two rings}
  \end{subfigure}\hfill
  \begin{subfigure}[b]{0.13\linewidth}
    \centering
    \begin{tikzpicture}[scale=0.30, font=\scriptsize]
      \foreach \p in {(-2,0),(-1,0),(0,0),(1,0),(2,0)}
        \fill[black!75] \p +(-0.5,-0.5) rectangle +(0.5,0.5);
      \draw[black!50] (-2.5,-2.5) grid (2.5,2.5);
      \draw[green!60!black, very thick] (-2,0) circle (0.42);
      \draw[green!60!black, very thick] (2,0) circle (0.42);
    \end{tikzpicture}
    \caption{$d{=}8$}
  \end{subfigure}\hfill
  \begin{subfigure}[b]{0.13\linewidth}
    \centering
    \begin{tikzpicture}[scale=0.30, font=\scriptsize]
      \foreach \p in {(2,0),(1,0),(0,0),(0,1),(0,2)}
        \fill[black!75] \p +(-0.5,-0.5) rectangle +(0.5,0.5);
      \draw[black!50] (-2.5,-2.5) grid (2.5,2.5);
      \draw[green!60!black, very thick] (2,0) circle (0.42);
      \draw[green!60!black, very thick] (0,2) circle (0.42);
    \end{tikzpicture}
    \caption{$d{=}4$}
  \end{subfigure}\hfill
  \begin{subfigure}[b]{0.13\linewidth}
    \centering
    \begin{tikzpicture}[scale=0.30, font=\scriptsize]
      \foreach \p in {(-2,0),(-1,0),(0,0)}
        \fill[black!75] \p +(-0.5,-0.5) rectangle +(0.5,0.5);
      \draw[black!50] (-2.5,-2.5) grid (2.5,2.5);
      \draw[green!60!black, very thick] (-2,0) circle (0.42);
    \end{tikzpicture}
    \caption{$n{=}1$}
  \end{subfigure}\hfill
  \begin{subfigure}[b]{0.13\linewidth}
    \centering
    \begin{tikzpicture}[scale=0.30, font=\scriptsize]
      \foreach \p in {(-2,0),(-1,0),(0,0),(1,0),(2,0),(0,1),(0,2)}
        \fill[black!75] \p +(-0.5,-0.5) rectangle +(0.5,0.5);
      \draw[black!50] (-2.5,-2.5) grid (2.5,2.5);
      \draw[green!60!black, very thick] (-2,0) circle (0.42);
      \draw[green!60!black, very thick] (2,0) circle (0.42);
      \draw[green!60!black, very thick] (0,2) circle (0.42);
    \end{tikzpicture}
    \caption{$n{=}3$}
  \end{subfigure}\hfill
  \begin{subfigure}[b]{0.13\linewidth}
    \centering
    \begin{tikzpicture}[scale=0.30, font=\scriptsize]
      \foreach \p in {(-2,0),(-1,0),(0,0),(1,0),(2,0)}
        \fill[black!75] \p +(-0.5,-0.5) rectangle +(0.5,0.5);
      \fill[black!75] (-2,2) +(-0.5,-0.5) rectangle +(0.5,0.5);
      \draw[black!50] (-2.5,-2.5) grid (2.5,2.5);
      \draw[green!60!black, very thick] (-2,0) circle (0.42);
      \draw[green!60!black, very thick] (2,0) circle (0.42);
      \draw[red!70, very thick] (-2.38,1.62) -- (-1.62,2.38);
      \draw[red!70, very thick] (-2.38,2.38) -- (-1.62,1.62);
    \end{tikzpicture}
    \caption{clip}
  \end{subfigure}
  \caption{The endpoint-candidate stencil. (a)~The $5\times5$ window of an
  edge pixel (dark center) read as two rings: the 8 inner pixels (blue) carry
  \emph{support}, the 16 outer pixels (orange, indexed 0--15 along the
  perimeter) are where the center's structure \emph{exits} the window; the
  arrow marks the direction of increasing index.
  (b)--(e)~Classification by surviving exits (green): (b)~a straight pass
  (two exits at ring distance 8) is the one non-candidate outcome; (c)~a
  corner (distance 4), (d)~a free end, and (e)~a branch all mark the center.
  (f)~An unrelated pixel clipping the window corner has no inner-ring
  support and is discarded before counting; the center remains interior.
  Rasterized staircases and 2-px-thick spots are likewise collapsed to one
  exit each by the thinning rule.}
  \label{fig:ring}
\end{figure}

For each edge pixel, the $5\times5$ neighborhood is read as two concentric
rings (Figure~\ref{fig:ring}a): the 8 \emph{inner} pixels and the 16
\emph{outer} perimeter pixels, indexed 0--15 along the window border. The
outer ring is where the center pixel's own structure exits the window; the
inner ring supplies the connectivity evidence. The test proceeds in three
fixed steps, all pure boolean/counting logic:
\begin{enumerate}
\item \textbf{Thinning \& support pruning.} A single fixed boolean reduction
  over the two rings (given in full in
  Appendix~\ref{app:ring}) produces two effects at once. It discards
  outer-ring pixels that are not 8-connected toward the center through the
  inner ring (\emph{support pruning}: unrelated structure that merely clips
  the window is never counted as an exit, Figure~\ref{fig:ring}f), and it
  collapses each connected arc of the remaining pixels to a single
  representative (\emph{thinning proper}: a line crossing the window counts
  once per crossing, regardless of its thickness or rasterization).
\item \textbf{Count} the surviving exits; call the count $n$.
\item \textbf{Classify.} $n = 0$ (an isolated point or tiny cluster),
  $n = 1$ (a free line end),
  and $n \ge 3$ (a branch) all mark the center as an endpoint candidate. For
  $n = 2$ the decisive quantity is the \emph{shorter arc distance between
  the two exits along the 16-position ring}, written $d$ below (at most 8):
  $d \ge 7$
  means the exits leave on roughly opposite sides --- a straight line
  passing through, the one case that is \emph{not} a candidate --- while
  $d \le 6$ means they leave on the same side, a sharp corner or a
  hairpin (Figure~\ref{fig:ring}b--e).
\end{enumerate}
The whole decision compiles to a single branch-free boolean dataflow over
8-bit flags (Appendix~\ref{app:ring}), which is why the stage auto-vectorizes
on CPUs and costs a handful of gate levels in hardware.

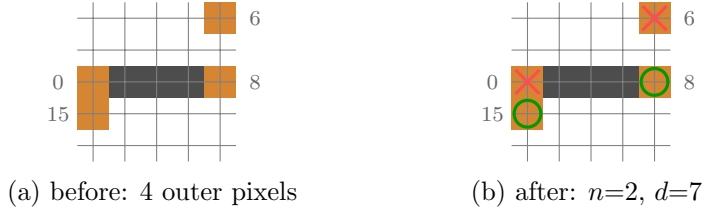
\begin{figure}[t]
  \centering
  \begin{subfigure}[b]{0.30\linewidth}
    \centering
    \begin{tikzpicture}[scale=0.42, font=\scriptsize]
      \fill[black!70] (0,0) +(-0.5,-0.5) rectangle +(0.5,0.5);
      \foreach \p in {(-1,0),(1,0)}
        \fill[black!70] \p +(-0.5,-0.5) rectangle +(0.5,0.5);
      \foreach \p in {(-2,0),(-2,-1),(2,0),(2,2)}
        \fill[orange!80!black!80] \p +(-0.5,-0.5) rectangle +(0.5,0.5);
      \draw[black!50] (-2.5,-2.5) grid (2.5,2.5);
      \node[black!60] at (-3.1, 0) {0};
      \node[black!60] at (-3.1,-1) {15};
      \node[black!60] at (3.1, 0) {8};
      \node[black!60] at (3.1, 2) {6};
    \end{tikzpicture}
    \caption{before: 4 outer pixels}
  \end{subfigure}\hspace{0.06\linewidth}
  \begin{subfigure}[b]{0.30\linewidth}
    \centering
    \begin{tikzpicture}[scale=0.42, font=\scriptsize]
      \fill[black!70] (0,0) +(-0.5,-0.5) rectangle +(0.5,0.5);
      \foreach \p in {(-1,0),(1,0)}
        \fill[black!70] \p +(-0.5,-0.5) rectangle +(0.5,0.5);
      \foreach \p in {(-2,0),(-2,-1),(2,0),(2,2)}
        \fill[orange!80!black!80] \p +(-0.5,-0.5) rectangle +(0.5,0.5);
      \draw[black!50] (-2.5,-2.5) grid (2.5,2.5);
      \draw[green!60!black, very thick] (-2,-1) circle (0.42);
      \draw[green!60!black, very thick] (2,0) circle (0.42);
      \draw[red!70, very thick] (-2.38,-0.38) -- (-1.62,0.38);
      \draw[red!70, very thick] (-2.38,0.38) -- (-1.62,-0.38);
      \draw[red!70, very thick] (1.62,1.62) -- (2.38,2.38);
      \draw[red!70, very thick] (1.62,2.38) -- (2.38,1.62);
      \node[black!60] at (-3.1, 0) {0};
      \node[black!60] at (-3.1,-1) {15};
      \node[black!60] at (3.1, 0) {8};
      \node[black!60] at (3.1, 2) {6};
    \end{tikzpicture}
    \caption{after: $n{=}2$, $d{=}7$}
  \end{subfigure}
  \caption{The reduction at work (given in full in Appendix~\ref{app:ring}).
  A straight run exits the window left with a rasterization double pixel
  ($O_0$ and $O_{15}$) and right cleanly ($O_8$), while an unrelated pixel
  clips the top-right corner ($O_6$). Read naively, four outer pixels look
  like a branch. Support pruning discards $O_6$ --- it has no inner-ring
  connection toward the center --- and thinning collapses the doubled left
  exit to a single representative, leaving $n = 2$ survivors at ring distance
  $\min(7, 9) = 7 \ge 7$: a straight line passing through, correctly not a
  candidate.}
  \label{fig:ringreduction}
\end{figure}

A worked example (Figure~\ref{fig:ringreduction}) exercises the whole test at
once: support pruning removes an unrelated corner clip, thinning collapses a
two-pixel-thick rasterized exit to a single representative, and the straight
crossing is counted once --- where a naive tally of occupied outer pixels
would have read a branch.

\begin{figure}[t]
  \centering
  \includegraphics[width=0.62\linewidth]{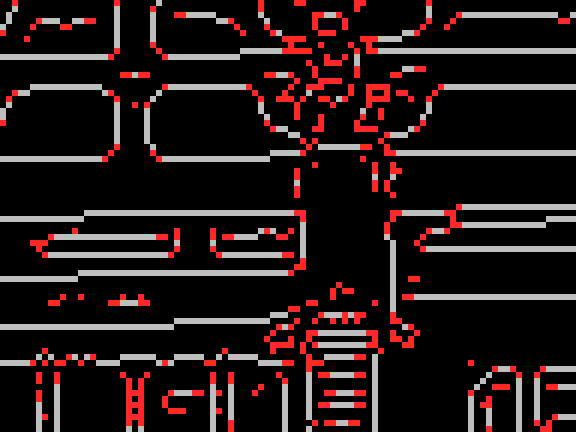}
  \caption{The candidate map on real structure ($6\times$ magnified crop of
  the stage-4 intermediate of Figure~\ref{fig:stages}): interior run pixels
  in gray, endpoint candidates in red. Long clean edges carry no candidates;
  red appears exactly at junctions, corners, and free ends --- the places
  where runs must be cut --- and each candidate cluster will cut the runs
  on either side of it.}
  \label{fig:featzoom}
\end{figure}

The result (Figure~\ref{fig:featzoom}) is that clean straight edges carry no
candidates in their interior; candidates appear exactly where runs must be
cut. Every accepted segment will ultimately be delimited by two candidate
contacts, which is what lets the labeler finalize segments in-stream without
ever tracing chains.

\subsection{Streaming labeling}
\label{sec:labeling}

The labeler is a single-pass connected-component analysis in the spirit
of~\cite{bailey2007}, specialized so that its components are endpoint-delimited
edge runs and its per-component payload is exactly what the line test needs.
It maintains two rows of label assignments (previous and current) and a table
of label records. Each record holds the running \emph{scatter moments}
$(N, \Sigma x, \Sigma y, \Sigma x^2, \Sigma xy, \Sigma y^2)$ in integers
(sub-pixel NMS offsets, when enabled, enter these sums in fixed point), a
recency stamp (last row and column touched), a union--find link, a count of
strong pixels for the hysteresis gate (Section~\ref{sec:edge}), the
pixels attaining the four bounding-box extremes for the endpoint
choice (Section~\ref{sec:judge}), and the run's \emph{contact state}: whether
a first endpoint contact has been seen and, if so, where. A component is
never represented by its pixels --- only by this constant-size record.
Everything downstream leans on that: merging two components is a handful of
integer additions, and judging a finished run costs $O(1)$ regardless of its
length.

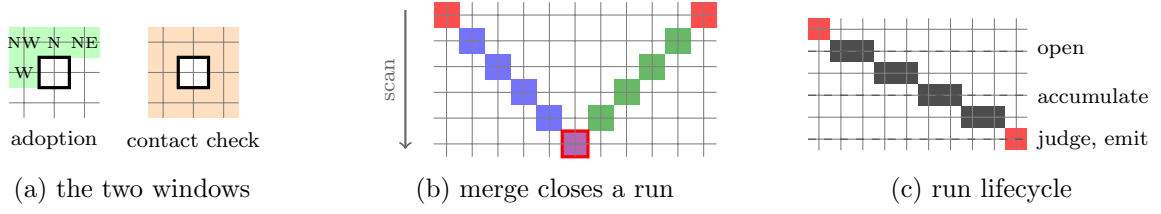
\begin{figure}[t]
  \centering
  \begin{subfigure}[b]{0.26\linewidth}
    \centering
    \begin{tikzpicture}[scale=0.40, font=\scriptsize]
      \foreach \p in {(-1,1),(0,1),(1,1),(-1,0)}
        \fill[green!25] \p +(-0.5,-0.5) rectangle +(0.5,0.5);
      \draw[black!50] (-1.5,-1.5) grid (1.5,1.5);
      \draw[very thick] (-0.5,-0.5) rectangle (0.5,0.5);
      \node at (-1,1) {\tiny NW}; \node at (0,1) {\tiny N};
      \node at (1,1) {\tiny NE}; \node at (-1,0) {\tiny W};
      \node at (0,-2.2) {adoption};
      \begin{scope}[xshift=4.6cm]
        \foreach \dx in {-1,0,1} \foreach \dy in {-1,0,1} {
          \ifnum\dx=0 \ifnum\dy=0 \else
            \fill[orange!25] (\dx-0.5,\dy-0.5) rectangle (\dx+0.5,\dy+0.5); \fi
          \else
            \fill[orange!25] (\dx-0.5,\dy-0.5) rectangle (\dx+0.5,\dy+0.5);
          \fi
        }
        \draw[black!50] (-1.5,-1.5) grid (1.5,1.5);
        \draw[very thick] (-0.5,-0.5) rectangle (0.5,0.5);
        \node at (0,-2.2) {contact check};
      \end{scope}
    \end{tikzpicture}
    \caption{the two windows}
  \end{subfigure}\hfill
  \begin{subfigure}[b]{0.40\linewidth}
    \centering
    \begin{tikzpicture}[scale=0.34, font=\scriptsize]
      \fill[red!70]   (0,0)  +(-0.5,-0.5) rectangle +(0.5,0.5);
      \fill[red!70]   (10,0) +(-0.5,-0.5) rectangle +(0.5,0.5);
      \foreach \p in {(1,-1),(2,-2),(3,-3),(4,-4)}
        \fill[blue!55] \p +(-0.5,-0.5) rectangle +(0.5,0.5);
      \foreach \p in {(9,-1),(8,-2),(7,-3),(6,-4)}
        \fill[green!55!black!60] \p +(-0.5,-0.5) rectangle +(0.5,0.5);
      \fill[violet!60] (5,-5) +(-0.5,-0.5) rectangle +(0.5,0.5);
      \draw[red, very thick] (5,-5) +(-0.5,-0.5) rectangle +(0.5,0.5);
      \draw[black!50] (-0.5,-5.5) grid (10.5,0.5);
      \draw[->, thick, black!60] (-1.6,0.2) -- (-1.6,-5.2)
        node[midway, left] {\rotatebox{90}{scan}};
    \end{tikzpicture}
    \caption{merge closes a run}
  \end{subfigure}\hfill
  \begin{subfigure}[b]{0.30\linewidth}
    \centering
    \begin{tikzpicture}[scale=0.29, font=\scriptsize]
      \fill[red!70] (0,0) +(-0.5,-0.5) rectangle +(0.5,0.5);
      \foreach \p in {(1,-1),(2,-1),(3,-2),(4,-2),(5,-3),(6,-3),(7,-4),(8,-4)}
        \fill[black!70] \p +(-0.5,-0.5) rectangle +(0.5,0.5);
      \fill[red!70] (9,-5) +(-0.5,-0.5) rectangle +(0.5,0.5);
      \draw[black!50] (-0.5,-5.5) grid (9.5,0.5);
      \draw[dashed, black!70] (-0.5,-1) -- (9.5,-1)
        node[right, black] {open};
      \draw[dashed, black!70] (-0.5,-3) -- (9.5,-3)
        node[right, black] {accumulate};
      \draw[dashed, black!70] (-0.5,-5) -- (9.5,-5)
        node[right, black] {judge, emit};
    \end{tikzpicture}
    \caption{run lifecycle}
  \end{subfigure}
  \caption{Labeling mechanics (endpoint candidates in red; the scan proceeds
  top to bottom). (a)~Label adoption reads only the causal half of the
  neighborhood; the contact check reads all eight neighbors --- the row
  below already exists because the labeler trails the front-end stages by
  one row of look-ahead. (b)~A V-shaped run: the two arms open at their own
  top contacts and grow as separate labels; at the bottom apex they merge ---
  the more recently extended half survives, the moments are summed, and
  since both halves had already opened, the joined run closes and is judged
  on the spot with the two recorded starts as endpoints. (c)~The common case:
  a run opens at its first contact, accumulates moments (and the strong flag
  and bounding-box extremes) as the sweep passes, and is judged and emitted
  at its second contact --- after which no pixel can touch its label again
  and the record expires by unreachability.}
  \label{fig:labeling}
\end{figure}

For each edge pixel at $(x, y)$ that is \emph{not} an endpoint candidate, the
labeler inspects the four causal neighbors that can already carry a label
--- NE, N, NW on the previous row and W on the current row
(Figure~\ref{fig:labeling}a, left) --- and resolves
each to its root label. Three cases:
\begin{itemize}
\item \textbf{No labeled neighbor}: allocate a fresh label, initialize its
  record.
\item \textbf{One distinct root}: adopt it.
\item \textbf{Two distinct roots}: \emph{merge}
  (Figure~\ref{fig:labeling}b). (At most two distinct roots can meet: of the
  four causal neighbors, only the NW/NE and W/NE pairs are not themselves
  8-adjacent, so any third root would already have been unified with one of
  the two.) The survivor is the half
  extended more recently (by lexicographic row--column recency
  stamp); the moments and strong counts are summed, the bounding-box
  extremes take the min/max, and the losing label records a union--find link
  to the survivor. The causal geometry keeps the link chains shallow --- in
  every corpus run the find depth never exceeded one.\footnote{The 2014
  implementation keeps no union--find: it uses the stored neighbor label
  directly and, on a merge, copies the combined moments into the survivor so the
  absorbed label is simply never referenced again. The present implementation
  instead keeps a true union--find with path compression and resolves on read. The two
  internal models are different; they produce bit-identical segments over the
  corpus.}
\end{itemize}
The pixel is then accumulated into the resulting label's record (moments,
recency, strong flag, extremes).

\textbf{Contact-driven judgment, implicit closure.} After labeling, all
eight neighbors of the pixel --- including the row \emph{below}, which the
labeler
can see because it trails the front-end stages by one row
(Figure~\ref{fig:labeling}a, right) --- are checked for endpoint
candidates. On a label's first contact, the run \emph{opens}: the contact
pixel (the interior pixel adjacent to the cut point) is recorded as the start
(Figure~\ref{fig:labeling}c). (The recorded pixel is the interior one ---
candidates are never part of the run --- so at a free tip, where two
consecutive pixels fire as candidates, the recovered tip retreats about two
pixels; Section~\ref{sec:endpoint} quantifies this extent bias and its
closed-form correction.)
On the second contact, the label's moments together with the two contact
pixels form a \emph{provisional segment}, which is judged immediately
(Section~\ref{sec:judge}). A merge of two halves that have both already opened
likewise submits the joined run, with the two recorded starts as its
endpoints; if only one half had opened, the survivor inherits that half's
recorded start and the run stays open; if neither had, the contact state
stays empty. There is deliberately \emph{no explicit close or terminate
operation} on the label: a run delimited by its two cut points simply stops
receiving pixels, so its label meets the same fate as that of a component
which never collects two contacts (a blob, a noise cluster, a dangling
scrap). Once its last touched
row has fallen more than one row behind the scan, no future pixel can reach
it, and it is dead and reclaimable. (One consequence, stated for
completeness: if rasterization exceptionally bridges past a cut --- the edge
map can be locally two pixels thick --- the component keeps growing and is
re-judged at each subsequent contact; emission is append-only, so an earlier
emission from the same label is not retracted.) The two contact pixels delimit the
run; the finalization step re-derives the endpoints from them and the
bounding-box extremes (Section~\ref{sec:judge}), so even components that
graze more than two candidates --- junction clusters, for instance --- get
extent-bracketing endpoints by construction.

\textbf{Capacity.} At most $\lceil W/2 \rceil$ labels can be alive at once on
a $W$-wide image: a label is live only while its most recent pixel is at most
one row behind the scan, and the most recent pixels of two distinct live
labels can never sit in the same or adjacent columns --- within the two live
rows such pixels are causally 8-adjacent, so adoption or merging would have
unified the labels --- so live labels occupy every other column at best (the
worst case, an interior/void alternation). A
fixed pool of that size therefore suffices (960 labels for $W = 1920$), managed
as a ring free-list that reclaims a label once the scan has passed beyond its
reach: after finishing row $y$, a label whose last-touched row is below $y$ can
neither receive another pixel (only row $y$'s labels reach row $y{+}1$) nor be
reached through a live union--find chain (every id stored in the current row was
accumulated this row, so live chains stay at recency $y$), so it is dead and its
slot is reclaimed. Reclaiming is therefore output-invariant (verified bit-identical
to an unbounded table over the corpus). Software and hardware
(Section~\ref{sec:fpga}) share this bounded pool, so label memory is $O(W)$ ---
tens of kilobytes, cache-resident --- in both, with a measured peak of 386 live
labels across the Full-HD corpus, well under the structural 960
bound.\footnote{As an engineering margin, the software library starts the
pool at a practical $W/4$ and grows it toward the structural
$\lceil W/2 \rceil$ (reporting the rare event) rather than dropping labels;
the hardware ring (a fixed 1024-label pool) instead sheds on overflow ---
a path never exercised on the corpus.}

\subsection{Line judgment and finalization}
\label{sec:judge}

A provisional segment with moments $(N, \Sigma x, \Sigma y, \Sigma x^2,
\Sigma xy, \Sigma y^2)$ is accepted if and only if all of the following hold:
\begin{enumerate}
\item \textbf{Pixel count}: $N \ge N_{\mathrm{th}}$ (default 15).
\item \textbf{Strong evidence}: at least 3 member pixels cleared the
  high gradient threshold --- the acceptance half of the streaming hysteresis
  of Section~\ref{sec:edge}.
\item \textbf{Straightness (relative)}: the scatter eigenvalue ratio satisfies
  $\lambda_{\min}/\lambda_{\max} \le \beta$ (default $0.05$), computed from
  the integer central moments $m_a = N\Sigma x^2 - (\Sigma x)^2$,
  $m_b = N\Sigma xy - \Sigma x \Sigma y$,
  $m_c = N\Sigma y^2 - (\Sigma y)^2$: the run must be thin and straight.
\item \textbf{Straightness (absolute)}: the perpendicular RMS spread
  of the member pixels about the fitted axis is at most 1\,px (the square
  root of the smaller eigenvalue of the per-pixel-normalized covariance, so
  the bound is in pixels). The ratio test alone is scale-free and admits long, gentle arcs:
  the ratio of a circular arc depends only on its subtended angle
  ($\approx 1/78$ at $50\degr$, $\approx 1/23$ at $90\degr$), so \emph{any}
  arc subtending $\lesssim 80\degr$ passes $\beta$ regardless of radius. A
  straight thin edge instead has a small \emph{constant} perpendicular
  variance; the absolute bound rejects the arcs and is the mechanism behind
  the ``0 segments on circles'' result of Section~\ref{sec:isotropy}.
\end{enumerate}
The straightness tests are expressible exactly in integer arithmetic by
cross-multiplication --- the hardware evaluates them that way (in its
configuration without sub-pixel NMS, whose fixed-point moments would push
the exact form past 128 bits; Section~\ref{sec:fpga}), and the
software's floating-point evaluation agrees with the exact form on every
judgment in the evaluation corpus.\footnote{For the record: the 2014 thesis
\emph{specifies} four acceptance criteria (adding an endpoint-distance and a
Chebyshev-span test to (1) and (3)), but its implementation exercises only
(1) and (3).}

Only after acceptance does floating point appear: the direction is
$\theta = \tfrac{1}{2}\operatorname{atan2}(2 m_b,\, m_a - m_c)$ (the
principal-axis angle); the endpoint pair is then chosen as the projection extremes
along that axis among six candidates --- the two contact pixels and
the four bounding-box extreme pixels --- which makes extent-bracketing
endpoints hold by construction where the plain first-contacts rule relies
on scan order (in practice the two rules coincide on straight runs; the
junction probe of Appendix~\ref{app:ablation} measures exactly that), and
the chosen pair is projected onto
the fitted axis through the centroid, yielding consistent sub-pixel
endpoints. The half-pixel lattice offset of the $2\times2$ gradient
(Section~\ref{sec:edge}) is applied here. This once-per-segment finalization
is the only floating-point code in the detector.

\subsection{The one-pass composition: memory and latency}
\label{sec:onepass}

The governing property is that \emph{every} stage above has bounded vertical
support: the Gaussian needs 5 rows, the gradient 2, NMS 3, the endpoint test 5,
and the labeler the previous row's label assignments plus one row of
look-ahead for contacts (the adaptive hysteresis threshold adds only an
$O(1)$ histogram). Chaining the lags, the labeler processes row $y-7$
while row $y$ is being read --- a fixed 7-row pipeline lag --- and nothing ever
requires revisiting an earlier row. Consequently each intermediate exists only
as a small ring of rows, and total intermediate memory is $O(\text{width})$:
for a 1920-wide image, a few tens of kilobytes including the label table,
independent of image height ($\approx$70\,KiB of detector state in the
hardware realization, Section~\ref{sec:fpga}).

Closure-driven emission gives the design a latency property that
frame-buffer detectors cannot have, and that we had not seen claimed
for line segment detection before (concurrently with this paper, a
line-segment ASIC design makes a comparable deterministic-latency
claim~\cite{jalilvand2026}, Section~\ref{sec:related}): a segment is judged the moment the labeler
reaches its terminating contact, and every closing event --- a second contact
seen with the one-row look-ahead, or a merge of two open halves --- fires
while the labeler processes one of the run's \emph{own} pixels, so every
segment is emitted within the \textbf{7-row pipeline lag} of its run's last
pixel entering the detector, independent of image height. (A run can outlive
its fitted segment's lower endpoint by a few rows where a merge or a bridged
cut extends it --- that, plus the fractional endpoint projection, is the
measured tail beyond 7 rows.) Measured over
the evaluation corpus (Section~\ref{sec:memlat}): median 6.5 rows, 99th
percentile 7.8, worst case 11.4. At 1080p30 video timing that is
$\approx$0.19\,ms of algorithmic latency, and segments of the upper image are
available long before the frame's last row has even been captured. Any
detector that must hold the image --- classical or learned --- has latency of
at least one frame period plus its processing time.

\section{Evaluation: speed and detection quality}
\label{sec:eval}

\textbf{Setup.} All experiments: 8-bit grayscale input, single thread,
Intel i7-8700K (Windows), all detectors built for AVX2. Baselines are the
\emph{original authors' implementations}, each built from that source at
the same ISA
target: canonical LSD 1.6~\cite{vongioi2012}, ED\_Lib
EDLines~\cite{akinlar2011}, and ELSED~\cite{suarez2022} (library
reimplementations such as OpenCV's \texttt{FastLineDetector} fall outside
this original-code criterion); all three run in
every study, and in the best-F ``F-max'' protocol of Section~\ref{sec:synth}
(which defines the term) ELSED's
\texttt{minLineLen} is swept
like EDLines' minimum length. Timing covers detection only --- image decode
and grayscale conversion are excluded, while a method's own input-format
requirement (LSD's byte-to-double conversion) is included; all detectors run
in one process, interleaved image-by-image, each image timed as the median of
5 runs. The OpenCV that ED\_Lib and ELSED link against is built without IPP
or a threading backend, so their internal filters run single-threaded, and
the baselines are pinned to specific upstream versions (ED\_Lib commit
\texttt{69b8d08}, ELSED commit \texttt{71893e4}; LSD is the fixed IPOL 1.6
release).
The benchmark harness,
synthetic-scene generators, and all protocols below ship in the repository;
LSD (AGPL-3.0) is fetched at configure time rather than vendored. Speed
ratios throughout are computed from unrounded medians, so they can differ in
the last digit from a ratio of the rounded table entries.

\textbf{Configuration.} \sweeplsd{} runs in its single shipped configuration
--- the library defaults, collected in Appendix~\ref{app:params} ---
throughout. Its one
optional feature exercised in this paper, the gap-tolerant collinear linker
(Section~\ref{sec:synth}), off by default,
is never silently enabled: every result that uses it says so. (A streaming
NFA validation also ships in the library, likewise off by default; it is not
used in any result here.)

\textbf{Development and evaluation data.} The core 2014 algorithm and its
original thresholds
predate every dataset in this paper by a decade; what was tuned in
2025--2026 are the five refinements of Section~\ref{sec:method} and the
judge/linker defaults. Three signals drove that tuning: the synthetic-GT
protocol of Section~\ref{sec:synth}, inspection of a private
150-photograph Full-HD corpus that appears nowhere in this paper, and the
downstream York Urban / NYU error of Section~\ref{sec:vp}. There was no
frozen holdout, so the synthetic-accuracy and York Urban / NYU numbers are
in-distribution for that selection. Every other dataset was adopted after
the shipped configuration was frozen and was never consulted during
development --- LIU4K-v2 (speed, dispersion, memory, repeatability), the
Blender attitude scenes, EuRoC, and TUM-VI --- and the orderings the
in-distribution numbers support are the ones that replicate there
(Sections~\ref{sec:attitude},~\ref{sec:realdata}).

\subsection{Speed}
\label{sec:speed}

\begin{table}[t]
  \centering
  \caption{Median per-image detection time over the 123 structure-rich
  (Building/Street) photographs of the public LIU4K-v2 corpus~\cite{liu2020}
  (CC0-licensed), each cropped to 16:9 and downscaled to Full-HD
  ($1920\times1080$) with a Lanczos filter whose kernel support
  scales with the downscale factor (i.e.\ properly anti-aliased); the
  corpus-preparation script ships in the repository. All four
  detectors are timed in one process under one toolchain (GCC 15.2, AVX2, single
  thread). Corpus-median segment counts sit in the same range --- \sweeplsd{}
  1{,}591, ELSED 1{,}377, EDLines 2{,}626, LSD 2{,}473 --- so the speed is not
  bought at the price of detecting fewer segments (EDLines and LSD return
  more at their defaults, consistent with the low-contrast coverage
  difference of Section~\ref{sec:limitations}).}
  \label{tab:speed}
  \begin{tabular}{lccc}
    \toprule
    detector & median time per frame & relative & intermediates \\
    \midrule
    \textbf{\sweeplsd{} (one-pass)} & \textbf{11.3\,ms} & \textbf{1$\times$} & $O(\text{width})$ \\
    \sweeplsd{} (multi-pass driver) & 14.6\,ms & 1.3$\times$ & $O(\text{pixels})$ \\
    ELSED & 51.4\,ms & 4.6$\times$ & $O(\text{pixels})$ \\
    EDLines (ED\_Lib) & 58.4\,ms & 5.2$\times$ & $O(\text{pixels})$ \\
    LSD & 277.8\,ms & 25$\times$ & $O(\text{pixels})$ \\
    \bottomrule
  \end{tabular}
\end{table}

Table~\ref{tab:speed} gives the headline numbers: a median 11.3\,ms per
Full-HD frame on a single thread --- $4.6\times$ faster than ELSED,
$5.2\times$ than EDLines, and $25\times$ than LSD. The multi-pass driver runs
the same kernels on the same arithmetic yet trails its one-pass sibling at
$1.3\times$: the $O(\text{width})$ working set of the one-pass composition
stays in cache. Where the time goes is broken down per stage in
Appendix~\ref{app:timing}; in brief, the cost is spread across the pipeline,
with the content-dependent streaming labeler the largest stage
($\approx$45\% on this high-detail corpus) and no front-end row kernel
exceeding a fifth of the stage sum.

\textbf{Scale behavior.} Because the LIU4K-v2 photographs are natively 4K--6K,
we downscale the \emph{same} scenes to five resolutions from $640\times360$ up to
$3840\times2160$ (4K) without ever upsampling (Table~\ref{tab:scale}).
\sweeplsd{} is $\sim$4.6--4.9$\times$ faster than ELSED and $\sim$5.1--5.7$\times$
faster than EDLines at \emph{every} scale, and the ratio is essentially flat
across the whole $36\times$ span of pixel counts --- the signature of
matched, near-linear scaling (Appendix~\ref{app:timing} fits \sweeplsd{}'s
own per-pixel cost directly). LSD scales roughly linearly in pixels
and stays $21$--$28\times$ behind throughout. (On these high-detail photographs ELSED and EDLines run close together
--- ELSED's drawing heuristic loses its usual lead on cluttered fine texture
--- an early sign of the content-dependence that the fourth caveat below
states in general; on the sparser imagery of Section~\ref{sec:realdata},
ELSED's usual lead over EDLines reappears at $\approx$1.6$\times$.)

\begin{table}[t]
  \centering
  \caption{Median per-image detection time across five resolutions, obtained by
  downscaling the \emph{same} 123 LIU4K-v2 photographs (natively 4K--6K, so no
  upsampling is needed at any rung; single thread, AVX2). All four detectors are
  timed in one process, built by the same compiler (GCC 15.2) and linked against
  an OpenCV built with it, so the comparison isolates the algorithms rather than
  the toolchains.}
  \label{tab:scale}
  \begin{tabular}{lcccc}
    \toprule
    resolution & \sweeplsd{} (one-pass) & ELSED & EDLines & LSD \\
    \midrule
    $640\times360$ (0.23\,MP)   & \textbf{1.5\,ms}  & 7.3\,ms  & 8.0\,ms  & 31.9\,ms \\
    $1280\times720$ (0.92\,MP)  & \textbf{5.3\,ms}  & 24.8\,ms & 27.5\,ms & 124\,ms \\
    $1920\times1080$ (2.07\,MP) & \textbf{11.3\,ms} & 51.4\,ms & 58.4\,ms & 278\,ms \\
    $2560\times1440$ (3.69\,MP) & \textbf{19.5\,ms} & 89.3\,ms & 102\,ms  & 535\,ms \\
    $3840\times2160$ (8.29\,MP) & \textbf{44.5\,ms} & 219\,ms  & 255\,ms  & 1253\,ms \\
    \bottomrule
  \end{tabular}
\end{table}

\begin{table}[t]
  \centering
  \caption{Frame-time dispersion over the 123 Full-HD photographs of
  Table~\ref{tab:speed}, each image timed as the median of 5 runs. The CV and
  the relative
  columns (p95 and worst frame, as multiples of each detector's own corpus
  median) are scale-free, so the four distributions compare on shape rather
  than on how fast the typical frame is.}
  \label{tab:spread}
  \begin{tabular}{lccccc}
    \toprule
    detector & median & SD & CV & p95 & worst frame \\
    \midrule
    \textbf{\sweeplsd{} (one-pass)} & 11.3\,ms & \textbf{$\pm$2.7\,ms} & \textbf{24.4\%} & \textbf{1.33$\times$} & \textbf{1.66$\times$} \\
    ELSED & 51.4\,ms & $\pm$21.4\,ms & 41.8\% & 1.64$\times$ & 2.14$\times$ \\
    EDLines (ED\_Lib) & 58.4\,ms & $\pm$22.5\,ms & 38.2\% & 1.70$\times$ & 2.55$\times$ \\
    LSD & 277.8\,ms & $\pm$102\,ms & 37.0\% & 1.59$\times$ & 2.33$\times$ \\
    \bottomrule
  \end{tabular}
\end{table}

\textbf{Frame-time predictability.} A median is not a frame budget: a
detector that is occasionally several times slower than typical still drops
frames. Table~\ref{tab:spread} therefore reports the dispersion of the 123
per-image times. \sweeplsd{} is the tightest on all three scale-free
measures --- CV, 95th percentile, and worst frame --- and its absolute
spread ($\pm2.7$\,ms against $\pm21$--$102$\,ms) is smaller by a wider
margin still, though that gap is partly arithmetic (a smaller total has less
room to vary), so the scale-free columns carry the claim. The reason is
structural: the four front-end stages touch every pixel exactly once,
regardless of content, so
roughly half of the Full-HD cost is spent identically on every image, and
only the labeler and the per-segment finalization scale with content
(Appendix~\ref{app:timing}). Two qualifiers. The advantage lives in the
tail, not the middle --- the interquartile bands largely overlap
($0.81$--$1.14\times$ for \sweeplsd{} against $0.71$--$1.26\times$ for
ELSED) and the separation opens above the 95th percentile, which is,
however, exactly the region a frame budget must absorb. And these are
between-image differences: run-to-run jitter on a fixed input is
$3$--$4\%$, and with 5 runs per image the fastest detector's CV is, if
anything, a mild overestimate.

Four caveats. First, LSD's official implementation is single-scale,
unoptimized C; its 278\,ms is in part the price of its a-contrario validation,
not carelessness. Second, the four detectors return comparable numbers of
segments (Table~\ref{tab:speed}), so the speed gap is not an artifact of
\sweeplsd{} detecting fewer of them. Third, the
no-intrinsics design makes \sweeplsd{}'s speed strongly compiler-dependent:
toolchains whose vectorizer handles all five stage kernels (GCC, both Clang
drivers) land within $1.38\times$ of one another, one that does not (MSVC's
\texttt{cl}) runs $4.1\times$ slower, and detections stay bit-identical
throughout. The full toolchain study, including the recommended compiler per
ABI, is in Appendix~\ref{app:compiler}. Fourth, the corpus is deliberately
structure-rich --- the Building and Street categories of LIU4K-v2, the
setting a line segment detector is for --- and the speed margins are
content-dependent: where there is little structure to draw, the edge-drawing
detectors speed up and the gap narrows (Section~\ref{sec:realdata}: to
$2.3$--$2.5\times$ over ELSED on sparse indoor imagery), while \sweeplsd{}'s
own cost barely moves.

\subsection{Memory and latency, measured}
\label{sec:memlat}

The $O(\text{width})$-memory and rows-of-latency claims of
Section~\ref{sec:onepass}
are design properties; here they are measured.

\textbf{Memory.} We measured each detector's peak working set (the Windows
process metric \texttt{PeakWorkingSetSize}; one detector per process, each
process streaming twelve corpus photographs at that resolution and reporting
its lifetime peak) across the five
resolutions of Table~\ref{tab:scale}
(Table~\ref{tab:mem}); memory is reported in MiB ($2^{20}$ bytes). \sweeplsd{}'s one-pass driver holds the smallest footprint
at every resolution, and --- the scaling property under test --- its peak grows only
$3.8\times$ while the pixel count grows $36\times$, because its intermediate
state is $O(\text{width})$ rather than $O(\text{pixels})$; the largest
single item in its 22\,MiB at 4K is the input image itself ($\approx$7.9\,MiB
at one byte per pixel), and the
other detectors sit at $6.5$--$11.4\times$ that footprint there. These are
whole-\emph{process} peaks --- input image, C runtime, and I/O buffers
included --- which is why the absolute floor is mebibytes where
Section~\ref{sec:onepass} counts the detector's own intermediate state in
tens of kilobytes; the algorithmic $O(\text{width})$ property shows in the
growth rate, not in the absolute footprint.

\begin{table}[t]
  \centering
  \caption{Peak working-set memory (MiB) per detector at each resolution
  (LIU4K-v2, \texttt{PeakWorkingSetSize}, one detector per process, twelve
  photographs per process). \sweeplsd{} one-pass's detector-owned
  intermediate state is $O(\text{width})$; its whole-process peak, which
  includes the $O(\text{pixels})$ input image, grows $3.8\times$ over this
  $36\times$ pixel span, while the others grow with pixel count.}
  \label{tab:mem}
  \begin{tabular}{lccccc}
    \toprule
    detector & 360p & 720p & 1080p & 1440p & 4K \\
    \midrule
    \textbf{\sweeplsd{} (one-pass)} & \textbf{5.9} & \textbf{7.4} & \textbf{10.0} & \textbf{13.3} & \textbf{22.4} \\
    \sweeplsd{} (multi-pass) & 7.6 & 14.3 & 25.8 & 41.2 & 85.5 \\
    ELSED & 9.9 & 22.4 & 37.7 & 69.1 & 146.7 \\
    EDLines (ED\_Lib) & 10.4 & 22.2 & 41.3 & 66.9 & 154.1 \\
    LSD & 12.6 & 33.6 & 67.3 & 114.1 & 255.0 \\
    \bottomrule
  \end{tabular}
\end{table}

\textbf{Latency.} Instrumenting the labeler's emission point over the Full-HD
photograph corpus (205{,}173 segments; latency measured as the sweep row
at emission minus the segment's lower endpoint row): median \textbf{6.5
rows}, 95th percentile 7.2, 99th percentile 7.8, worst case 11.4 rows. At
1080p30 video timing the median is $\approx$0.19\,ms between a segment's last
pixel arriving and the segment being emitted --- more than two orders of
magnitude below the one-frame floor ($\ge$33\,ms at 30\,fps) of any detector
that must hold the image. (The fractional values are a property of the measurement, not the
pipeline: endpoint coordinates carry the half-pixel lattice shift of
Section~\ref{sec:judge} --- hence the 6.5 median rather than 7 --- and,
being projections, can overshoot the last pixel row by a fraction; the
pipeline bound itself is the 7-row lag relative to the run's last pixel,
Section~\ref{sec:onepass}.)

\subsection{Accuracy on synthetic ground truth}
\label{sec:synth}

\begin{table}[t]
  \centering
  \caption{F-max under noise (synthetic scenes, $1280\times720$, 18 random
  segments, 20 images per condition, strict one-to-one matching; each
  detector's main knob swept and the best F reported). A broader 200-scene
  randomized validation with bootstrap confidence intervals follows in
  Table~\ref{tab:rand}.}
  \label{tab:fmax}
  \begin{tabular}{ccccc}
    \toprule
    noise $\sigma$ & \sweeplsd{} & LSD & EDLines (ED\_Lib) & ELSED \\
    \midrule
    0 (clean) & 0.958 & 0.936 & 0.955 & \textbf{0.991} \\
    5  & 0.959 & 0.938 & 0.950 & \textbf{0.992} \\
    10 & 0.948 & 0.788 & 0.952 & \textbf{0.990} \\
    20 & 0.905 & 0.510 & 0.951 & \textbf{0.990} \\
    \bottomrule
  \end{tabular}
\end{table}

\begin{figure}[t]
  \centering
  \begin{subfigure}[b]{0.19\linewidth}
    \includegraphics[width=\linewidth]{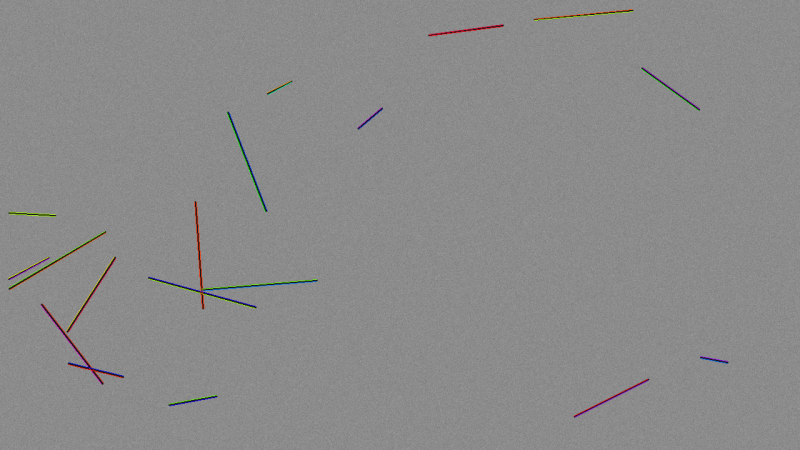}\caption{GT ($\sigma$10)}
  \end{subfigure}\hfill
  \begin{subfigure}[b]{0.19\linewidth}
    \includegraphics[width=\linewidth]{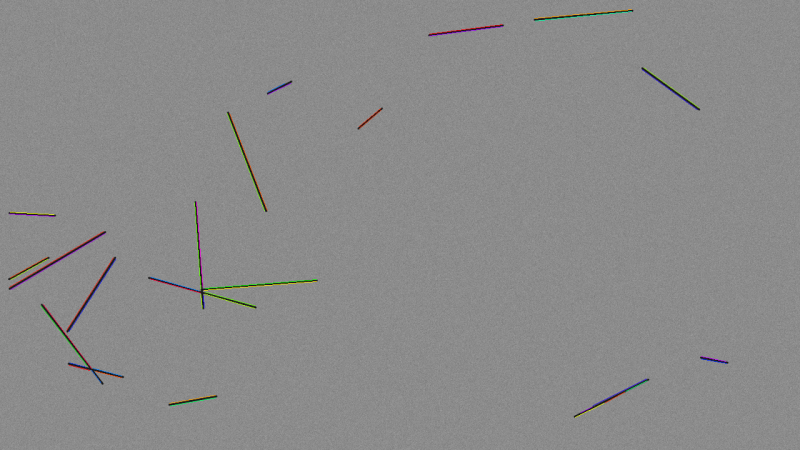}\caption{\sweeplsd{}}
  \end{subfigure}\hfill
  \begin{subfigure}[b]{0.19\linewidth}
    \includegraphics[width=\linewidth]{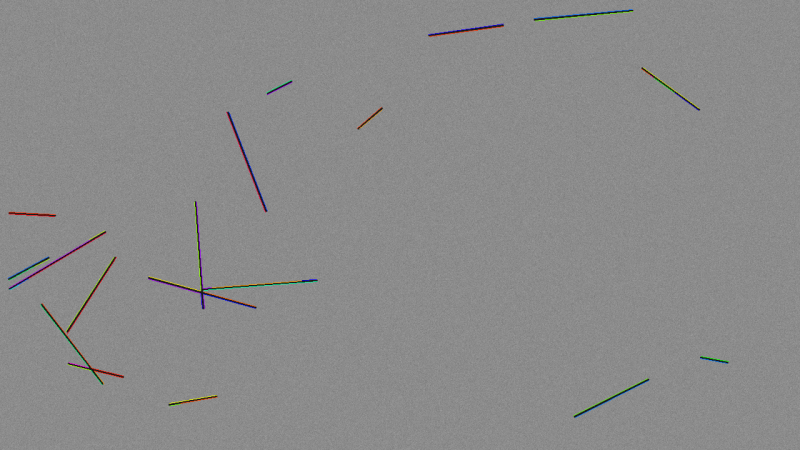}\caption{LSD}
  \end{subfigure}\hfill
  \begin{subfigure}[b]{0.19\linewidth}
    \includegraphics[width=\linewidth]{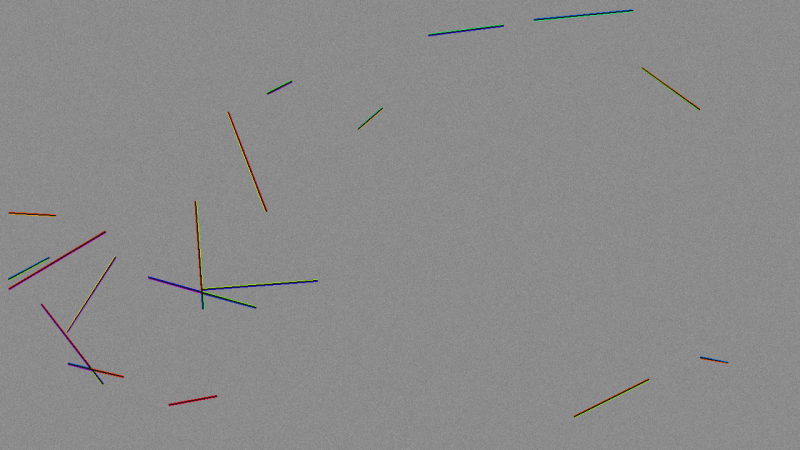}\caption{EDLines}
  \end{subfigure}\hfill
  \begin{subfigure}[b]{0.19\linewidth}
    \includegraphics[width=\linewidth]{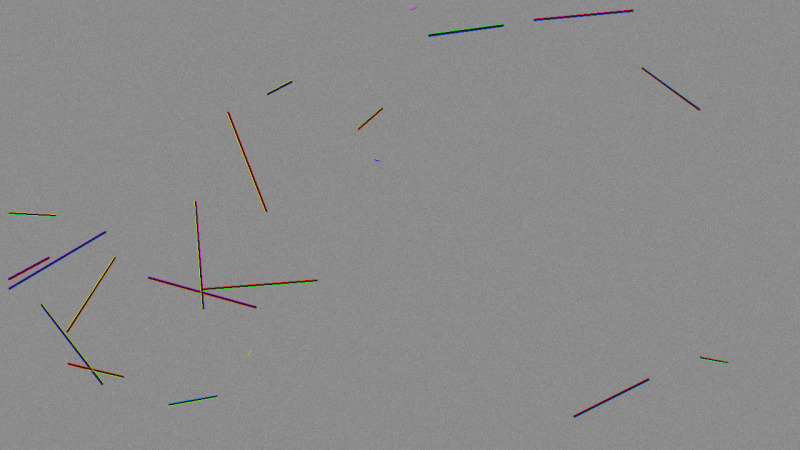}\caption{ELSED}
  \end{subfigure}
  \caption{Qualitative comparison on a $\sigma{=}10$ synthetic scene
  (EDLines = ED\_Lib; EDLines and ELSED at their swept minimum length 10,
  \sweeplsd{} at its swept $N_{\mathrm{th}} = 16$, LSD at its defaults).}
  \label{fig:eval}
\end{figure}

The scenes are uniform-background bar images: 3-px-wide dark bars (gray 40 on
210), anti-aliased, with each bar's two flank edges as ground truth; noise is
additive Gaussian with standard deviation $\sigma$ (in gray levels), clipped
to $[0,255]$.
The protocol uses \emph{strict one-to-one} matching: a detection may match a
ground-truth segment only if their orientations differ by at most 10\degr{},
both detection endpoints lie within 2\,px of the ground-truth line, and at
least half the detection's length projects onto the ground-truth extent;
candidates are then matched greedily by projected overlap, each ground-truth
segment matching at most one detection, so fragmentation is penalized. (Lenient
one-to-many matching flattered every method and inverted rankings; it is not
used.) Each detector's main operating knob --- minimum segment
length for the baselines, the judge's pixel count $N_{\mathrm{th}}$ for
\sweeplsd{} --- is swept, precision and recall are computed at each setting,
and \emph{F-max}
is the best F1 along the sweep: every detector is scored at its own best
operating point, in the best-F convention of boundary-detection benchmarks,
so the comparison measures the detectors rather than their default tunings.

Table~\ref{tab:fmax}, stated plainly: \textbf{ELSED leads F-max at
every noise level} on this protocol --- its fused drawing-and-fitting with
validation is extremely effective on synthetic bar scenes, and essentially
flat in noise. \sweeplsd{} is second on clean and $\sigma \le 5$ images;
ED\_Lib is similarly noise-stable at a slightly lower level; LSD degrades
fastest under heavy noise (Figure~\ref{fig:eval} compares the four on a
$\sigma{=}10$ scene). (One sweep asymmetry to disclose: ELSED's
implementation aborts below \texttt{minLineLen} $\approx$7--10, so its knob
sweep spans 10--40 where the others reach down to 5; its F-max points did not
lie at the clipped end.) Part of the residual gap to ELSED is
\emph{fragmentation} rather than spurious detections. The mechanism is
visible in the protocol: under strict one-to-one matching every extra
fragment of a ground-truth line counts as a false positive, and ELSED's
drawing pass jumps discontinuities natively where \sweeplsd{}'s endpoint
candidates cut runs at junctions. It can also be tested by intervention: the
optional gap-tolerant collinear linker (off by default; a finalization-level
feature that preserves the streaming property, specified in full in
Appendix~\ref{app:linker}) does nothing but re-assemble
collinear fragments across junction cuts and noise breaks, and by itself
recovers about a fifth of the $\sigma = 20$ gap but little of the
clean-scene one --- with it, F-max rises from 0.905 to 0.924 at
$\sigma = 20$ and only from 0.958 to 0.960 on clean scenes, so on this
protocol fragmentation is a noisy-scene effect (its downstream
vanishing-point effect is reported in Section~\ref{sec:vp}).

\begin{table}[t]
  \centering
  \caption{Per-segment geometric accuracy on matched segments, on the
  \emph{synthetic} scenes of Table~\ref{tab:fmax} with exact ground truth
  (mean over the matched set; ranges span that table's noise levels; a single
  value did not vary across noise levels at this precision).
  Section~\ref{sec:vpdecomp} measures per-line direction quality on real
  photographs instead, against a different reference --- ground-truth Manhattan
  axes, whose own imperfection the residual includes --- and so at a different
  scale; the numbers there and here are not directly comparable.}
  \label{tab:geom}
  \begin{tabular}{lcc}
    \toprule
    detector & lateral error (px) & direction error (\degr{}) \\
    \midrule
    \textbf{\sweeplsd{}} & 0.12--0.13 & \textbf{0.02--0.04} \\
    ELSED & 0.12--0.13 & 0.07--0.09 \\
    EDLines (ED\_Lib) & 0.12--0.13 & 0.13--0.14 \\
    LSD & 0.15 & 0.02--0.12 \\
    \bottomrule
  \end{tabular}
\end{table}

On matched segments, per-segment \emph{direction} is \sweeplsd{}'s strongest
axis (Table~\ref{tab:geom}): the scatter-moment fit over every member pixel
plus sub-pixel NMS gives it the best direction accuracy of the four detectors
at every noise level --- $2.2$--$4.4\times$ better than ELSED and
$3.5$--$8.7\times$ better than ED\_Lib. Lateral error does not separate
them: \sweeplsd{}, ELSED, and ED\_Lib sit within $0.013$\,px of one another
with no ordering stable across noise levels, and only LSD is clearly behind.
(One conditioning caveat:
Table~\ref{tab:geom} scores each detector on the segments it matched, and the
matched sets differ --- most for LSD under heavy noise --- so part of a gap
could in principle reflect \emph{which} lines a detector recovers rather than
how well it fits them. The real-photo decomposition of
Section~\ref{sec:vpdecomp}, which measures every detector's lines against
common ground-truth axes, finds the same ordering for the three detectors it
covers; the randomized companion below adds a direct control on identical
segments, matched by all four detectors.) The synthetic story is therefore two-sided: on raw F-max
ELSED is the strongest detector here, while \sweeplsd{}'s edge is the
direction accuracy of each matched segment --- which is exactly the property
the downstream study of Section~\ref{sec:vp} shows to matter indoors.

\textbf{A randomized companion study.} The fixed protocol above varies only
noise; a companion run randomizes the scene itself --- 200 scenes, each
drawing its bar count (6--40), bar width ($1.5$--$5$\,px), contrast
(log-uniform over 10--170 gray levels), pre-noise Gaussian blur
($\sigma_b \in [0, 1.5]$\,px), noise ($\sigma \in [0, 20]$), and close
parallel neighbors (30\% of bars gain one at 4--10\,px) --- and scores
every detector at a \emph{single} pool-best operating point rather than a
per-condition best (Table~\ref{tab:rand}; the generator and analysis ship
in the repository). The verdicts survive with confidence intervals
attached: ELSED leads, ED\_Lib is close behind, \sweeplsd{} sits third
(the linker recovers part of the gap), LSD is last when held to one fixed
operating point across this noise range --- and \sweeplsd{} keeps the best
matched-segment geometry under every randomization ($0.32$\,px,
$0.13$\degr{}). The conditioning caveat of the fixed protocol can be closed
directly here: recomputing direction error on the \emph{intersection} of
ground-truth segments matched by all four detectors --- $6{,}657$ segments,
$54\%$ of the pool, identical for every method --- the margins over LSD and
ED\_Lib survive intact (paired mean gaps $+0.047$\degr{} [bootstrap 95\%:
$0.034$, $0.061$] and $+0.040$\degr{} [$0.023$, $0.056$]; \sweeplsd{} is
the more accurate on $61\%$ and $69\%$ of individual segments), so those
gaps are not artifacts of which segments each detector recovers. Against
ELSED the same control splits the claim by statistic: \sweeplsd{} remains
the more accurate on $59\%$ of the identical segments and keeps the lower
median ($0.033$\degr{} vs.\ $0.047$\degr{}), but its \emph{mean} advantage
closes to a statistical tie ($-0.013$\degr{} [$-0.028$, $+0.001$]) --- a
small set of hard segments costs \sweeplsd{} more than it costs ELSED.
Two slices are worth naming. Under heavy noise ED\_Lib is
the most stable, as in Table~\ref{tab:fmax}. At low contrast ($<$20 gray
levels) every detector loses most of its F, but \sweeplsd{} holds up
\emph{better} than ED\_Lib and LSD --- at a fixed configuration the
adaptive hysteresis of Section~\ref{sec:edge} is what keeps
sharp-but-faint edges alive, and only ELSED does better. This does not
contradict the coverage deficit of Section~\ref{sec:limitations}, item i:
that deficit concerns soft, \emph{wide} luminance ramps on real
photographs, which never form a gradient peak; the bars here are sharp at
every contrast.

\begin{table}[t]
  \centering\small
  \caption{Randomized companion protocol: 200 scenes with randomized bar
  count, width, contrast, blur, noise, and parallel neighbors; every
  detector at its single pool-best operating point (its knob swept once
  over the whole pool). CI = bootstrap 95\% over scenes; slice columns are
  pooled F within the subset at the same operating point; ``dir.''\ is the
  mean direction error on matched segments.}
  \label{tab:rand}
  \begin{tabular}{lcccc}
    \toprule
    detector & pooled F (95\% CI) & F, contrast $<$20 & F, $\sigma \ge 10$ & dir.\ (\degr) \\
    \midrule
    ELSED & \textbf{0.807} [0.769, 0.840] & \textbf{0.457} & 0.701 & 0.15 \\
    EDLines (ED\_Lib) & 0.776 [0.739, 0.809] & 0.303 & \textbf{0.719} & 0.19 \\
    \sweeplsd{} $+$ linker & 0.699 [0.664, 0.732] & 0.420 & 0.567 & \textbf{0.13} \\
    \sweeplsd{} & 0.654 [0.617, 0.686] & 0.382 & 0.519 & \textbf{0.13} \\
    LSD & 0.600 [0.565, 0.633] & 0.248 & 0.455 & 0.20 \\
    \bottomrule
  \end{tabular}
\end{table}

\textbf{False detections on line-free images.} The protocol above measures
quality where lines exist; a negative probe measures what the detectors
invent where none do. On uniform-noise images ($\sigma \in \{5, 10, 20,
40\}$, eight $1280{\times}720$ images per level, every detector at its
defaults) \sweeplsd{} detects essentially nothing --- at most $0.1$ segments
per image at any level --- even though its default configuration has no
a-contrario validation (Section~\ref{sec:limitations}, item iv): the
contrast gate, the strong-evidence gate, and the two straightness tests act
as an effective false-positive filter on noise. LSD and ED\_Lib, whose NFA
control targets exactly this, are comparably clean ($\le$1 and $\le$1.5
per image on every probe). Smooth value-noise ``clouds'' --- line-free by
construction --- also stay clean ($0.2$ per image); the one probe that
leaks is clouds \emph{plus} noise ($\sigma{=}10$), where \sweeplsd{} emits
$3.2$ segments per image and enabling the optional streaming NFA cuts that
to $0.8$. ELSED is the least controlled on every negative probe
($3.6$--$10.8$ per image). The probe generator and harness ship in the
repository.

\subsection{Orientation isotropy and curve rejection}
\label{sec:isotropy}

\begin{table}[t]
  \centering
  \caption{Synthetic probes ($1024\times1024$). CV = coefficient of variation
  of the length-weighted orientation histogram (lower = more isotropic).}
  \label{tab:iso}
  \setlength{\tabcolsep}{3.6pt}
  \begin{tabular}{lcccccccc}
    \toprule
    & \multicolumn{2}{c}{\sweeplsd{}} & \multicolumn{2}{c}{LSD} & \multicolumn{2}{c}{EDLines (ED\_Lib)} & \multicolumn{2}{c}{ELSED} \\
    \cmidrule(lr){2-3}\cmidrule(lr){4-5}\cmidrule(lr){6-7}\cmidrule(lr){8-9}
    pattern & segs & CV & segs & CV & segs & CV & segs & CV \\
    \midrule
    Siemens star (straight spokes) & 110 & 0.74 & 108 & 0.51 & 108 & 0.59 & 108 & \textbf{0.50} \\
    Angular fan (straight lines)   & 198 & 0.21 & 302 & \textbf{0.18} & 193 & 0.20 & 192 & 0.22 \\
    Concentric circles (no lines) & \textbf{0} & --- & 1{,}796 & 0.13 & 1{,}768 & 0.17 & 1{,}443 & 0.29 \\
    Zone plate (curved everywhere) & \textbf{27} & 1.58 & 859 & 0.18 & 675 & 0.24 & 475 & 0.31 \\
    \bottomrule
  \end{tabular}
\end{table}

\begin{figure}[t]
  \centering
  \begin{subfigure}[b]{0.19\linewidth}
    \includegraphics[width=\linewidth]{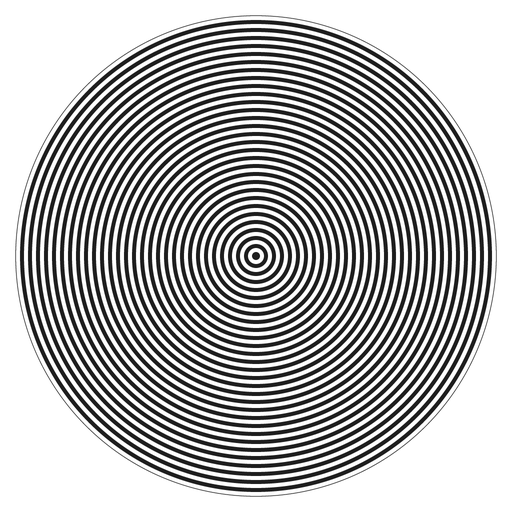}\caption{input}
  \end{subfigure}\hfill
  \begin{subfigure}[b]{0.19\linewidth}
    \includegraphics[width=\linewidth]{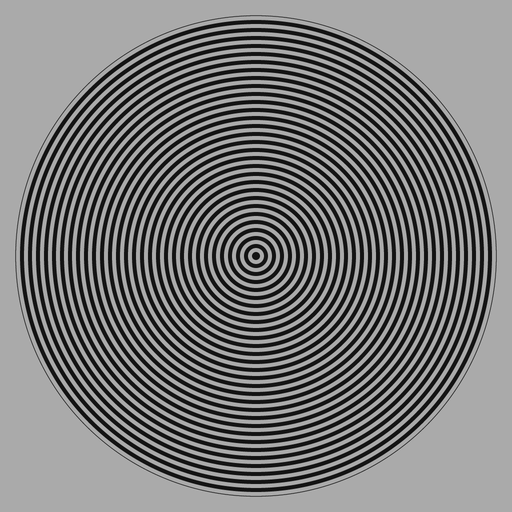}\caption{\sweeplsd{}: 0}
  \end{subfigure}\hfill
  \begin{subfigure}[b]{0.19\linewidth}
    \includegraphics[width=\linewidth]{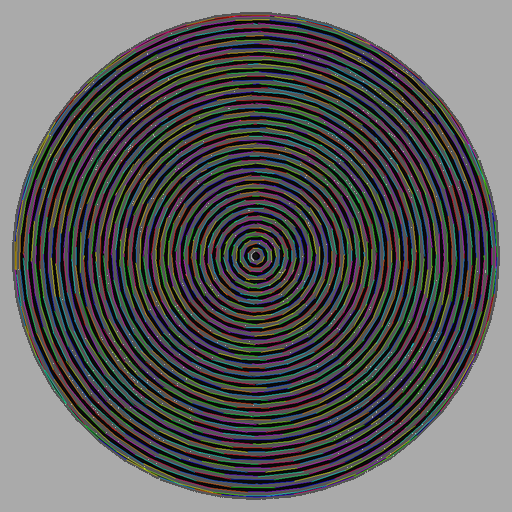}\caption{LSD: 1{,}796}
  \end{subfigure}\hfill
  \begin{subfigure}[b]{0.19\linewidth}
    \includegraphics[width=\linewidth]{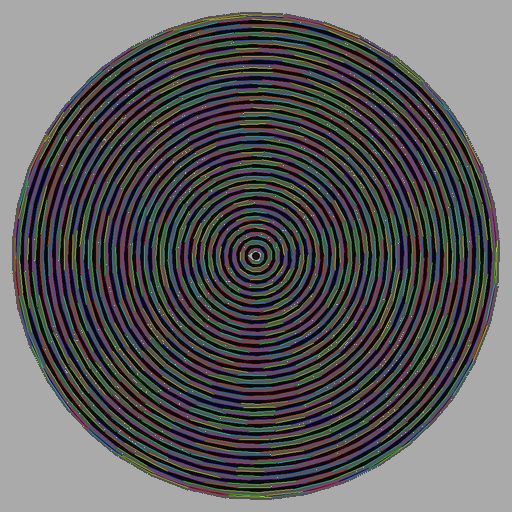}\caption{EDLines: 1{,}768}
  \end{subfigure}\hfill
  \begin{subfigure}[b]{0.19\linewidth}
    \includegraphics[width=\linewidth]{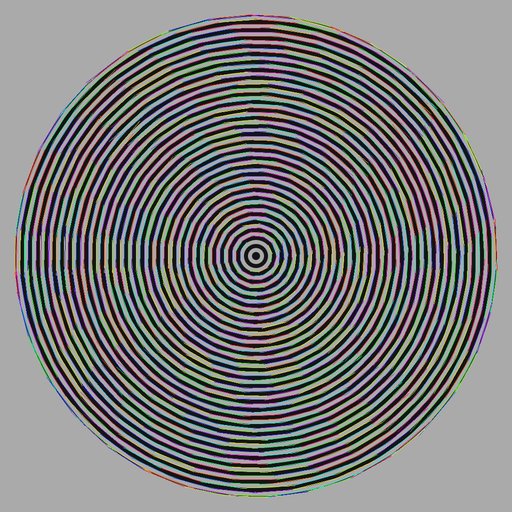}\caption{ELSED: 1{,}443}
  \end{subfigure}
  \caption{Concentric circles contain no straight lines. \sweeplsd{}'s curve
  rejection returns zero segments; LSD, EDLines, and ELSED shred the curves
  into $\sim$1{,}400--1{,}800 short straight fragments.}
  \label{fig:circles}
\end{figure}

\begin{figure}[t]
  \centering
  \begin{subfigure}[b]{0.19\linewidth}
    \includegraphics[width=\linewidth]{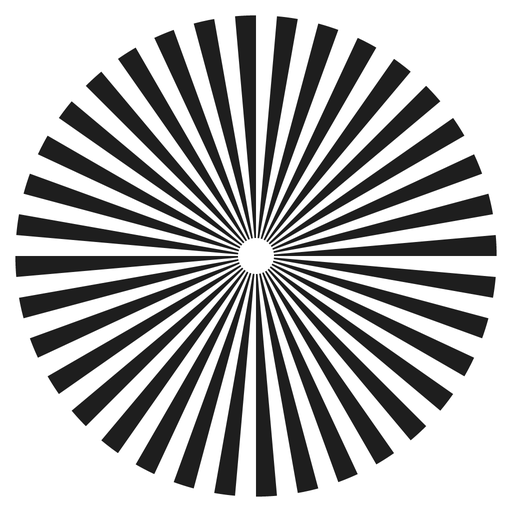}\caption{input}
  \end{subfigure}\hfill
  \begin{subfigure}[b]{0.19\linewidth}
    \includegraphics[width=\linewidth]{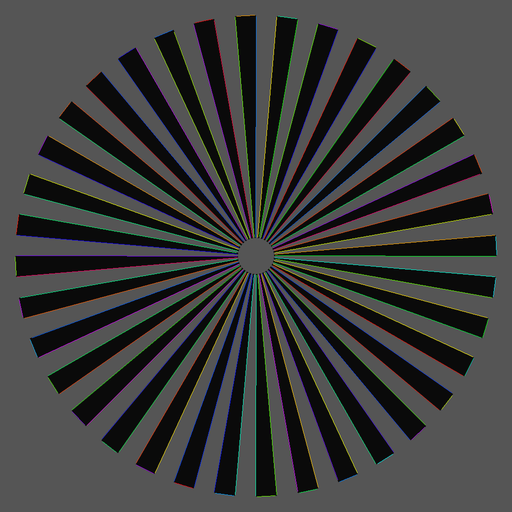}\caption{\sweeplsd{}: 110}
  \end{subfigure}\hfill
  \begin{subfigure}[b]{0.19\linewidth}
    \includegraphics[width=\linewidth]{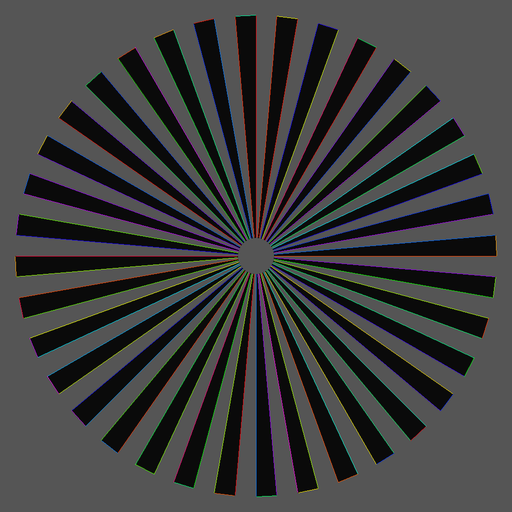}\caption{LSD: 108}
  \end{subfigure}\hfill
  \begin{subfigure}[b]{0.19\linewidth}
    \includegraphics[width=\linewidth]{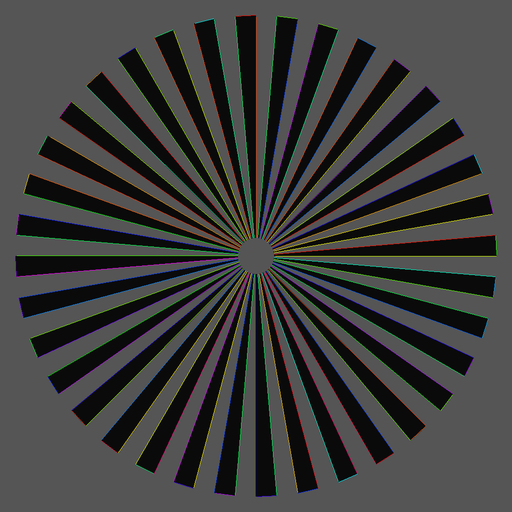}\caption{EDLines: 108}
  \end{subfigure}\hfill
  \begin{subfigure}[b]{0.19\linewidth}
    \includegraphics[width=\linewidth]{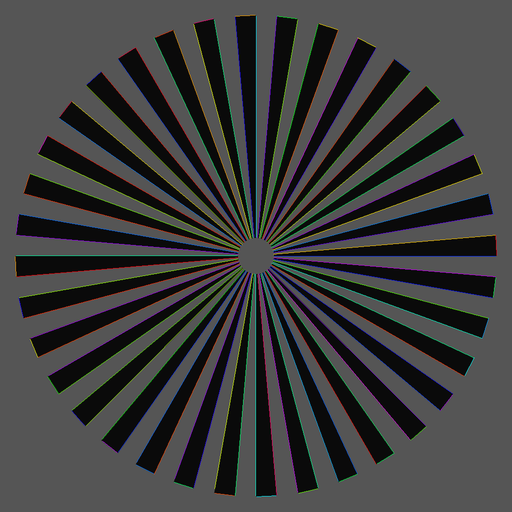}\caption{ELSED: 108}
  \end{subfigure}
  \caption{The Siemens-star probe behind the first row of
  Table~\ref{tab:iso}: input and each detector's segments (counts in the
  subcaptions).}
  \label{fig:star}
\end{figure}

The probes are noise-free, anti-aliased $1024\times1024$ charts --- a
72-sector Siemens star, a 48-line angular fan, concentric rings at 16-px
spacing, and a 26-cycle zone plate --- and CV is computed on a 36-bin
($5$\degr{}) length-weighted orientation histogram.
On straight-line probes, \sweeplsd{}'s orientation isotropy depends on the
probe: on the dense Siemens star it is the weakest of the four (CV $0.74$
vs.\ $0.50$--$0.59$ --- its two-way direction quantization shows), while on
the sparser angular fan it is at parity ($0.21$ vs.\ $0.18$--$0.22$).
Figure~\ref{fig:star} shows what the star number means in practice: all four
detectors recover every spoke and the overlays are close to indistinguishable
by eye --- the CV gap is a statistical property of the length-weighted
orientation histogram, not a visible failure. The circles row of
Table~\ref{tab:iso} separates the designs by intent: a \emph{line segment} detector
asked to describe circles should say ``no lines here''. With the
curve-rejection bound, \sweeplsd{} returns 0 segments on concentric circles
and 27 on a zone plate, while LSD, EDLines, and ELSED fragment the same curves
into $\sim$1{,}400--1{,}800 and $\sim$470--860 short straight segments
respectively (Figure~\ref{fig:circles}). Downstream geometry --- vanishing
points, wireframes, line-based SLAM --- is precisely where such spurious straight
fragments hurt.

The 27 zone-plate survivors delimit what per-segment curve rejection can and
cannot do. Tracing them back through the pipeline shows they are short
tangent chords (14--52\,px) of the widest, most blurred rings, cut where the
rasterization of the two-way-quantized NMS happens to fire endpoint
candidates; their perpendicular deviation from a true line is below one pixel
(RMS spread $0.03$--$0.9$\,px, ratio down to $1/989$). They are, locally,
genuinely straight at pixel scale --- no straightness criterion evaluated on
one run can reject them, and every detector in Table~\ref{tab:iso} emits its
own, more numerous, version of them. Rejecting even these would require
curvature evidence integrated \emph{across} runs. Their conspicuous
orientation CV has the same root: the accidental cut points fall in
four-fold-symmetric azimuth families rather than uniformly --- and 27
segments is a small sample for a CV in any case.

\subsection{Endpoint accuracy: a small, correctable extent bias}
\label{sec:endpoint}

Direction accuracy (Sections~\ref{sec:synth},~\ref{sec:vpdecomp}) is one part
of a segment's geometry; another is where its \emph{endpoints} land.
Against exact synthetic ground truth we measured the signed inward offset
(positive = short of the true tip) of
each detector's endpoint from the true tip, using only \emph{free} tips ---
endpoints well inside the frame and not coincident with another segment's
endpoint, so that neither border clipping nor junctions confound the
measurement --- across two scenes: the building verticals of a synthetic
Manhattan-style scene, and a dedicated controlled scene of isolated
thin high-contrast segments spanning all orientations (which isolates the
endpoint geometry from the contrast gate).

\begin{figure}[t]
  \centering
  \begin{subfigure}[b]{0.98\linewidth}
    \includegraphics[width=\linewidth]{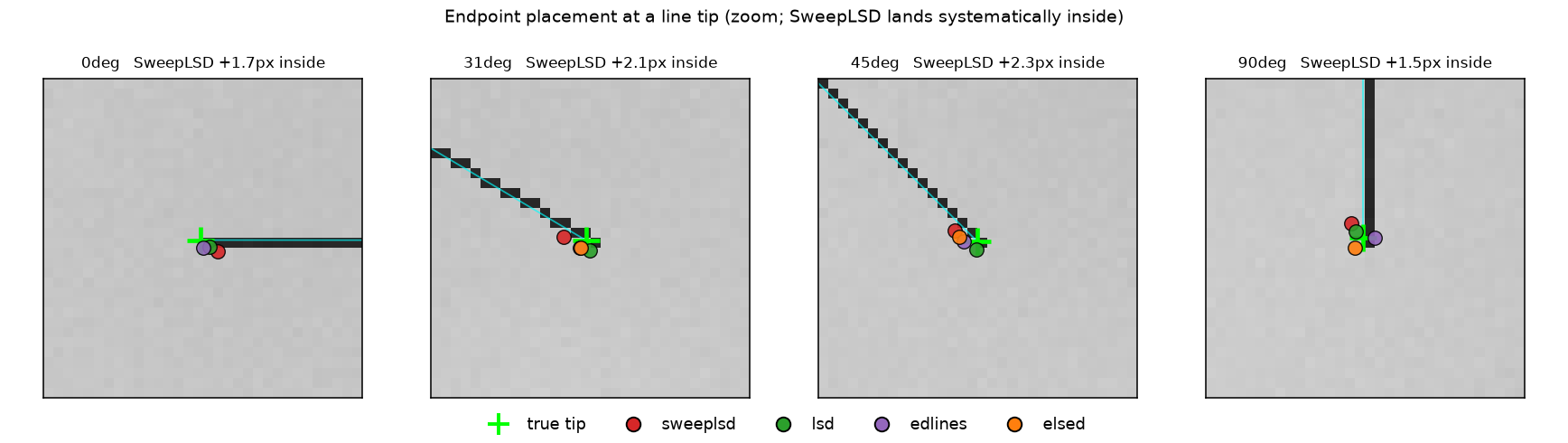}
    \caption{Zoom on a line tip: the true endpoint (green $+$) against each
    detector's endpoint. \sweeplsd{} (red) lands systematically inside.}
  \end{subfigure}
  \\[4pt]
  \begin{subfigure}[b]{0.98\linewidth}
    \includegraphics[width=\linewidth]{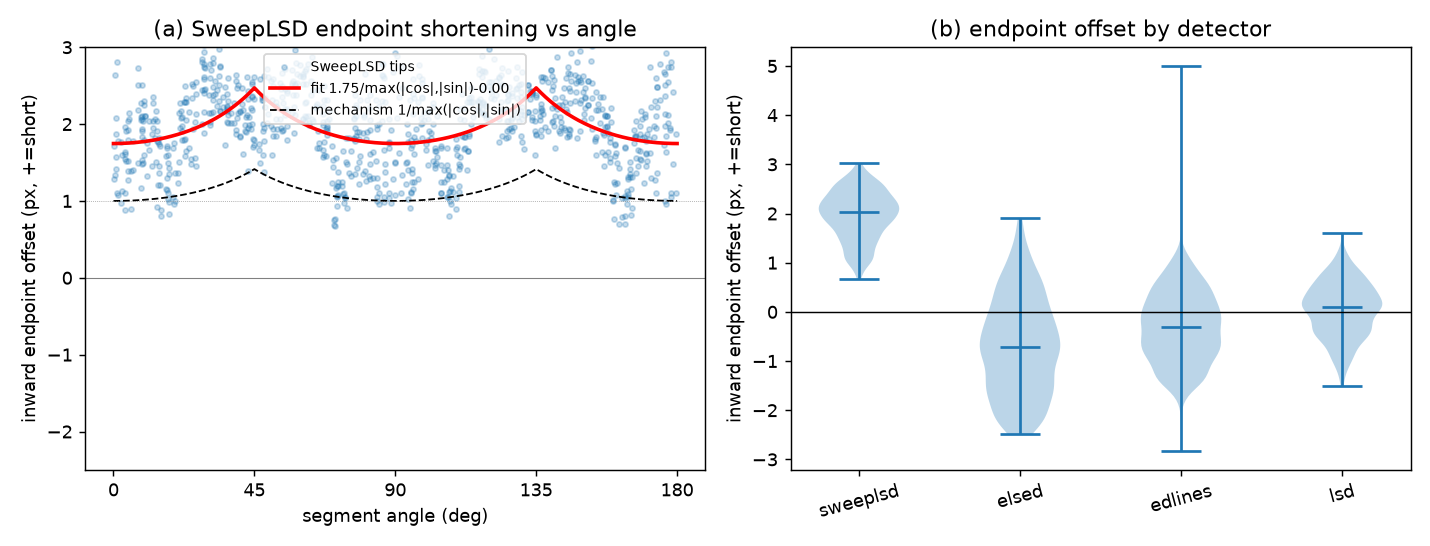}
    \caption{Left: \sweeplsd{}'s inward offset vs.\ angle follows
    $1/\max(|\cos\theta|,|\sin\theta|)$ (dashed), fitted at $1.75\times$ (red).
    Right: offset distribution per detector; only \sweeplsd{} is systematically
    short.}
  \end{subfigure}
  \caption{Endpoint extent bias. \sweeplsd{}'s endpoints are a deterministic
  $\approx$1.7--2.0\,px short (100\% of free tips) --- an angle-dependent,
  correctable \emph{extent} bias that leaves direction, and hence vanishing
  points and attitude, unaffected.}
  \label{fig:endpoint}
\end{figure}

\sweeplsd{} places its endpoints \emph{systematically inside} the true tip:
\textbf{100\% of free tips are short}, by a median of $\approx$1.7--2.0\,px, in
both scenes and at every orientation. The other detectors reach the tip (LSD,
median $+0.1$\,px) or run slightly past it (ELSED $-0.7$, EDLines $-0.3$\,px),
so \sweeplsd{}'s endpoints sit $\approx$1--3\,px inside theirs
(Figure~\ref{fig:endpoint}b). The cause is deterministic and exact: at a free
tip the ring test marks \emph{two consecutive pixels} as endpoint candidates,
and candidates are never labeled (Section~\ref{sec:endpoints}) --- the run's
recorded contact is the interior pixel adjacent to them
(Section~\ref{sec:labeling}). The two candidate pixels rightfully belong to
the segment, so the recovered tip sits roughly two pixel steps inside the
true tip (junction cuts lose their candidate pixels the same way). Along the
segment axis one pixel step spans $1/\max(|\cos\theta|,|\sin\theta|)$ ---
one pixel at $0/90$\degr{}, $\sqrt{2}$ at $45$\degr{} --- so the shortening
follows that angular shape. A least-squares fit to 960 tips gives
$\mathrm{offset} = 1.75/\max(|\cos\theta|,|\sin\theta|)$ with a
constant term of zero to three decimals (Figure~\ref{fig:endpoint}b, left): the bias is
purely this angular shape, at a scale just under the nominal two steps; the
quarter-pixel difference is a sub-pixel effect of the smoothed tip and the
finalization's lattice handling (Section~\ref{sec:judge}).

Two consequences follow. First, this is an \emph{extent} bias, not a
\emph{direction} bias --- it is measured along the fitted axis and leaves
orientation untouched --- so vanishing-point and attitude estimates
(Sections~\ref{sec:vp},~\ref{sec:attitude}) are unaffected, consistent with
\sweeplsd{}'s best-in-class per-segment direction (Section~\ref{sec:vpdecomp}).
Second, because the bias is a closed-form function of angle it is
\emph{correctable}: subtracting the fitted law drives the median absolute
residual to
$0.32$\,px, from $2.0$\,px. The fit generalizes rather than overfits:
trained on half of the rendered frames and scored on the held-out half, the
out-of-sample median absolute residual is $0.33$\,px (500 random
frame-level splits; 2.5--97.5\% range $0.29$--$0.42$\,px). Applications that need absolute endpoint position
(precise measurement, junction detection, line-based metric reconstruction) can
apply this correction or snap endpoints to nearby gradient maxima;
direction-based applications need nothing.

\subsection{Repeatability under rotation and reflection}
\label{sec:equiv}

A detector should return the same segments --- geometrically transformed --- when
the image is rotated by a multiple of $90$\degr{} or mirrored. We test this
directly: each of the 123 photographs of the Full-HD corpus
(Section~\ref{sec:speed}) is transformed by five of the seven
non-identity elements of the dihedral group --- horizontal and vertical
flips, $180$\degr{} rotation, and $\pm90$\degr{} rotations; the two diagonal
reflections, compositions of the maps tested, are omitted --- every detector
is run on each version, its segments are
mapped back to the original frame, and we report the fraction of the original
segments reproduced within 6\,px and 5\degr{} (Table~\ref{tab:equiv}).
\sweeplsd{} is the \emph{most} repeatable of the four --- median $96.8\%$ against
ELSED $70\%$, EDLines $77\%$, and LSD $70\%$, pooled over the $123 \times 5$
image--transform pairs (Table~\ref{tab:equiv} gives per-transform medians).
The fraction-reproduced metric is one-sided --- segments invented on the
transformed frame would not lower it --- so we also computed a symmetric
one-to-one F1 (each mapped-back segment may serve one original segment;
extras count against precision): the ordering and the gap survive,
\sweeplsd{} $93.0\%$ against ELSED $67.6\%$, EDLines $75.5\%$, and LSD
$68.3\%$.
The reason is structural: the
dihedral transforms permute the horizontal and vertical axes among themselves,
and \sweeplsd{}'s deterministic, integer, two-way pipeline maps consistently
under that permutation, whereas the greedy anchor-drawing and region-growing of
the edge-based detectors follow contours in a scan order that the transform
changes, so they re-segment differently. (\sweeplsd{}'s own residual
$\approx$3\% is consistent with its few scan-order asymmetries: the one-sided
NMS tie-break, the merge recency rule, and the adaptive threshold's running
histogram.) Detect time is nearly
orientation-invariant (portrait / $90$\degr{} frames run $\sim$5--7\% slower than
landscape for \emph{every} detector, from having more but shorter rows), and peak
memory does not change with orientation at a fixed resolution (the input image
dominates the footprint, and \sweeplsd{}'s $O(\text{width})$ scratch is far below
that floor). This answers the exact-transform repeatability question
(Section~\ref{sec:limitations}, item vii): under flips and
$90$\degr{}-multiple rotations \sweeplsd{} is the steadiest of the four. One
qualifier is owed, and we measure it rather than leave it open: these
transforms preserve the two-way quantization axes exactly, so they are the
family most favorable to \sweeplsd{}. Repeating the protocol with
\emph{arbitrary-angle} rotations ($15/30/45$\degr{}, bilinear resampling,
same-size canvas; the reference set restricted to segments fully visible
after rotation, with a 6\,px border guard on both sides) inverts the
ordering: every detector drops sharply, and \sweeplsd{} drops the most ---
median $48/44/39\%$ at $15/30/45$\degr{} --- while LSD is steadiest
($64/62/62\%$), with ELSED ($57/54/49\%$) and EDLines ($57/54/54\%$)
between. The symmetric one-to-one F1 leaves this inverted ordering intact
as well (median $44/41/36\%$ for \sweeplsd{} vs.\ $60/59/59\%$ for LSD at
$15/30/45$\degr{}, ELSED and EDLines between), with mapped-back precision
tracking recall for every detector --- none inflates its segment count
after rotation. Under exact axis-preserving transforms \sweeplsd{} is the
steadiest of the four; under axis-moving rotation it is the weakest ---
two faces of the same two-way quantization, and the second is the
price of the first.

\begin{table}[t]
  \centering
  \caption{Repeatability under flips and $90$\degr{}-multiple rotations:
  percentage of a photograph's
  segments reproduced (within 6\,px and 5\degr{}) after the image is flipped or
  rotated by a multiple of $90$\degr{} and the detections are mapped back (median
  over the 123 Full-HD LIU4K-v2 photographs of Section~\ref{sec:speed}).}
  \label{tab:equiv}
  \begin{tabular}{lccccc}
    \toprule
    detector & hflip & vflip & $180$\degr{} & $+90$\degr{} & $-90$\degr{} \\
    \midrule
    \textbf{\sweeplsd{}} & \textbf{97.7} & \textbf{96.3} & \textbf{96.4} & \textbf{97.0} & \textbf{96.9} \\
    ELSED & 65.7 & 81.7 & 64.4 & 70.4 & 70.2 \\
    EDLines (ED\_Lib) & 79.2 & 80.6 & 71.4 & 78.4 & 77.1 \\
    LSD & 71.2 & 71.3 & 66.6 & 70.9 & 71.5 \\
    \bottomrule
  \end{tabular}
\end{table}

\section{Applications: do the architectural advantages reach the task?}
\label{sec:apps}

The streaming design buys three things a frame-buffer detector cannot offer at
once: $O(\text{width})$ memory, a per-pixel cost that barely moves with image
content, and segments emitted a few rows behind the scan line
(Sections~\ref{sec:onepass},~\ref{sec:memlat}). Those are micro-architectural
properties; the question for an application is whether they survive to
\emph{task} accuracy. We answer it on calibrated-camera geometry along three
axes: single-frame camera attitude, where synthesis lets us hold accuracy fixed
while sweeping resolution to 4K (Section~\ref{sec:attitude}); the generalization
of that same attitude task to two real visual-inertial datasets with metric
ground truth (Section~\ref{sec:realdata}); and vanishing-point recovery, where we
compare the detectors head-to-head on real urban imagery (Section~\ref{sec:vp}).
A closing scope subsection (Section~\ref{sec:scope}) states where the method fits
and where it does not.

\subsection{Attitude from a single frame, from 540p to 4K}
\label{sec:attitude}

\textbf{Protocol.} Real datasets with line-task ground truth top out near
2\,MP, and upsampling only
manufactures false edges, so the resolution claim of Section~\ref{sec:speed}
cannot be tested downstream on real data. We instead render two photorealistic
scenes --- the Blender demo assets Classroom (indoor) and Barcelona Pavilion
(outdoor) --- at 40 camera poses per scene (Blender Cycles, no added sensor
noise), from 540p to 4K with the poses \emph{identical} across resolutions
(only the sampling density changes) and \emph{exact} attitude ground truth;
the medians below are over the 40 poses. Each detector runs at its detection
defaults; a common resolution-scaled minimum segment length (12\,px at
480-row scale) filters every detector's segments before estimation. Each
detector's segments feed the \emph{same} calibrated Manhattan-frame (gravity)
estimator --- a fixed common estimator, used here to test accuracy parity
across resolutions; the per-detector fairness question of
Section~\ref{sec:fair} is taken up on the real data of
Section~\ref{sec:realdata} --- and we report the angle between the recovered
and true gravity directions. In brief: with known intrinsics each segment votes for the
vanishing directions it is consistent with, a multi-start search recovers the
orthogonal (Manhattan) direction triad with the strongest inlier support, and
the attitude is taken from the triad's vertical axis --- then re-fitted against
that axis' own inlier lines. The refinement matters here: as a triad member the
vertical axis is constrained orthogonal to the other two, and outdoors those are
weakly determined (an architectural exterior offers few horizontal Manhattan
cues), so the constraint transfers their error onto gravity. Releasing it
improves \emph{every} detector on these scenes by $3$--$20\times$. The trade is
scene-dependent, and we do not take it on the real data of
Section~\ref{sec:realdata}: in cluttered rooms all three axes are well
supported, the constraint is a useful regularizer rather than a liability, and
releasing it costs $0.07$--$0.1$\degr{} of mean error on EuRoC. The estimator is
detector-agnostic; Section~\ref{sec:vp} specifies the evaluation protocol and
the estimator-variant menu in full, the estimator itself ships as working
code in the repository, and Section~\ref{sec:realdata} reuses the
cross-validated best-estimator-per-detector protocol.

\begin{figure}[t]
  \centering
  \includegraphics[width=\linewidth]{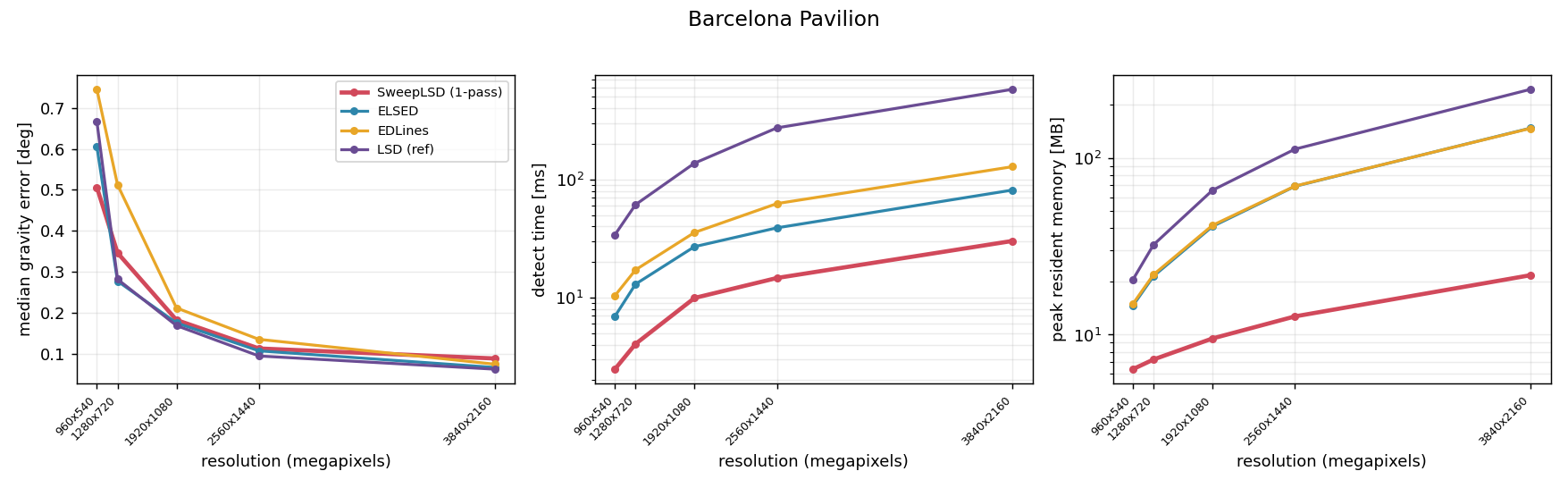}
  \caption{Single-frame attitude on the Barcelona Pavilion, 540p--4K at 40
  poses held identical across resolutions, with exact ground truth. Left:
  median gravity error --- the four detectors converge to $0.06$--$0.09$\degr{}
  at 4K, within $0.03$\degr{} of one another. Middle/right: detect time and peak memory ---
  with only $O(\text{width})$ detector state atop the shared input image,
  \sweeplsd{}'s whole-process 4K peak sits at $\approx$1/7 of
  the edge-drawing detectors and 1/11 of the reference LSD.}
  \label{fig:attitude}
\end{figure}

\textbf{Accuracy is preserved, not traded.} Across the sweep the four detectors
are within each other's noise on attitude, and they converge as resolution
rises. Outdoors the median gravity error falls from $0.51$--$0.75$\degr{} at
540p to $0.06$--$0.09$\degr{} at 4K, where all four sit within $0.03$\degr{} of
one another (\sweeplsd{} $0.09$, ELSED $0.07$, EDLines $0.07$, LSD
$0.06$\degr{}); indoors the picture is the same with no consistent ordering
($0.60$--$1.01$\degr{} at 540p, $0.08$--$0.12$\degr{} at 4K; the classroom
curves are not plotted in Figure~\ref{fig:attitude}). \sweeplsd{} leads at 540p
outdoors and trails at 4K by $0.03$\degr{} --- a gap far below what 40 poses
resolve: bootstrap 95\% confidence intervals on the four 4K medians are
$0.03$--$0.09$\degr{} wide and mutually overlapping. Attitude accuracy is therefore not a discriminator between these
detectors; the speed and memory that follow are not paid for in task accuracy.
What the sweep does show sharply is that \emph{resolution} pays: \sweeplsd{}
improves $5.7\times$ from 540p to 4K ($0.51 \to 0.09$\degr{},
Figure~\ref{fig:attitude}, left).

\textbf{Memory and speed diverge with resolution.} At 4K (Barcelona Pavilion)
\sweeplsd{} detects in $\approx$30\,ms against ELSED's 81\,ms ($2.7\times$),
EDLines' 128\,ms, and the reference LSD's $\approx$580\,ms, and holds a
peak resident set of
$\approx$22\,MiB against ELSED's and EDLines' $\approx$148\,MiB ($6.8\times$) and
the reference LSD's 245\,MiB ($11\times$) (Figure~\ref{fig:attitude}, middle and
right). All four detectors run under the single toolchain of
Section~\ref{sec:speed}. \sweeplsd{}'s memory grows only $3.4\times$ (6.4\,$\to$\,21.7\,MiB) as
the pixel count grows $16\times$, and the largest single item in the 22\,MiB
is the $\approx$7.9\,MiB input image: the detector's own working set is
$O(\text{width})$, exactly as Section~\ref{sec:memlat}
measured, now shown at the task level and at a resolution real ground truth
cannot reach. (ELSED's medians track the others, but its error \emph{tail}
grows at 4K --- p90 $10.1$\degr{}: its defaults target $640{\times}480$, so at
4K it over-splits foliage into short lines that survive the shared minimum
length and pollute the inlier vote. At its design minimum length --- a length
cut downstream of the same detections, so its $81$\,ms detection time is
unchanged by construction --- the tail returns to parity: p90
$10.1 \to 3.3$\degr{} against the other detectors' $2.4$--$8.2$\degr{}, with
the median unchanged. The memory and speed gaps are independent of it.) Downstream cost does not change this picture: the
Manhattan-frame estimation that consumes the segments adds $5$--$7$\,ms at 4K
for \emph{every} detector --- it scales with the number of surviving lines,
not with pixels (the horizon-lock demonstration below filters to
$\approx$87 lines, which is why its estimation costs only $\approx$1\,ms) ---
so including it compresses the ratios only slightly
(\sweeplsd{} $\approx$36\,ms against ELSED's $\approx$88\,ms end-to-end,
$2.4\times$ instead of $2.7\times$).

\begin{figure}[t]
  \centering
  \includegraphics[width=\linewidth]{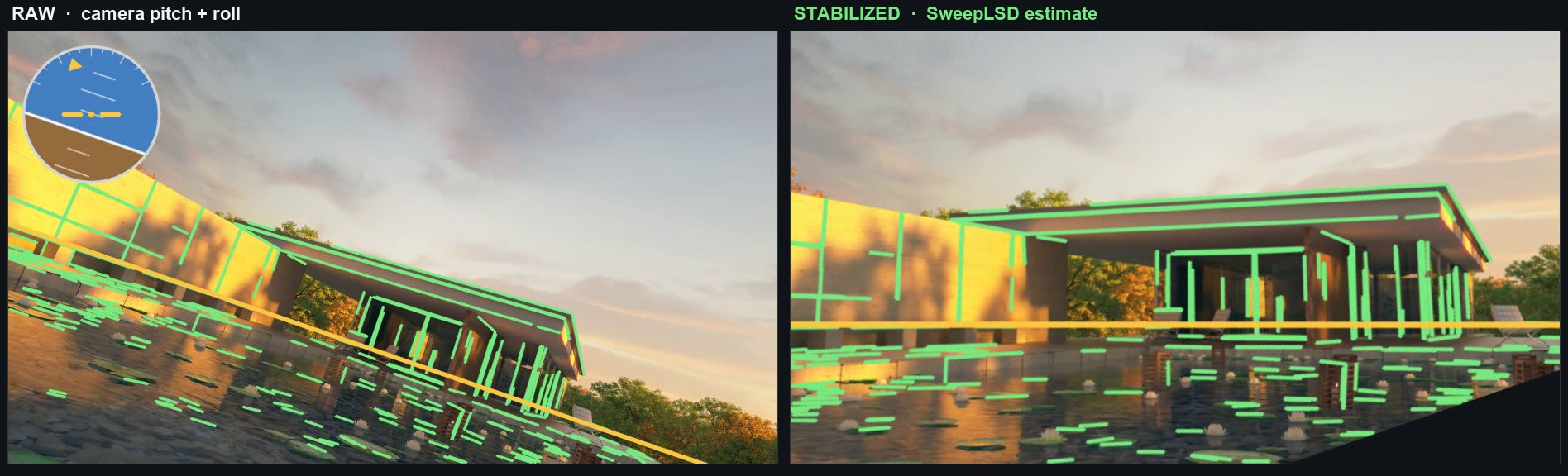}
  \caption{Real-time horizon lock, one 4K frame near an attitude extreme
  (estimated roll $-19.1$\degr{}, pitch $+9.8$\degr{}; error against truth
  $0.03$\degr{}). Left: raw frame with detected segments (green) and the
  \sweeplsd{}-estimated horizon (amber) over ground truth (blue) --- at this
  error the two are indistinguishable; the attitude-indicator inset (top left)
  shows the estimated roll and pitch. Right: the same frame counter-rotated and
  re-centered by the estimate. Median attitude error $0.06$\degr{} over the
  150-frame flight, with no temporal filtering.}
  \label{fig:horizon}
\end{figure}

\textbf{Real-time horizon lock.} As an end-to-end demonstration we stabilize a
4K forward flight through the Pavilion (camera rolling $\pm20$\degr{} and
pitching $\pm12$\degr{}) by counter-rotating and re-centering each frame with
\sweeplsd{}'s per-frame attitude estimate --- the estimator above
with a resolution-scaled minimum segment length and \emph{no} temporal
filtering, so every frame is independent. The attitude error has median
$0.06$\degr{} over the 150-frame flight (bootstrap 95\% CI
$0.05$--$0.07$\degr{}; mean $0.06$, worst frame
$0.15$\degr{}); the estimated horizon sits on the true one closely enough that
the two are hard to tell apart, while the raw feed tilts and pitches
(Figure~\ref{fig:horizon}). The complete
per-frame compute --- detection \emph{plus} attitude estimation --- has a
median of 31.9\,ms (detection 30.7, estimation 1.2 on the $\approx$87
length-filtered lines; worst frame 45.2\,ms; per-quantity medians, so the
parts need not sum), so the whole loop, not just the
front end, meets the 30\,fps budget in the median on a single desktop core.
Temporal smoothing is the obvious addition to a stabilizer, and we deliberately
omit it: at these rotation rates an exponential moving average \emph{lags}, and
the lag dominates whatever jitter it suppresses. Smoothing the same estimates
with $\alpha = 0.65$ costs an order of magnitude --- $0.06 \to 0.63$\degr{}
median. Per-frame geometry this accurate is what makes the temporal term
unnecessary, and omitting it is what keeps the demonstration single-frame, and
so bounded-latency, end to end.
By the standard of Section~\ref{sec:speed} a median is not a deadline: the
worst frame overruns the 33.3\,ms budget, so a deployment that must never
drop a frame needs a one-frame buffer or a drop policy at this resolution on
this CPU. The
$O(\text{width})$, low-latency front end is what makes per-frame 4K attitude
affordable in the first place. Only rendering the stabilized output falls
outside this budget: it is presentation, and in a deployed system (a gimbal or crop window) it is
not part of the estimate's critical path.

\subsection{Generalization to real data: EuRoC and TUM-VI}
\label{sec:realdata}

Synthesis let us isolate the resolution effect; two real visual-inertial
datasets let us check that the single-frame attitude estimator survives real
optics, MAV and handheld motion, and imperfect scene verticality. We reuse the estimator of
Section~\ref{sec:attitude} unchanged --- single frame, monocular, geometry only,
with \emph{no temporal smoothing} --- and score the recovered gravity direction
against each dataset's ground-truth poses. With \sweeplsd{} segments,
\textbf{EuRoC MAV}~\cite{burri2016}
($752{\times}480$, all 11 sequences, 26{,}540 frames) gives an overall median
gravity error of $0.73$\degr{} ($<3$\degr{} on 87.8\% of frames) at
$\approx$284\,fps on one core; \textbf{TUM-VI}~\cite{schubert2018} (handheld
fisheye reprojected to a $512^2$ pinhole, six rooms, 15{,}393 frames) gives
$0.96$\degr{} median ($<3$\degr{} on 88.5\%) at $\approx$511\,fps. Both
throughput figures are single-frame, geometry-only, one-core, and cover
detection \emph{plus} estimation.

\textbf{The error floor is the scene, not the detector.} The rigid Machine Hall
sequences reach $0.33$--$0.44$\degr{} median, while the instrumented
Vicon/mocap rooms
(EuRoC V-series, all TUM-VI rooms) sit near $1$\degr{}. An IMU cross-check ---
recovering gravity dynamically from accelerometer and pose --- shows the
ground-truth world-$z$ axis is itself tilted by $\le 0.006$\degr{} in Machine
Hall but by $0.06$--$0.47$\degr{} in the mocap rooms, so a single visual frame
can only return the scene's \emph{dominant pseudo-vertical}, and its floor is set
by how truly vertical the room was built. This is a property of monocular
single-frame gravity, not of the line detector, and we record it as an
intrinsic limit of the task.

A second measurement reinforces the resolution story of
Section~\ref{sec:attitude} from the opposite side. TUM-VI ships both a native
$1024^2$ capture and an official $512^2$ downscale of the same footage --- a
\emph{real} $4\times$ detail difference, not upsampling. Feeding the $4\times$
finer imagery leaves the pooled six-room accuracy unchanged; the $512^2$
version is in fact marginally better (median $0.96$ vs.\ $1.01$\degr{}),
because the floor here is the scene's pseudo-vertical ($\approx$0.8\degr{}) plus
the ground-truth tilt ($\approx$0.1\degr{}), not line localization. Resolution
pays off only when the task is localization-limited --- the synthetic sweep of
Section~\ref{sec:attitude}, where rendered detail genuinely accrues --- not on a
task whose accuracy is capped by scene geometry.

Finally, the detector-agnostic protocol of Section~\ref{sec:fair} --- every
detector's segments into the identical estimator, with estimator variants
selected per detector on one room and evaluated on the remaining five ---
reproduces the vanishing-point verdict on
attitude. This comparison runs on TUM-VI's native $1024^2$ frames (unlike the
$512^2$ headline above). At standard field of view (test rooms 2--6)
\sweeplsd{} has the best median ($0.96$\degr{}, $<1$\degr{} on 51.3\% of
frames) against ELSED $1.19$\degr{}, EDLines $1.06$\degr{}, and reference LSD
$1.06$\degr{}, while detecting fastest ($3.75$\,ms one-pass at $1024^2$,
$2.5\times$ ELSED and $3.9\times$ EDLines); at a wide $139$\degr{} field of
view the four converge to a near-tie ($0.89$--$0.95$\degr{}), with the ELSED
detection gap at $2.3\times$. As on the York Urban / NYU study of
Section~\ref{sec:vp}, task accuracy is at parity and the streaming design's
speed is the discriminator --- now on handheld real data.

\subsection{Vanishing points: a head-to-head detector study}
\label{sec:vp}

\textbf{Protocol.} As noted in Section~\ref{sec:related}, hand-labeled line
annotations are incomplete and coarse, so detectors are not scored against
them. Instead each detector's segments feed the
\emph{same} calibrated Manhattan-frame estimator, and the recovered triplet of
orthogonal vanishing directions is compared against rotation ground truth ---
York Urban~\cite{denis2008} (102 images of urban scenes --- 45 indoor, 57
outdoor) and NYU-VP~\cite{kluger2020}
(1{,}033 indoor images from NYU Depth v2~\cite{silberman2012} --- the subset
of NYU-VP's 1{,}449 scenes whose ground-truth vanishing points supply a clean
orthogonal triad, which a rotation-error protocol requires). Per image, the
error is the mean angular deviation of the three recovered directions from
the ground-truth axes; we report medians over each dataset.

\begin{table}[t]
  \centering
  \caption{Fixed common estimator: median Manhattan-frame error per dataset
  (every detector's segments into the identical estimator; hand-labeled
  ground-truth lines as the ceiling --- that row is this estimator's reading
  of the annotation limit, not a property of the corpus alone, and a different
  Manhattan-frame estimator reads it differently). The fixed estimator is the menu's
  length-weighted, $2$\degr{}-band, base-search variant
  (Section~\ref{sec:fair}). The ELSED and \sweeplsd{}+linker rows
  are from a verification re-run with the same estimator, which reproduces
  the other rows to within 0.2\degr{}.}
  \label{tab:fixedest}
  \begin{tabular}{lcc}
    \toprule
    segments into the fixed estimator & York Urban & NYU (indoor) \\
    \midrule
    LSD & 0.92\degr{} & 10.7\degr{} \\
    EDLines (ED\_Lib) & 0.98\degr{} & 10.4\degr{} \\
    ELSED & 0.96\degr{} & 10.4\degr{} \\
    \sweeplsd{} & 1.04\degr{} & 12.7\degr{} \\
    \sweeplsd{} + linker & 0.99\degr{} & 10.7\degr{} \\
    \midrule
    hand-labeled lines (ceiling) & 0.59\degr{} & 6.0\degr{} \\
    \bottomrule
  \end{tabular}
\end{table}

\textbf{Fixed common estimator.} Table~\ref{tab:fixedest} gives the medians
under a single fixed estimator: the three baselines sit in one band on both
datasets, \sweeplsd{} trails the band, and the optional linker of
Section~\ref{sec:synth} lifts \sweeplsd{} to its edge --- the fragmentation
repair feeds directly through to downstream support. ELSED lands in that
band with
the fewest segments per image of the three baselines --- in line with the
support-per-line picture of Section~\ref{sec:vpdecomp}, though the
decomposition there covers \sweeplsd{}, LSD, and ED\_Lib. Read the indoor
medians with care: the indoor error distribution is \emph{bimodal} --- an image
either succeeds ($\approx$1.5\degr{}) or the frame search locks onto the wrong
structure ($\approx$27\degr{}) --- so the median is only the mixing ratio of the
two modes, and even the hand-labeled ceiling sits at 6\degr{}. NYU is a
clutter-robustness stress test, not a precision benchmark.

\subsubsection{Where the differences come from}
\label{sec:vpdecomp}

A per-line decomposition against the ground-truth axes --- per-line residual
angles and per-axis inlier support, computed for \sweeplsd{}, LSD, and
ED\_Lib --- locates the source of the gaps:

\begin{itemize}
\item \textbf{Per-line direction quality is best for \sweeplsd{}} on real
  photographs too: median residual against the ground-truth axes 0.51\degr{}
  $<$ LSD 0.57\degr{} $<$ ED\_Lib 0.65\degr{} on York Urban. The synthetic-GT
  geometry advantage of Section~\ref{sec:synth} is not an artifact.
\item \textbf{The ranking is driven by inlier \emph{support}, especially on the
  weakest axis.} \sweeplsd{}'s contrast-gated edge model brings $\sim$20\% less
  supporting line length. Subsampling LSD's segments down to \sweeplsd{}'s
  count reproduces \sweeplsd{}'s York Urban score almost exactly
  (0.92\degr{} $\to$ 1.05\degr{}, vs.\ \sweeplsd{}'s 1.04\degr{}): the York
  Urban gap is quantity, not quality.
\item \textbf{Indoors, the three fail on largely the same images.} Images in
  the bottom quartile of weak-axis support fail for each of them; on
  the 360 images where all three succeed, they are statistically
  indistinguishable (1.56\degr{} / 1.53\degr{} / 1.48\degr{}).
\end{itemize}

\subsubsection{Best estimator per detector, cross-validated}
\label{sec:fair}

Because the differences are support-driven, they depend strongly on how the
estimator weighs lines --- which raises a fairness problem: a single fixed
estimator implicitly favors whichever detector its weighting suits. We
therefore fixed, in advance, a menu of twelve estimator variants, identical
for every
detector: the cross of line weighting \{length, unit\} with inlier band
\{1\degr{}, 2\degr{}, 3\degr{}\}, plus, at the two wider bands, variants with
a stronger multi-start search, and, at 2\degr{}, with an additional
vertical-prior seed on the stronger search ($6+4+2=12$). We let \emph{each detector pick its best}, with
\textbf{selection and evaluation on separate data} (split-half
cross-validation within each dataset: variants are selected by median error
on one half and scored on the other, so every image is evaluated under a
variant chosen without it). One asymmetry remains and should be weighed
alongside the table: the separation covers estimator selection only ---
the \sweeplsd{} detector configuration itself was developed with York
Urban / NYU downstream error among its tuning signals
(Section~\ref{sec:eval}), whereas the baseline detectors run their
upstream defaults. The menu harness and the aggregation ship in the repository;
independent re-runs of the same build reproduce Table~\ref{tab:fair} within
the multi-start search's scatter (on York Urban within 0.03\degr{}).

\begin{table}[t]
  \centering
  \caption{Best-estimator-per-detector protocol, split-half cross-validation
  (selection on one half, evaluation on the other; pooled test halves). All
  four detectors use the identical 12-variant menu. Time columns are medians
  at $640{\times}480$, one toolchain and one measurement window for all
  detectors, interleaved image-by-image (detection: median of 15 runs;
  estimation: median of 5 runs of each image's cross-validated pick, timed on
  the exact segments behind the accuracy columns). All time columns are
  per-image medians, so detect and estimate need not sum exactly to total.
  VP-estimation time is a property of the selected variant rather than of the
  detector and is not bolded.}
  \label{tab:fair}
  \small\setlength{\tabcolsep}{3.4pt}
  \begin{tabular}{llcccccc}
    \toprule
    dataset & detector & median err (cross-val.) & $<2$\degr{} & $<5$\degr{}
      & \multicolumn{3}{c}{time (ms)} \\
    \cmidrule(lr){6-8}
    & & & & & detect & VP est. & total \\
    \midrule
    \multirow{4}{*}{NYU (indoor)}
      & \textbf{\sweeplsd{}} & \textbf{6.08\degr} & \textbf{33.2\%} & \textbf{48.6\%}
        & \textbf{1.5} & 2.1 & \textbf{3.7} \\
      & ELSED & 7.67\degr & 32.4\% & 47.1\% & 4.0 & 1.6 & 5.7 \\
      & LSD & 7.87\degr & 29.2\% & 46.4\% & 27.1 & 3.0 & 30.1 \\
      & EDLines (ED\_Lib) & 8.09\degr & 30.5\% & 45.7\% & 5.8 & 2.9 & 8.9 \\
    \midrule
    \multirow{4}{*}{York Urban}
      & ELSED & \textbf{0.82\degr} & 84.3\% & 96.1\% & 5.5 & 2.8 & 8.4 \\
      & LSD & 0.83\degr & \textbf{93.1\%} & 96.1\% & 32.0 & 3.7 & 35.7 \\
      & \sweeplsd{} & 0.94\degr & 85.3\% & 95.1\% & \textbf{1.8} & 2.4 & \textbf{4.2} \\
      & EDLines (ED\_Lib) & 1.00\degr & 89.2\% & \textbf{98.0\%} & 7.3 & 3.5 & 11.0 \\
    \bottomrule
  \end{tabular}
\end{table}

Table~\ref{tab:fair}: \textbf{on indoor NYU the cross-validated \sweeplsd{}
detector--estimator pair leads on every
metric} --- median, $<2$\degr{}, and $<5$\degr{} --- now including over
ELSED's; the selected variant counts each line as one vote. The mechanism is the support picture of
Section~\ref{sec:vpdecomp}: length-weighted voting lets long clutter contours
hijack the frame search, whereas \sweeplsd{}'s support consists of many short
but accurate lines. On York Urban the result is a metric-dependent near-tie:
ELSED and LSD share the best median (0.82\degr{} and 0.83\degr{}), LSD
clearly leads recall at 2\degr{}, and \sweeplsd{} is 0.1\degr{} behind on the
median. The York Urban medians are also near the protocol's reproducibility
floor: recompiling the harness with a different GCC generation perturbs the
baselines' double-precision detections at the last-bit level, which moves
LSD's cross-validated York Urban median between 0.83\degr{} and 0.94\degr{}
(ELSED's and \sweeplsd{}'s by $\le$0.02\degr{}; every NYU median reproduces
to 0.01\degr{}) --- so LSD's outdoor lead over \sweeplsd{} is within build
jitter, and the stable outdoor read is ELSED $\approx$0.1\degr{} ahead of
both, while the indoor ordering is unaffected. (\sweeplsd{}'s own detections
are exempt from this jitter: its integer pipeline reproduces bit-identically
across those same compiler generations, Appendix~\ref{app:compiler}.)

The split itself is a second sensitivity, quantified the same way:
re-running the identical selection rule over 1{,}000 random half splits (in
place of the deterministic even/odd split) leaves every ordering conclusion
intact --- ELSED and LSD trade the best York Urban median almost evenly
(54\%/46\% of splits), both lead \sweeplsd{} in $\approx$97\% of splits, and
on NYU \sweeplsd{} leads ELSED in 92\% of splits and LSD and ED\_Lib in all
of them --- while exposing one split-sensitive cell: ELSED's NYU median. Its
selection flips between two variants, and its typical (median-over-splits)
value is $6.35$\degr{}; the even/odd split of Table~\ref{tab:fair} happens
to land in that distribution's sparse upper tail ($7.67$\degr{}), so the
table's $1.6$\degr{} \sweeplsd{} margin overstates the typical
$\approx$0.3\degr{} margin. The seven other medians move by
$\le$0.03\degr{}.

The optional linker (Section~\ref{sec:synth}) compounds the indoor lead.
Under the identical menu and cross-validation it scores \textbf{5.23\degr{}}
on NYU; its York Urban cross-validated median, 0.93\degr{}, is
indistinguishable from the linker-less 0.94\degr{}, and the linker-less
configuration, run alongside as a control, reproduces Table~\ref{tab:fair}
exactly, so the two
configurations are directly comparable.

One estimator was best overall --- unit-weight voting with
vertical-prior seeding and the stronger multi-start search --- and ships as
working code in the repository. With
it, \sweeplsd{} scores 5.98\degr{} median on NYU, and \textbf{5.25\degr{}
with the optional linker --- the best of all detector/estimator combinations
measured}; the best non-\sweeplsd{} combination is ELSED at 6.53\degr{}.
(These single-estimator numbers are selected on the full data rather than
cross-validated --- given for practical use; their cross-validated
counterparts are the 6.08\degr{} of Table~\ref{tab:fair} and the linker's
5.23\degr{}.)

Because the task is the vanishing points, not the segments, we report the
task's time the same way we report its accuracy (Table~\ref{tab:fair}, time
columns). Detection keeps its usual ordering --- \sweeplsd{}
$2.6\times$/$3.8\times$/$17.6\times$ faster than ELSED/ED\_Lib/LSD on NYU
($3.1\times$/$4.2\times$/$18.3\times$ on York Urban).

The estimation stage was profiled and optimized for this measurement
(its scoring hotspot vectorized, verified bit-identical over all 68{,}100
menu evaluations), yet still costs $1.6$--$3.7$\,ms per frame at
$640{\times}480$: the remaining work is the multi-start refinement's serial
per-seed eigensolves, which cannot be batched without changing the result.
For the fastest detector, that estimation cost exceeds \sweeplsd{}'s own
$1.5$--$1.8$\,ms detection. End-to-end, \sweeplsd{} has
the fastest pipeline on both datasets (3.7\,ms on NYU against ELSED's 5.7),
but the margin compresses to $1.5\times$--$2.0\times$ over ELSED: at this
resolution the shared estimator, not the detector, bounds the frame time,
and that is a property of the task, not of an unoptimized implementation.
The picture inverts at high resolution, where estimation stays a few
milliseconds ($5$--$7$\,ms at 4K; it scales with surviving lines, not
pixels) while detection grows with pixel count --- Section~\ref{sec:attitude}
measures $30$--$580$\,ms of detection at 4K (the 580 being LSD).

The linker adds
$\approx$0.6\,ms of detection on NYU (2.1 vs.\ 1.5\,ms) and leaves
estimation unchanged, so its accuracy gain (5.23\degr{}) is not bought with
meaningful time. ED\_Lib's detect time is measured on a same-toolchain
rebuild whose output differs from the accuracy run's segments by $\pm$1--2
segments per frame (its edge chaining is build-sensitive; the other
detectors reproduce to float rounding or exactly).

Within this study, ``which detector is best for vanishing points?'' therefore
has no detector-only answer: the ordering above is a property of the (detector
$\times$ estimator) pair, and the fixed-estimator ordering of
Table~\ref{tab:fixedest} is in part a ranking of estimator compatibility. Two
single-image corpora and one twelve-variant menu do not settle how far that
generalizes; here the protocol is the fairness device this comparison needed.

\subsection{Scope: where \sweeplsd{} fits, and where it does not}
\label{sec:scope}

The applications above map out a sweet spot. \sweeplsd{} is at its best when the
scene is \emph{structure-dominant} (architecture, indoor Manhattan geometry,
urban imagery), when the task needs \emph{line direction} rather than absolute
endpoint position (attitude, vanishing points), and --- above all ---
when the deployment is \emph{high-resolution or memory-/power-constrained}, where
the $O(\text{width})$ working set and the content-stable, tight-tailed
per-pixel cost (Sections~\ref{sec:speed},~\ref{sec:memlat},
Appendix~\ref{app:timing}) are worth the most. That is precisely the embedded, streaming-sensor, FPGA-oriented regime the
original thesis targeted.

Four conditions push outside it. First, \textbf{curved and textured content}:
the two-way gradient quantization fragments curves into short chords, and the
judge's absolute straightness bound rejects them
(Section~\ref{sec:isotropy}), so \sweeplsd{} is a poor fit
for lane-marking or contour tasks in their raw form (see the outlook below for a
possible remedy) and for heavily textured scenes. Second, \textbf{low edge
density or low resolution}: the same content-flat cost that keeps the frame
time predictable
is also why the speed margin over edge-drawing detectors narrows when there is
little to draw --- on the sparse mocap rooms of Section~\ref{sec:realdata}
the ELSED gap falls from the
$\approx$4.6$\times$ of dense high-detail imagery to $2.3$--$2.5\times$, and at low
resolution the memory and latency advantages simply have less to act on. Third,
\textbf{absolute endpoint position}: the raw endpoints are deterministically
short (Section~\ref{sec:endpoint}) and the contrast gate can clip faint line
ends (Section~\ref{sec:limitations}); direction-only tasks are insensitive to
both, but metrology or junction-precise tasks should apply the angle-based
correction of Section~\ref{sec:endpoint} for the extent bias --- the
gate-clipped ends, being missing evidence, cannot be recovered by a
correction. Fourth, \textbf{coverage-hungry tasks on soft, low-contrast
structure}: the contrast-gated edge model misses $\sim$20\% of supporting
line length on real scenes
(Sections~\ref{sec:vpdecomp},~\ref{sec:limitations}); applications that must
recover every faint edge are better served by LSD or EDLines.

\textbf{Outlook (untested).} Several directions follow naturally from the
scan-line front end but are left as future work. (i)~The row-streaming pipeline is
a structural match for \emph{line-scan} industrial sensors, which deliver one
image row at a time --- the detector could run without ever buffering a frame.
(ii)~The curve limitation could be turned into a \emph{domain specialization}:
inserting a blank edge row periodically in $y$ (say every 50 rows) would cut long
curves into short quasi-straight chords that the labeler recovers as a chain of
segments --- a candidate lane/curve extractor built by \emph{removing} work rather
than adding it. (iii)~The $O(\text{width})$ memory makes very large images cheap,
suggesting high-DPI document and engineering-drawing line extraction, or
whole-slide and aerial mosaics, where a frame-buffer detector's footprint is the
binding constraint. (iv)~Room-layout and wireframe estimation would reuse the same
short-but-accurate-segment support that won the indoor vanishing-point study.
All four are prospective; none is measured here.

\section{Hardware realization}
\label{sec:fpga}

The 2014 thesis proposed the algorithm \emph{for} hardware but implemented it
only in software; the FPGA form was left as future work. We close that gap
with two independent hardware expressions of Section~\ref{sec:method}, both
held to a single contract: \textbf{software, HLS, and RTL must produce
bit-identical segments}. The contract is decided at the emitted integer run
records --- the hardware data path ends there, and the once-per-segment
floating-point finalization of Section~\ref{sec:judge} is the same host code
for all three expressions, so record-level equality is segment-level
equality.

\textbf{HLS.} The detector was re-expressed as synthesizable HLS C++ (Vitis
HLS). On an Artix-7 (XC7A35T) target the reports show every front-end stage at
initiation interval II${=}1$ at 100\,MHz --- one pixel per clock, as the
streaming design intends --- and C/RTL co-simulation passes. (The labeling
back end, in both hardware expressions, is not a pixel-synchronous pipeline
but an event-driven unit of variable latency behind the event FIFO --- a
measured $\approx$6 cycles per interior event on the heaviest Full-HD
photograph profiled;
the overload paragraph below quantifies its budget.) The same sources
compile as ordinary C++, so the model is parity-tested against
the multi-pass software reference without any vendor tools in the loop.

\textbf{RTL.} A hand-written, portable Verilog core (streaming front end
$\rightarrow$ sparse event FIFO $\rightarrow$ event-driven labeling back end
with a single shared sequential multiplier for the judge) runs \emph{live} on
a Digilent Atlys --- a 2009-era Spartan-6 XC6SLX45: HDMI in, detect, green
overlay, HDMI out, at 1080p30 and 720p60, entirely in the single recovered
pixel-clock domain at 74.25\,MHz. There is \textbf{no frame buffer and no
external memory}: detector state occupies $\approx$70\,KiB of on-chip block
RAM (matching the on-chip budget the thesis projected), and with the demo
overlay mask and video I/O the
whole system uses $\approx$140\,KiB of the device's 261\,KiB. Segments emerge
a median 6.5 rows of sweep progress after their last pixel
(Sections~\ref{sec:onepass},
\ref{sec:memlat}), $\approx$0.19\,ms at 1080p30 --- at no point does a frame
exist in memory.

\textbf{Resource usage.} Table~\ref{tab:fpga} gives both realizations. The HLS
core proves the streaming schedule --- front-end stages at II${=}1$, one pixel
per clock --- but, instantiating a moment multiplier per product, it spends 79
DSP blocks and 95\% of the small Artix-7's LUTs on the detector alone. The
hand-written RTL folds every judge product onto \emph{one} shared sequential
multiplier: the detector's arithmetic drops to $\approx$7 DSP blocks (4 in
the shared multiplier, 3 in the moment accumulators), and the \emph{complete}
live system --- detector, HDMI receive and transmit, and
the overlay drawer --- uses 14 DSP blocks rather than the HLS core's 79 and
fits the 2009-era Spartan-6 within budget (46\% of
LUTs, 60\% of slices, $\approx$75\% of block RAM), with every timing constraint
met (critical path 13.37\,ns, i.e.\ 74.8\,MHz against the 74.25\,MHz pixel
clock).

\begin{table}[t]
  \centering
  \caption{Hardware resource usage of the two realizations. ``HLS core'' is the
  detector alone as scheduled by Vitis HLS; ``RTL system'' is the complete design
  that runs on the board --- detector plus HDMI receive/transmit and the overlay
  drawer --- after place-and-route. Percentages are of the target device. Block
  RAM is counted as allocated primitives (the logical footprint is
  $\approx$140\,KiB of the device's 261\,KiB, see text). The RTL's shared
  sequential multiplier collapses the detector's DSP demand to $\approx$7 of
  the system's 14 (the rest is video I/O and overlay), against 79 in the HLS
  core.}
  \label{tab:fpga}
  \begin{tabular}{lcc}
    \toprule
     & HLS core & RTL system \\
     & (Artix-7 XC7A35T) & (Spartan-6 XC6SLX45) \\
    \midrule
    Clock           & 100\,MHz, II${=}1$    & 74.25\,MHz pixel clock \\
    Slice LUTs      & 19{,}789 (95\%)       & 12{,}664 (46\%) \\
    Slice registers & 14{,}684 (35\%)       & 8{,}813 (16\%) \\
    Occupied slices & ---                   & 4{,}109 (60\%) \\
    Block RAM       & 79 (79\%)             & 74 RAMB16 $+$ 26 RAMB8 (75\%) \\
    DSP blocks      & 79 (87\%)             & 14 (24\%) \\
    Timing          & met (slack 0.02\,ns)  & all met (74.8\,MHz) \\
    \bottomrule
  \end{tabular}
\end{table}

\textbf{Verification.} The acceptance gate is three-way bit-exactness:
204{,}759 segments over the full 123-photograph Full-HD corpus, matched
segment-for-segment on the emitted integer run records (the floating-point
finalization is the same host code for all three), in the hardware
configuration (the
shipped configuration of Section~\ref{sec:method} minus sub-pixel NMS, which
exceeds the judge's 128-bit exact-integer envelope and remains software-only;
the shipped configuration itself yields 205{,}173 segments on the same
corpus). On a corpus photograph, enabling sub-pixel NMS moves endpoints by a
mean $0.064$\,px (max $0.43$) and changes the accepted count by one. The
quality cost of the omission is quantified in Appendix~\ref{app:ablation}
(the ``$-$ sub-pixel NMS'' row): the hardware configuration's synthetic
F-max and detection counts are \emph{identical} to the shipped
configuration's at every noise level; the cost is confined to per-segment
geometry --- mean direction error $0.05$--$0.08$\degr{} against the shipped
$0.02$--$0.04$\degr{}, still comparable to the best baseline's
(Table~\ref{tab:geom}) --- and, downstream, a York Urban fixed-estimator
median of $1.27$\degr{} against $1.02$\degr{} in the same run. The hardware
configuration is also the faster one in software ($8.1$ vs.\ $11.3$\,ms at
Full-HD): sub-pixel NMS is the one costly refinement.
Golden-vector parity testbenches hold every stage and the full chain to the
HLS C model and the C++ reference. The parity gate exercises the functional
pipeline with the event stream fully drained; sustained real-time throughput
is a separate question, quantified under the overload limit below. The
discipline paid off beyond its remit:
a synthesis-tool fault (an FSM re-encoding that mis-synthesized the labeling
back end while passing timing analysis and RTL simulation) was isolated by
replaying the same vectors through a gate-level simulation of the netlist and
comparing per-pass counters against the bit-exact model.

\textbf{Overload limit.} The event-driven back end labels edge pixels, not raster
pixels, so a pathologically dense frame can outrun it; the event FIFO then
sheds. After a sequence of individually measured drain optimizations (each
output-invariant in the absence of overload), a burst simulation replaying
the corpus's recorded event streams at real 1080p30 pixel timing puts the
loss at 0.48\% of corpus segments, concentrated in two frames --- no other
frame fills even a quarter of the 2{,}048-event FIFO; at $8\times$ that depth
the corpus replays losslessly, and the residue vanishes entirely on parts
with more block RAM than this 2009 device. To state the claims separately:
continuous 1080p30 video I/O with no frame buffer is demonstrated
unconditionally, and lossless detection at that rate is demonstrated on all
but the two densest corpus frames; guaranteed lossless throughput for
\emph{arbitrary} input is not claimed on this device. The service rate makes
the boundary concrete: at the measured $\approx$6 cycles per interior event,
the 1080p30 frame-cycle budget ($\approx$2.5M cycles at 74.25\,MHz) covers
$\approx$0.4M interior events per frame --- roughly a fifth of the active
pixels --- beyond which a \emph{sustained} density would shed regardless of
FIFO depth. The heaviest Full-HD photograph profiled consumed $\approx$51\%
of that budget, and the corpus replay above bounds how far real photographs
sit from the limit.

\section{Limitations}
\label{sec:limitations}

Stated plainly: (i) \sweeplsd{}'s edge model is \emph{contrast-gated}: a pixel
becomes an edge only if its gradient peak clears a threshold. Soft, wide
luminance ramps --- dim wall corners, defocused structure --- never form a
peak, so LSD and EDLines, which link pixels by \emph{orientation coherence},
recover low-contrast structure \sweeplsd{} misses; lowering the threshold does
not retrieve it (the misses are architectural, and threshold sweeps confirm
this). This is the mechanism behind the $\sim$20\% support deficit of
Section~\ref{sec:vpdecomp}, and coverage-hungry applications should weigh it.
(ii) On the synthetic F-max protocol ELSED leads at \emph{every} noise level,
and ED\_Lib is also more noise-stable than \sweeplsd{} at $\sigma \ge 10$
(Table~\ref{tab:fmax}); \sweeplsd{}'s synthetic edge is per-segment
direction, not raw F-max. (iii) The two-way direction quantization costs some
orientation isotropy: on the dense Siemens-star probe \sweeplsd{} is the
weakest of the four detectors (Table~\ref{tab:iso}). (iv)
There is no a-contrario false-detection control in the default configuration;
an optional streaming NFA validation exists but is off by default. The
line-free negative probe of Section~\ref{sec:synth} bounds the practical
consequence: near-zero false positives on pure noise at the defaults, $3.2$
per image on smooth texture plus noise --- which the optional NFA cuts to
$0.8$. (v) Speed
depends entirely on compiler auto-vectorization: of the toolchains tested,
only MSVC's \texttt{cl} fails --- one kernel, about $4\times$ slower,
identical detections --- and clang-cl covers that ABI without the cost
(Appendix~\ref{app:compiler}). (vi) On the
2009-era FPGA of Section~\ref{sec:fpga}, pathologically dense frames can
overrun the event FIFO and shed a small fraction of segments ($\approx$0.5\%
of the corpus segments, concentrated in two frames); larger parts absorb this
entirely. (vii) Repeatability
under flips and $90$\degr{}-multiple rotations \emph{is} evaluated and is a
\sweeplsd{} strength, but under arbitrary-angle rotation --- also now
evaluated (Section~\ref{sec:equiv}) --- \sweeplsd{} is the \emph{weakest}
of the four (median $39$--$48\%$ against LSD's $62$--$64\%$): the two-way
quantization that makes the axis-preserving family stable is exactly what
arbitrary rotation disturbs. Repeatability under viewpoint and
illumination change (homography-based, cf.\ ELSED's HPatches protocol)
remains unevaluated here. (viii) The raw segment endpoints are a
deterministic $\approx$1.7--2.0\,px short (Section~\ref{sec:endpoint}); this is
an extent bias, not a direction bias, so vanishing-point and attitude accuracy
are unaffected, but applications that need absolute endpoint position should
apply the angle-based correction of Section~\ref{sec:endpoint} or snap
endpoints to nearby evidence.

\section{Conclusion}
\label{sec:conclusion}

This paper gave the first complete public description of \sweeplsd{} --- a line
segment detector, designed in 2014, that runs as a true raster stream: one pass,
a few rows of state, an integer-only per-pixel core, segments emitted a median
6.5 rows after their last pixel. The description is normative for an
open-source implementation whose measured refinements --- among them
sub-pixel NMS, streaming hysteresis, and curve rejection --- all preserve
the streaming property. Streaming FPGA line detectors have since been published
(Section~\ref{sec:related}), but by a different mechanism and at lower
resolution; against the established detectors the design point holds up ---
$4.6\times$/$5.2\times$/$25\times$
faster than the original authors' ELSED/EDLines/LSD on Full-HD CPU detection, the
tightest
frame-time distribution of the four (worst frame $1.66\times$ the median),
the best per-segment direction accuracy of the four detectors evaluated, and
curve rejection by design --- with a quantified trade-off: a coverage deficit
on soft, low-contrast structure. The 2014 design's hardware claim is no longer a claim:
the same algorithm, bit-exact in its hardware configuration
(Section~\ref{sec:fpga}), detects segments in live 1080p30 video on a
2009-era FPGA
with no frame buffer and no external memory, lossless on all but the two
densest corpus frames. The downstream study contributes
an error-source decomposition that separates per-line quality from inlier
support, and evaluates every detector under a selection/evaluation-separated
best-estimator-per-detector protocol. The
implementation, the HLS and RTL sources, all baseline glue, and every
evaluation harness are open source (MIT) at
\url{https://github.com/yosh-shimizu/sweeplsd}, with documentation at
\url{https://yosh-shimizu.github.io/sweeplsd/}.

\section*{Acknowledgments}

The algorithm was designed in the author's 2014 master's thesis at Waseda
University~\cite{shimizu2014}. The reimplementation, the refinements, the
evaluation suite, and the HLS/RTL hardware implementations presented here were
developed by the author in collaboration with Claude (Anthropic) in
AI-assisted pair-work sessions; all algorithmic claims were verified against
the measurements reported in this paper, all of which are reproducible from
the repository.

\section*{Declarations}

\textbf{Funding.} The author received no financial support for the research,
authorship, or publication of this article.

\textbf{Competing interests.} The author declares no competing interests.

\textbf{Use of AI assistance.} The implementation and the evaluation suite
were developed, and this manuscript was drafted, with AI assistance (Claude,
Anthropic) in author-directed pair-work sessions, as described in the
Acknowledgments; the author verified all claims against the reported
measurements and takes full responsibility for the content.

\textbf{Data and code availability.} The implementation, the HLS and RTL
sources, all baseline glue, and every evaluation harness are openly available
under the MIT license at \url{https://github.com/yosh-shimizu/sweeplsd}. The
evaluation uses the public LIU4K-v2~\cite{liu2020} (v2 release:
\url{https://structpku.github.io/LIU4K_Dataset/}, CC0-licensed), York
Urban~\cite{denis2008}, NYU-VP~\cite{kluger2020}, EuRoC MAV~\cite{burri2016},
and TUM-VI~\cite{schubert2018} datasets, each obtained and used under its own
terms; their images are not redistributed. The synthetic attitude scenes are
the Blender demo assets ``Classroom'' (Christophe Seux, CC0) and ``Barcelona
Pavilion'' (eMirage / Hamza Cheggour, CC-BY).

\appendix

\section{The outer-ring thinning reduction}
\label{app:ring}

This appendix states in full the boolean reduction behind step 1 of the
endpoint-candidate test (Section~\ref{sec:endpoints}); the compound example of
Figure~\ref{fig:ringreduction} there is worked from these
rules.\footnote{The reduction is the thinning algorithm proposed in the
2014 thesis~\cite{shimizu2014}.} Index the outer ring
$O_0 \ldots O_{15}$ as in Figure~\ref{fig:ring}a, and the inner ring
$A_0 \ldots A_7$ the same way starting at the left-middle pixel, so that
inner position $k$ faces outer position $2k$. All values are 0/1 edge flags.
By role, the outer positions are \emph{midpoints} ($O_0, O_4, O_8, O_{12}$,
facing an inner pixel squarely), \emph{corners} ($O_2, O_6, O_{10}, O_{14}$),
and \emph{flanks} (the odd positions). The rules below are written for the
left midpoint, its flanks, and the top-left corner; the other quadrants follow by
rotation (outer indices mod 16, inner mod 8). Updates apply rule by rule ---
each rule is evaluated at all four rotations before the next, reading the
values left by the previous rule:
\begin{align*}
&\text{(1) midpoint support:} & O_0 &\leftarrow O_0 \wedge (A_7 \vee A_0 \vee A_1)\\
&\text{(2) flank support:}    & O_1 &\leftarrow O_1 \wedge (A_0 \vee A_1 \vee O_0),
                              \quad O_{15} \leftarrow O_{15} \wedge (A_7 \vee A_0 \vee O_0)\\
&\text{(3) midpoint yield:}   & O_0 &\leftarrow O_0 \wedge \neg(O_{15} \vee O_1)\\
&\text{(4) corner rule:}      & O_2 &\leftarrow O_2 \wedge \bigl(A_1 \vee (O_1 \wedge O_3)\bigr),\\
&                              &     &\quad\text{then } O_1 \leftarrow O_1 \wedge \neg O_2,\;
                              O_3 \leftarrow O_3 \wedge \neg O_2.
\end{align*}
Rules (1)--(2) are the \emph{support pruning} effect: an exit exists only
where the inner ring (or an adjacent, already-supported midpoint) connects
the outer pixel toward the center, so structure that merely clips the window
never counts. Rules (3)--(4) are the \emph{thinning} effect: within a
connected arc the midpoint yields to its flanks and a surviving corner
absorbs its flanks, leaving one representative per arc (the first clause of
rule (4) doubles as the corner's own support test; a supported arc spanning
three or more positions --- two strands leaving the same side --- keeps its
two flanks, one per strand). The count $n$ of
Section~\ref{sec:endpoints} is the sum of the sixteen reduced flags.

For $n = 2$ the classifier needs the shorter arc distance between the two
survivors on the 16-cycle. The implementation reads it off without a table:
each position $p$ carries an 8-bit \emph{thermometer mask} $T_p$ --- empty at
$p = 0$, one bit added per step around the ring until all eight are set at
$p = 8$, then one bit removed per step in the same order --- chosen so that
$\operatorname{popcount}(T_p \oplus T_q) = \min(|p-q|,\, 16-|p-q|)$. The
distance is the popcount of a single XOR. The entire reduction is evaluated
branch-free in 8-bit flags ($a \wedge \neg b$ as \texttt{a \& (b\^{}1)}),
identically in the C++ row kernel (where it auto-vectorizes), the HLS model,
and the RTL (where it is four-fold-replicated combinational logic); the
three are held bit-identical by the parity gates of Section~\ref{sec:fpga}.

\section{Per-stage timing and a two-term cost model}
\label{app:timing}

Table~\ref{tab:speed} of Section~\ref{sec:speed} reports the total per-frame
time; this appendix breaks it down by stage. The figures are measured on the
multi-pass driver, whose stages run as separable full-image passes and so can
be timed independently (the one-pass driver interleaves them by construction).
Over the LIU4K-v2 Full-HD photographs the median per-image cost splits as
in Table~\ref{tab:perstage}.

\begin{table}[h]
  \centering
  \caption{Per-stage median time (multi-pass driver, $1920\times1080$ LIU4K-v2
  photographs, single thread, AVX2). Shares are of the 13.3\,ms stage sum,
  which differs from the one-pass 11.3\,ms of Table~\ref{tab:speed} because
  the drivers have different working-set behavior, not different arithmetic.}
  \label{tab:perstage}
  \begin{tabular}{lcc}
    \toprule
    stage & median time & share \\
    \midrule
    Gaussian $5\times5$ (separable) & 1.7\,ms & 13\% \\
    gradient $2\times2$ & 2.3\,ms & 17\% \\
    thresholding + sub-pixel NMS + adaptive hysteresis & 1.4\,ms & 11\% \\
    endpoint candidates & 1.9\,ms & 14\% \\
    streaming labeler (CCA + judgment) & 6.0\,ms & 45\% \\
    \bottomrule
  \end{tabular}
\end{table}

The labeling-and-judgment stage is the largest ($\approx$45\%) and the one
that grows with scene content: it accumulates a moment fit per edge run and fires
the judge per closed run, so it scales with segment density (on sparser
imagery its share falls toward 30\%). The four front-end
row kernels (Gaussian, gradient, thresholding, endpoint) are pure byte/integer
sweeps that auto-vectorize and stay near-constant at $\approx$1.4--2.3\,ms
regardless of content --- none exceeds a fifth of the stage sum. The labeler
carries the scalar per-pixel bookkeeping (union--find links, moment
accumulation); the endpoint test carries the heaviest per-pixel logic, the
16-flag ring reduction of Appendix~\ref{app:ring}, which vectorizes --- or
fails to (Appendix~\ref{app:compiler}). Because the software stages run sequentially, these shares are
a CPU-implementation property and do not map to the hardware pipeline, where
the stages run concurrently and throughput is set by the slowest single stage
rather than their sum; the balance above simply indicates that no stage is
grossly over- or under-provisioned for that intended form.

The stage split says where a frame's time goes; a two-term model says how it
scales. Fitting the per-image one-pass time over five downscale rungs of the
corpus (615 measurements: $640\times360$ to $3840\times2160$, 123 photographs
each, per-image median of 9 runs) against pixel count and edge-pixel count
gives Table~\ref{tab:costmodel}.

\begin{table}[h]
  \centering
  \caption{Two-term frame-cost model over 615 per-image measurements (5
  resolution rungs $\times$ 123 photographs, one-pass driver; $t$ in ms).
  Standard errors: $a \pm 0.07$, $b \pm 0.5$. Freeing an
  intercept moves it to $-0.02$\,ms, so the two-term form is kept; the
  MP-only row shows the content term is not optional (its own intercept fits
  to $0.08$\,ms).}
  \label{tab:costmodel}
  \begin{tabular}{lccc}
    \toprule
    model & $a$ [ms/MP] & $b$ [$\mu$s per $10^3$ edge px] & $R^2$ \\
    \midrule
    $t = a\cdot\text{MP} + b\cdot\text{edge}$ & 2.84 & 16.0 & 0.963 \\
    $t = a\cdot\text{MP} + c$ & 5.12 & --- & 0.891 \\
    \bottomrule
  \end{tabular}
\end{table}

The two terms are the two halves of the stage table: the pixel-proportional
term covers the four front-end row kernels, which touch every pixel
regardless of content, and the content term is the labeler-and-judge half,
whose work scales with the edge pixels that reach it. (Label lifetime events
--- creates, merges, judgments --- fit marginally better at $R^2 = 0.967$,
but edge count is externally observable, so the model is stated in it.)
Pooling five rungs inflates $R^2$; within a fixed rung the same fit still
explains $0.55$--$0.89$ of the between-image variance (most at 720p--1440p,
least at the extremes), and the dispersion check that follows is the
practical form of that within-rung test. The
model also accounts for the frame-time dispersion of Table~\ref{tab:spread}:
from the corpus's edge-pixel distribution alone it predicts a CV of
$\approx$20\% and a worst frame $\approx$1.5$\times$ the median at Full-HD
--- most of the measured 24.4\% and 1.66$\times$. The spread is the content
term at work, not timing noise.

\section{Compiler sensitivity of the raster sweep}
\label{app:compiler}

Section~\ref{sec:speed} noted that the no-intrinsics design makes
\sweeplsd{}'s speed strongly compiler-dependent; this appendix records the
toolchain study. The detector was built with five toolchains from identical
sources at the same ISA target and run interleaved image-by-image in one
window
(Table~\ref{tab:toolchains}); only MSVC's \texttt{cl} fails to vectorize the
endpoint-candidate kernel.

\begin{table}[h]
  \centering
  \caption{The five-toolchain study: identical sources, same ISA target, run
  interleaved image-by-image in one measurement window. Times relative to
  GCC 15.2 ($\sim$12\,ms one-pass at Full-HD in this five-binary interleaved
  window; cf.\ the 11.3\,ms of Table~\ref{tab:speed}); detections are
  bit-identical across all toolchains.}
  \label{tab:toolchains}
  \begin{tabular}{llc}
    \toprule
    toolchain & vectorizes & relative time \\
    \midrule
    GCC 15.2 & all five stage kernels & 1$\times$ \\
    GCC 8.1 & all five, less well & 1.11$\times$ \\
    Clang 22.1 & all five & 1.35$\times$ \\
    clang-cl 22.1 & all five & 1.38$\times$ \\
    MSVC \texttt{cl} 19.34 (VS 2022) & endpoint kernel stays scalar & 4.1$\times$ \\
    \bottomrule
  \end{tabular}
\end{table}

That single kernel accounts for essentially the whole gap, and the
evidence points at the vectorizer rather than the formulation: \texttt{cl}
declines even a reduced probe of the same dataflow shape, and ISA flags
change nothing. The one failure that proved fixable is recorded because it
is actionable: Clang originally paid a similar penalty for a different
reason --- its inliner kept the ring test as a per-pixel call, which blocked
vectorization of the row loop outright --- and a single force-inline
annotation recovers full speed with bit-identical output and no effect on
GCC. In our measurements the baselines are insensitive to compiler
generation (their anchor-chaining inner loops are branch-bound), so the
sensitivity is specific to the raster-sweep design. Detections are unaffected throughout:
the output is bit-identical across all five toolchains, verified
float-for-float in exact hexadecimal form (410{,}346 segment records --- the
205{,}173 segments of Section~\ref{sec:memlat} from each of the two drivers
--- over the 123 photographs). Practically, we recommend GCC $\ge$ 15 for
the MinGW ABI and clang-cl for the MSVC ABI.

\section{The refinements, ablated}
\label{app:ablation}

Contribution 2 claims that each refinement of the 2014 design was adopted
only after measuring its effect; this appendix consolidates those
measurements into one table. Every row is a \sweeplsd{} configuration built
from the library's individual refinement switches
(Table~\ref{tab:ablation}): the upper block adds the refinements
cumulatively onto the 2014 pipeline, the lower block removes each one from
the shipped configuration (leave-one-out). Columns: F-max and, on matched
segments, mean direction error, under the protocol of
Section~\ref{sec:synth} (the same generator, matcher, knob sweep, and
20 scenes per condition as Table~\ref{tab:fmax}, whose \sweeplsd{} row this
block's shipped row reproduces exactly); the fixed-estimator York Urban median of
Section~\ref{sec:vp}, all rows from one run (the shipped row's
$1.02$\degr{} reproduces Table~\ref{tab:fixedest}'s $1.04$\degr{} within
the re-run scatter noted there); and the corpus-median segment count and
one-pass detection time on the 123 Full-HD photographs of
Section~\ref{sec:speed} (the shipped row reproduces the headline: 1{,}591
segments, $11.3$\,ms). The switches and the harness ship in the repository.

\begin{table}[h]
  \centering\small
  \caption{Refinement ablation. Upper block: cumulative from the 2014
  pipeline; lower block: shipped configuration minus one refinement.
  ``dir.'' and ``lat.'' are mean direction / lateral error on matched
  segments (ranges span $\sigma = 0$--$20$); ``York U.''\ is the
  fixed-estimator York Urban median of Section~\ref{sec:vp}; segs / time are
  corpus medians at Full-HD (one-pass
  driver, 5 runs per image). $^{\dagger}$``2014 pipeline'' includes the
  half-pixel lattice bookkeeping of Section~\ref{sec:edge} (a coordinate
  convention, not a refinement); the as-implemented 2014 configuration
  differs from that row only in a $0.46$\,px lateral bias --- the same bias
  the ``$-$ lattice'' row isolates.}
  \label{tab:ablation}
  \footnotesize
  \setlength{\tabcolsep}{3.4pt}
  \begin{tabular}{lccccccc}
    \toprule
    configuration & \multicolumn{2}{c}{F-max} & dir.\ (\degr) & lat.\ (px)
      & York U.\ (\degr) & segs & time \\
    \cmidrule(lr){2-3}
    & $\sigma{=}0$ & $\sigma{=}20$ & & & & & (ms) \\
    \midrule
    2014 pipeline$^{\dagger}$        & 0.951 & 0.898 & 0.05--0.08 & 0.10--0.11 & 1.25 & 1{,}646 & 7.8 \\
    $+$ strict NMS tie-break         & 0.951 & 0.901 & 0.05--0.08 & 0.10--0.11 & 1.24 & 1{,}621 & 7.6 \\
    $+$ streaming hysteresis         & 0.951 & 0.901 & 0.05--0.08 & 0.10--0.11 & 1.18 & 1{,}621 & 8.1 \\
    $+$ sub-pixel NMS                & 0.951 & 0.901 & 0.02--0.04 & 0.12--0.13 & 1.18 & 1{,}621 & 11.2 \\
    $+$ proj.-extreme endpoints      & 0.951 & 0.901 & 0.02--0.04 & 0.12--0.13 & 1.12 & 1{,}621 & 11.3 \\
    $+$ curve rejection $=$ \textbf{shipped} & \textbf{0.958} & \textbf{0.905} & \textbf{0.02--0.04} & 0.12--0.13 & \textbf{1.02} & 1{,}591 & 11.3 \\
    \midrule
    \multicolumn{8}{l}{\emph{shipped configuration minus one refinement:}} \\
    $-$ tie-break                    & 0.958 & 0.903 & 0.02--0.04 & 0.12--0.13 & 1.10 & 1{,}623 & 11.3 \\
    $-$ hysteresis                   & 0.958 & 0.905 & 0.02--0.04 & 0.12--0.13 & 1.03 & 1{,}591 & 10.7 \\
    $-$ sub-pixel NMS (= HW config.) & 0.958 & 0.905 & 0.05--0.08 & 0.10--0.11 & 1.27 & 1{,}586 & 8.1 \\
    $-$ proj.-extreme endpoints      & 0.958 & 0.905 & 0.02--0.04 & 0.12--0.13 & 1.02 & 1{,}591 & 11.2 \\
    $-$ curve rejection              & 0.951 & 0.901 & 0.02--0.04 & 0.12--0.13 & 1.12 & 1{,}621 & 11.4 \\
    $-$ lattice bookkeeping$^{\dagger}$ & 0.958 & 0.905 & 0.02--0.04 & 0.46 & 1.05 & 1{,}591 & 11.3 \\
    \bottomrule
  \end{tabular}
\end{table}

Read by axis, the attribution is clean. \textbf{Curve rejection} is the
only F-max mover on the standard bars ($+0.007$ clean) and worth
$0.10$\degr{} downstream --- besides being the ``0 segments on circles''
mechanism of Section~\ref{sec:isotropy} --- at zero cost.
\textbf{The strict tie-break} adds $0.003$ F-max under heavy noise and
$0.08$\degr{} downstream, and is free (thinner plateaus mean marginally
fewer labels). \textbf{Sub-pixel NMS} never moves F-max; it is a pure
geometry refinement --- direction error $0.05$--$0.08 \to 0.02$--$0.04$\degr{}
--- and the single largest downstream contributor ($+0.25$\degr{} when
removed), and it is also the one costly refinement ($+3.2$\,ms, $\approx$28\%
of the frame: the per-edge-pixel interpolation and the wider fixed-point
moment arithmetic it forces). The hardware configuration
(Section~\ref{sec:fpga}) omits exactly this row.
\textbf{Projection-extreme endpoints} show no effect on these protocols,
and a dedicated junction probe explains why. On 200 scenes per condition
of a long subject bar crossed by wide occluding bars (X), abutted by them
(T), or grazed by bar tips stopping $1$--$2.5$\,px short of its flank ---
the more-than-two-candidate contact case of Section~\ref{sec:judge} ---
with and without noise, the endpoint extent error of the two rules is
\emph{identical} on all but one of $\approx$5{,}800 paired flank
intervals: on a monotone straight run the first and last recorded
contacts already bracket the extent, and both rules project the chosen
pair onto the same fitted axis. The refinement is thus a guard for
components whose contact pair does \emph{not} bracket their extent
(merge-path emissions, jagged runs at junction clusters --- present on
real photographs, unexercised by straight-bar probes); it costs
$0.2$\,ms, and its only measured trace is cumulative on York Urban
($1.18 \to 1.12$\degr{}, not reproduced leave-one-out). The probe
harness ships in the repository (\texttt{--junctions}). \textbf{The lattice bookkeeping} is purely
lateral ($0.46 \to 0.10$\,px).

Two effects deserve explicit note. First, \textbf{the streaming hysteresis
is invisible in every column above} --- the standard bars sit 170 gray
levels above the background, so the low threshold never binds. Its axis is
faint structure: re-running the same protocol with the bar contrast lowered
to 12 gray levels, the clean-scene F-max is $0.960$ with hysteresis against
$0.898$ without (recall $0.975$ vs.\ $0.885$), while at $\sigma = 5$ the two
configurations tie ($0.582$ vs.\ $0.588$) --- precisely the ``raises recall
on clean images without flooding noisy ones'' behavior claimed in
Section~\ref{sec:edge}, bought for $0.5$\,ms. Second, the two views
disagree instructively about sub-pixel NMS downstream ($+0.00$\degr{} when
added before curve rejection, $+0.25$\degr{} when removed from the shipped
configuration): while arc fragments remain in the segment set, their
misdirection dominates the estimator's error budget and masks the moment
precision; once curve rejection removes them, that precision is
load-bearing. The refinements compose --- their joint downstream value
($1.25 \to 1.02$\degr{}) exceeds the sum of their visible single steps.

\section{The gap-tolerant collinear linker}
\label{app:linker}

The optional linker (off by default; exercised only where
Sections~\ref{sec:synth} and~\ref{sec:vp} say so) re-assembles collinear
fragments that junction cuts and noise breaks leave behind. It is a
finalization-level stage in the strict sense: it consumes only segments the
judge has already accepted, in emission order, and never touches pixels or
rows, so the streaming property of Section~\ref{sec:onepass} is untouched
by construction. This appendix specifies it completely; the four
parameters and their defaults are in Table~\ref{tab:params}.

\textbf{State and window.} The linker holds a list of \emph{active
chains}, each still a candidate for extension: the chain's current span (a
segment), its unit direction, the row of its lowest endpoint, and a
provisional flag (below). When the judge emits a segment at sweep row $y$,
every active chain whose lowest endpoint is more than
$\lceil \mathrm{gap} \rceil + 1$ rows above $y$ is first flushed to the
output --- nothing that closes below the sweep can reach it anymore --- and
the new segment then tries to extend the survivors. A chain thus becomes
immutable once the sweep is $\lceil \mathrm{gap} \rceil + 1$ rows past its
lowest endpoint (it is physically written out with the next emission, at
the latest when the frame ends). The list holds at most the chains closed
within that sliding window: $O(\text{width})$ state at the fixed default
gap, bounded the same way the label pool is
(Section~\ref{sec:labeling}).

\textbf{Link test.} A new segment $s$ may absorb an active chain $t$ iff
(i) their directions differ by at most $4$\degr{}; (ii) \emph{mutual
lateral consistency}: every endpoint of each lies within $1$\,px of the
\emph{other's} infinite line --- being parallel is not enough, and without
this test a gap larger than the bar width fuses the two parallel flanks of
a thin bar into one diagonal; (iii) their smallest endpoint-to-endpoint
distance is at most $9$\,px; and (iv) the \emph{merged} span again lies
within $4$\degr{} of both parts.

\textbf{Merge, transitivity, conflicts.} The merged span is the
farthest-apart pair among the four endpoints, and the absorbed chain
leaves the list. After each absorption the (now longer) segment re-scans
the list, so linking is transitive --- $k$ collinear fragments assemble in
$k-1$ absorptions --- and conflicts resolve greedily and deterministically
in emission order: the first eligible candidate in list order is taken.

\textbf{Two-stage admission.} With the linker on, the judge admits
fragments down to $5$ pixels (instead of $N_{\mathrm{th}} = 15$), so
noise-broken runs can reassemble \emph{before} the length test rather than
dying piecemeal --- the streaming analogue of ELSED's gap jumping. Chains
carry a provisional flag: a chain none of whose members cleared
$N_{\mathrm{th}}$ on its own is kept at flush only if its Chebyshev
\emph{span} clears $N_{\mathrm{th}}$. Span rather than accumulated pixel
count is deliberate --- a pixel-count gate re-admits dense, wiggly chains
of noise fragments (many pixels, short span) and measurably floods
precision at high noise. A segment accepted at the full threshold is never
dropped.

Costs and effects are reported where they occur: the F-max recovery in
Section~\ref{sec:synth}, the downstream gain (NYU $5.23$\degr{}) and the
$\approx$0.6\,ms detection cost in Section~\ref{sec:vp}.

\section{Default configuration}
\label{app:params}

For reference and reproducibility, Table~\ref{tab:params} collects the
algorithmic defaults of the shipped detector in one place, each with the
section that describes it. \sweeplsd{} runs in this single configuration throughout
the paper; the linker rows apply only when the optional gap-tolerant linker of
Section~\ref{sec:synth} is enabled (off by default; specified in
Appendix~\ref{app:linker}).

\begin{table}[h]
  \centering
  \caption{Default parameters of the shipped configuration (8-bit
  grayscale input).}
  \label{tab:params}
  \small
  \begin{tabular}{llll}
    \toprule
    stage & parameter & default & \S \\
    \midrule
    edge & gradient power threshold $g_{\mathrm{th}}$ & 256 at the $64\times$ scale ($\approx$4 gray levels) & \ref{sec:edge} \\
    edge & hysteresis low threshold & adaptive ($\approx 2\times$ the 80th-percentile power), & \ref{sec:edge} \\
         &                          & \quad clamped to $[120,\, g_{\mathrm{th}}]$ & \\
    edge & adaptive-histogram decay & $\times(1-2^{-8})$ per row & \ref{sec:edge} \\
    edge & sub-pixel NMS resolution & $1/16$\,px & \ref{sec:edge} \\
    edge & border exclusion & 3\,px & \ref{sec:edge} \\
    labeling & label pool & starts $W/4$, grows to $\lceil W/2 \rceil$ & \ref{sec:labeling} \\
    judge & pixel count $N_{\mathrm{th}}$ & 15 & \ref{sec:judge} \\
    judge & strong-pixel minimum & 3 & \ref{sec:judge} \\
    judge & eigenvalue ratio $\beta$ & 0.05 & \ref{sec:judge} \\
    judge & perpendicular RMS bound & 1\,px & \ref{sec:judge} \\
    linker & direction difference & $\le 4$\degr{} & \ref{sec:synth} \\
    linker & endpoint gap & $\le 9$\,px & \ref{sec:synth} \\
    linker & lateral consistency & $\le 1$\,px & \ref{sec:synth} \\
    linker & fragment admission & $\ge 5$\,px ($N_{\mathrm{th}}$ applies to the linked chain) & \ref{sec:synth} \\
    \bottomrule
  \end{tabular}
\end{table}

\bibliographystyle{plain}
\bibliography{references}

\begin{thebibliography}{10}

\bibitem{akinlar2011}
Cuneyt Akinlar and Cihan Topal.
\newblock {EDLines}: A real-time line segment detector with a false detection
  control.
\newblock {\em Pattern Recognition Letters}, 32(13):1633--1642, 2011.

\bibitem{almazan2017}
Emilio~J. Almaz{\'a}n, Ron Tal, Yiming Qian, and James~H. Elder.
\newblock {MCMLSD}: A dynamic programming approach to line segment detection.
\newblock In {\em IEEE Conference on Computer Vision and Pattern Recognition
  (CVPR)}, pages 5854--5862, 2017.

\bibitem{bailey2011}
Donald~G. Bailey.
\newblock {\em Design for Embedded Image Processing on {FPGAs}}.
\newblock John Wiley \& Sons, 2011.

\bibitem{bailey2007}
Donald~G. Bailey and Christopher~T. Johnston.
\newblock Single pass connected components analysis.
\newblock In {\em Image and Vision Computing New Zealand (IVCNZ)}, pages
  282--287, 2007.

\bibitem{bailey2019}
Donald~G. Bailey and Michael~J. Klaiber.
\newblock Zig-zag based single-pass connected components analysis.
\newblock {\em Journal of Imaging}, 5(4):45, 2019.

\bibitem{burns1986}
J.~Brian Burns, Allen~R. Hanson, and Edward~M. Riseman.
\newblock Extracting straight lines.
\newblock {\em IEEE Transactions on Pattern Analysis and Machine Intelligence},
  8(4):425--455, 1986.

\bibitem{burri2016}
Michael Burri, Janosch Nikolic, Pascal Gohl, Thomas Schneider, Joern Rehder,
  Sammy Omari, Markus~W. Achtelik, and Roland Siegwart.
\newblock The {EuRoC} micro aerial vehicle datasets.
\newblock {\em The International Journal of Robotics Research},
  35(10):1157--1163, 2016.

\bibitem{canny1986}
John Canny.
\newblock A computational approach to edge detection.
\newblock {\em IEEE Transactions on Pattern Analysis and Machine Intelligence},
  8(6):679--698, 1986.

\bibitem{caprile1990}
Bruno Caprile and Vincent Torre.
\newblock Using vanishing points for camera calibration.
\newblock {\em International Journal of Computer Vision}, 4(2):127--139, 1990.

\bibitem{cho2018}
Nam-Gyu Cho, Alan~L. Yuille, and Seong-Whan Lee.
\newblock A novel linelet-based representation for line segment detection.
\newblock {\em IEEE Transactions on Pattern Analysis and Machine Intelligence},
  40(5):1195--1208, 2018.

\bibitem{denis2008}
Patrick Denis, James~H. Elder, and Francisco~J. Estrada.
\newblock Efficient edge-based methods for estimating {Manhattan} frames in
  urban imagery.
\newblock In {\em European Conference on Computer Vision (ECCV)}, pages
  197--210, 2008.

\bibitem{desolneux2000}
Agn{\`e}s Desolneux, Lionel Moisan, and Jean-Michel Morel.
\newblock Meaningful alignments.
\newblock {\em International Journal of Computer Vision}, 40(1):7--23, 2000.

\bibitem{gomezojeda2019}
Ruben Gomez-Ojeda, Francisco-Angel Moreno, David Zu{\~n}iga-No{\"e}l, Davide
  Scaramuzza, and Javier Gonzalez-Jimenez.
\newblock {PL-SLAM}: A stereo {SLAM} system through the combination of points
  and line segments.
\newblock {\em IEEE Transactions on Robotics}, 35(3):734--746, 2019.

\bibitem{gu2022}
Geonmo Gu, Byungsoo Ko, SeoungHyun Go, Sung-Hyun Lee, Jingeun Lee, and Minchul
  Shin.
\newblock Towards light-weight and real-time line segment detection.
\newblock In {\em AAAI Conference on Artificial Intelligence (AAAI)}, pages
  726--734, 2022.

\bibitem{huang2018}
Kun Huang, Yifan Wang, Zihan Zhou, Tianjiao Ding, Shenghua Gao, and Yi~Ma.
\newblock Learning to parse wireframes in images of man-made environments.
\newblock In {\em IEEE/CVF Conference on Computer Vision and Pattern
  Recognition (CVPR)}, pages 626--635, 2018.

\bibitem{jalilvand2026}
Amir~Hossein Jalilvand, Parsa Hassani~Shariat Panahi, and M.~Hassan Najafi.
\newblock A low-latency {ASIC} architecture for real-time line segment
  detection, 2026.
\newblock arXiv:2608.06439.

\bibitem{janampa2024}
Sebastian Janampa and Marios Pattichis.
\newblock {DT-LSD}: Deformable transformer-based line segment detection, 2024.
\newblock arXiv:2411.13005.

\bibitem{ke2025}
Zeran Ke, Bin Tan, Xianwei Zheng, Yujun Shen, Tianfu Wu, and Nan Xue.
\newblock {ScaleLSD}: Scalable deep line segment detection streamlined.
\newblock In {\em IEEE/CVF Conference on Computer Vision and Pattern
  Recognition (CVPR)}, 2025.

\bibitem{kluger2020}
Florian Kluger, Eric Brachmann, Hanno Ackermann, Carsten Rother, Michael~Ying
  Yang, and Bodo Rosenhahn.
\newblock {CONSAC}: Robust multi-model fitting by conditional sample consensus.
\newblock In {\em IEEE/CVF Conference on Computer Vision and Pattern
  Recognition (CVPR)}, pages 4634--4643, 2020.

\bibitem{lin2024}
Xinyu Lin, Yingjie Zhou, Yipeng Liu, and Ce~Zhu.
\newblock A comprehensive review of image line segment detection and
  description: Taxonomies, comparisons, and challenges.
\newblock {\em IEEE Transactions on Pattern Analysis and Machine Intelligence},
  2024.

\bibitem{liu2020}
Jiaying Liu, Dong Liu, Wenhan Yang, Sifeng Xia, Xiaoshuai Zhang, and Yuanying
  Dai.
\newblock A comprehensive benchmark for single image compression artifact
  reduction.
\newblock {\em IEEE Transactions on Image Processing}, 29:7845--7860, 2020.

\bibitem{lu2013}
Xiaofeng Lu, Li~Song, Sumin Shen, Kang He, Songyu Yu, and Nam Ling.
\newblock Parallel {Hough} transform-based straight line detection and its
  {FPGA} implementation in embedded vision.
\newblock {\em Sensors}, 13(7):9223--9247, 2013.

\bibitem{lu2015}
Xiaohu Lu, Jian Yao, Kai Li, and Li~Li.
\newblock {CannyLines}: A parameter-free line segment detector.
\newblock In {\em IEEE International Conference on Image Processing (ICIP)},
  pages 507--511, 2015.

\bibitem{manabe2022}
Taito Manabe, Taichi Katayama, and Yuichiro Shibata.
\newblock {FPGA} implementation of a stream-based real-time hardware line
  segment detector.
\newblock {\em IEICE Transactions on Fundamentals of Electronics,
  Communications and Computer Sciences}, E105.A(3):468--477, 2022.

\bibitem{ossimitz2021}
Christoph Ossimitz and Nima Taherinejad.
\newblock A fast line segment detector using approximate computing.
\newblock In {\em IEEE International Symposium on Circuits and Systems
  (ISCAS)}, pages 1--5, 2021.

\bibitem{panahi2026}
Parsa Hassani~Shariat Panahi, Amir~Hossein Jalilvand, and M.~Hassan Najafi.
\newblock {MiLSD}: A micro line-segment detector for resource-constrained
  devices, 2026.
\newblock arXiv:2607.06600.

\bibitem{pautrat2023}
R{\'e}mi Pautrat, Daniel Barath, Viktor Larsson, Martin~R. Oswald, and Marc
  Pollefeys.
\newblock {DeepLSD}: Line segment detection and refinement with deep image
  gradients.
\newblock In {\em IEEE/CVF Conference on Computer Vision and Pattern
  Recognition (CVPR)}, pages 17327--17336, 2023.

\bibitem{pautrat2021}
R{\'e}mi Pautrat, Juan-Ting Lin, Viktor Larsson, Martin~R. Oswald, and Marc
  Pollefeys.
\newblock {SOLD2}: Self-supervised occlusion-aware line description and
  detection.
\newblock In {\em IEEE/CVF Conference on Computer Vision and Pattern
  Recognition (CVPR)}, pages 11368--11378, 2021.

\bibitem{schubert2018}
David Schubert, Thore Goll, Nikolaus Demmel, Vladyslav Usenko, J{\"o}rg
  St{\"u}ckler, and Daniel Cremers.
\newblock The {TUM} {VI} benchmark for evaluating visual-inertial odometry.
\newblock In {\em IEEE/RSJ International Conference on Intelligent Robots and
  Systems (IROS)}, pages 1680--1687, 2018.

\bibitem{shimizu2014}
Yoshiyasu Shimizu.
\newblock A frame-buffer-free one-pass line segment extraction method.
\newblock Master's thesis, Waseda University, 2014.
\newblock In Japanese; title translated by the author.

\bibitem{silberman2012}
Nathan Silberman, Derek Hoiem, Pushmeet Kohli, and Rob Fergus.
\newblock Indoor segmentation and support inference from {RGBD} images.
\newblock In {\em European Conference on Computer Vision (ECCV)}, pages
  746--760, 2012.

\bibitem{suarez2022}
Iago Su{\'a}rez, Jos{\'e}~M. Buenaposada, and Luis Baumela.
\newblock {ELSED}: Enhanced line {SE}gment drawing.
\newblock {\em Pattern Recognition}, 127:108619, 2022.

\bibitem{topal2012}
Cihan Topal and Cuneyt Akinlar.
\newblock Edge drawing: A combined real-time edge and segment detector.
\newblock {\em Journal of Visual Communication and Image Representation},
  23(6):862--872, 2012.

\bibitem{vongioi2010}
Rafael~Grompone von Gioi, J{\'e}r{\'e}mie Jakubowicz, Jean-Michel Morel, and
  Gregory Randall.
\newblock {LSD}: A fast line segment detector with a false detection control.
\newblock {\em IEEE Transactions on Pattern Analysis and Machine Intelligence},
  32(4):722--732, 2010.

\bibitem{vongioi2012}
Rafael~Grompone von Gioi, J{\'e}r{\'e}mie Jakubowicz, Jean-Michel Morel, and
  Gregory Randall.
\newblock {LSD}: a line segment detector.
\newblock {\em Image Processing On Line}, 2:35--55, 2012.

\bibitem{wang2024}
Zikai Wang, Baojiang Zhong, Xueyuan Chen, and Hangjia Zheng.
\newblock {MPG-LSD}: A high-quality line segment detector based on multi-scale
  perceptual grouping.
\newblock {\em Pattern Recognition}, 149:110286, 2024.

\bibitem{xu2021}
Yifan Xu, Weijian Xu, David Cheung, and Zhuowen Tu.
\newblock Line segment detection using transformers without edges.
\newblock In {\em IEEE/CVF Conference on Computer Vision and Pattern
  Recognition (CVPR)}, 2021.

\bibitem{xue2020}
Nan Xue, Tianfu Wu, Song Bai, Fu-Dong Wang, Gui-Song Xia, Liangpei Zhang, and
  Philip H.~S. Torr.
\newblock Holistically-attracted wireframe parsing.
\newblock In {\em IEEE/CVF Conference on Computer Vision and Pattern
  Recognition (CVPR)}, 2020.

\bibitem{xue2023}
Nan Xue, Tianfu Wu, Song Bai, Fu-Dong Wang, Gui-Song Xia, Liangpei Zhang, and
  Philip H.~S. Torr.
\newblock Holistically-attracted wireframe parsing: From supervised to
  self-supervised learning.
\newblock {\em IEEE Transactions on Pattern Analysis and Machine Intelligence},
  2023.

\bibitem{zhang2021}
Yongjun Zhang, Dong Wei, and Yansheng Li.
\newblock {AG3line}: Active grouping and geometry-gradient combined validation
  for fast line segment extraction.
\newblock {\em Pattern Recognition}, 113:107834, 2021.

\bibitem{zhou2018}
Fuqiang Zhou, Yu~Cao, and Xinming Wang.
\newblock Fast and resource-efficient hardware implementation of modified line
  segment detector.
\newblock {\em IEEE Transactions on Circuits and Systems for Video Technology},
  28(11):3262--3273, 2018.

\bibitem{zhou2019}
Yichao Zhou, Haozhi Qi, and Yi~Ma.
\newblock End-to-end wireframe parsing.
\newblock In {\em IEEE/CVF International Conference on Computer Vision (ICCV)},
  2019.

\end{thebibliography}

\end{document}